\documentclass[11pt,a4paper]{article}
\usepackage{jcappub}

\usepackage{amssymb,amsmath,amstext,amsfonts}
\usepackage{bbold}
\usepackage{bm}
\usepackage{graphicx}
\usepackage{xcolor}
\usepackage[normalem]{ulem}
\usepackage{overpic}
\usepackage{subcaption}
\usepackage{tikz}

\newcommand{\beq}{\begin{equation}}
\newcommand{\eeq}{\end{equation}}
\newcommand{\bal}{\begin{aligned}}
\newcommand{\eal}{\end{aligned}}

\newcommand{\Pzero}{\mathcal{P}_{0}}
\newcommand{\Pzeta}{\mathcal{P}_{\zeta}}
\newcommand{\PCMB}{{\cal P}_{{\textsc{cmb}}}}
\newcommand{\Pbar}{\overline{\mathcal{P}}_{\zeta}}

\newcommand{\omegalin}{\omega_{\textrm{lin}}}
\newcommand{\omegagwlin}{\omega_{\textrm{lin}}^{\textsc{gw}}}
\newcommand{\omegalog}{\omega_{\textrm{log}}}

\newcommand{\omegalogc}{\omega_{\textrm{log,c}}}
\newcommand{\omegalogd}{\omega_{\textrm{log,d}}}
\newcommand{\Alog}{A_{\textrm{log}}}
\newcommand{\thetalog}{\vartheta_{\textrm{log}}}

\newcommand{\OGW}{\Omega_\textrm{GW}}
\newcommand{\kref}{k_\textrm{ref}}
\newcommand{\TRD}{\mathcal{T}}

\title{Resonant features in the stochastic gravitational wave background}

\author{Jacopo Fumagalli,}
\author{S\'{e}bastien Renaux-Petel}
\author{and Lukas~T.~Witkowski}

\affiliation{Institut d'Astrophysique de Paris, GReCO, UMR 7095 du CNRS et de Sorbonne Universit\'{e},\\ 98bis
boulevard Arago, 75014 Paris, France}

\emailAdd{jacopo.fumagalli@iap.fr}
\emailAdd{renaux@iap.fr}
\emailAdd{lukas.witkowski@iap.fr}

\abstract{We analyse the post-inflationary contribution to the stochastic gravitational wave background due to a resonant feature in the scalar power spectrum, which is characterised by an oscillation in $\log(k)$, complementing our previous work arXiv:2012.02761 on sharp features. Primordial features signal departures of inflation from the single-field slow-roll paradigm and are motivated by embeddings of inflation in high energy physics. We find that the oscillation in the scalar power spectrum leads to a corresponding modulation in the gravitational wave spectrum that can be understood as a superposition of two oscillatory pieces, one with the original frequency of the scalar oscillations, and one with double frequency. For oscillations with slowly-varying amplitude this oscillatory part can be computed semi-analytically. Our results can be used as templates for the reconstruction of the signal from future data and permit extracting information about the small scale scalar power spectrum from measurements of the stochastic gravitational wave background.}

\begin{document}

\maketitle

\section{Introduction}
\label{sec:intro}
Cosmological inflation refers to the era of accelerated expansion in the very early universe that sets the initial conditions for the subsequent hot big bang phase. Initially proposed to solve conceptual problems of the hot big bang model, inflation is supported by observations of the Cosmic Microwave background (CMB) and Large-scale-structure (LSS) surveys and has thus become the standard paradigm for the very early universe. 

Inflation is tested experimentally through its predictions for primordial fluctuations, which are constrained by their effects on the CMB and LSS. For scalar fluctuations, observations require these to be effectively Gaussian-distributed and to possess a nearly scale-invariant power spectrum $\Pzeta(k)$ with amplitude $\PCMB \sim 10^{-9}$. What is important to note is that  current data is only effective in constraining primordial scalar fluctuations at large scales ($\gtrsim 1 \textrm{ Mpc}$), corresponding to modes that exited the horizon some 50-60 $e$-folds before the end of inflation. At smaller scales ($\ll 1 \textrm{ Mpc}$) the tight CMB and LSS constraints do not apply and less is known experimentally about the scalar power spectrum. Inflation can be modelled theoretically in terms of a scalar field, the inflaton, coupled to gravity, and slowly rolling down a very flat potential. In this scenario, the sourced primordial fluctuations are well in agreement with large scales observations. However, embeddings of inflation in theories of high energy physics suggest a much richer phenomenology. Inflation typically operates at energies where on general grounds one expects new particles and interactions beyond the Standard Model ones to be present. This is for example the case in string theory compactifications, where one generically obtains a multitude of scalar fields that can be dynamical during inflaton. 
The effect of these additional fields and interactions is that inflation in these setting deviates episodically from single-field slow-roll. 

Such departures from single-field slow-roll inflation often give rise to corresponding effects in the scalar power spectrum, also known as `primordial features' (see the reviews \cite{Chen:2010xka,Chluba:2015bqa,Slosar:2019gvt}). These manifest themselves as an oscillation in $\Pzeta(k)$ that can be either periodic in $k$ (`sharp feature') or $\log(k)$ (`resonant feature') or a combination of the two. At small scales these oscillations are not constrained and their relative amplitudes can potentially be $\mathcal{O}(1)$. A sharp feature is observed to arise whenever there is some sharp transition that occurs during inflation. This can be due to sudden changes in the potential or other Lagrangian parameters in single-field inflation \cite{Starobinsky:1992ts,Adams:2001vc, Bean:2008na,Adshead:2011jq,Miranda:2012rm,Bartolo:2013exa, Palma:2014hra, Ballesteros:2018wlw, Kefala:2020xsx}, or the result of brief transient phenomena in multi-field settings such as sharp turns \cite{Achucarro:2010da,Shiu:2011qw, Gao:2012uq, Palma:2020ejf, Fumagalli:2020adf, Fumagalli:2020nvq, Braglia:2020taf}. In contrast, a resonant feature is typically induced by an oscillation in the background with a frequency larger than the Hubble scale \cite{Chen:2008wn}. This e.g.~arises in models of axion monodromy inflation \cite{Silverstein:2008sg,Flauger:2009ab}, but also due to resonance effects in multi-field models with multiple turns \cite{Gao:2015aba}. Oscillations with $\log(k)$-periodicity typical for a resonant feature also frequently occur after the initial sudden perturbation of a sharp feature \cite{Chen:2011zf,Chen:2014joa,Chen:2014cwa,Braglia:2020taf} (e.g.~when a heavy field begins oscillating after a sharp turn in the inflationary trajectory).

Common to all these feature models is that over some range of scales the scalar power spectrum can be captured by the following templates:
\begin{align}
    \label{eq:P_of_k_sharp}
    \textrm{Sharp:} \quad & \Pzeta(k)=\Pbar(k) \Big[1 + A_{\textrm{lin}} \cos \big(\omega_{\textrm{lin}} k + \vartheta_{\textrm{lin}} \big) \Big] \, , \\
    \label{eq:P_of_k_resonant}
    \textrm{Resonant:} \quad & \Pzeta(k)=\Pbar(k) \Big[1 + \Alog \cos \big(\omegalog \log(k/k_\textrm{ref}) + \thetalog \big) \Big] \, ,
\end{align}
i.e.~the feature manifests itself as a sinusoidal modulation of a smooth envelope $\Pbar$. For a sharp feature the frequency of oscillation $\omegalin$ is related to the scale $k_\textrm{f}$ of the feature as $\omegalin \simeq 2 / k_\textrm{f}$, where $k_\textrm{f}$ should be understood as the wavenumber of the mode exiting the horizon at the time of the sudden transition. For a resonant feature one has $\omegalog \sim M /H$, where $M$ is the frequency of the background oscillations. Thus, in general $\omegalog \gtrsim 1$ and in models with a large hierarchy of masses one obtains $\omegalog \gg 1$.

An important question then is how inflation models with small scale features can be tested experimentally. Firstly, if scalar fluctuations are significantly enhanced ($\Pzeta \sim 10^7 \PCMB$) this induces overdensities that collapse into primordial black holes (PBHs) in the post-inflationary epoch, \cite{Hawking:1971ei, Carr:1974nx}, and which can make up a significant fraction of dark matter today. The abundance of PBHs can then be constrained through various effects depending on the mass of the PBHs produced, see \cite{Carr:2020gox} for a recent review. Secondly, scalar fluctuations induce gravitational waves (GWs) through non-linear effects, \cite{Acquaviva:2002ud, Mollerach:2003nq, Ananda:2006af, Baumann:2007zm}, and hence features model can be constrained through their GW signal. These GWs are sourced both during inflation and again in the post-inflationary era, which contribute to the stochastic gravitational wave background (SGWB) today (see e.g.~the relevant chapter in the review \cite{Caprini:2018mtu}). This is typically expressed in terms of the energy density fraction in GWs today, denoted by $\OGW(k)$. Depending on the level of enhancement and the scale $k$ of maximal enhancement, this scalar-induced contribution to $\OGW(k)$ is potentially detectable by current and future ground- and space-based GW observatories. 

As the post-inflationary contribution to the SGWB mainly depends on primordial physics through the scalar power spectrum of fluctuations, the language of features is highly economical for our purposes, allowing for an investigation of the effects of deviations from single-field slow-roll inflation on the SGWB in a largely model-independent fashion. The most model-dependent part of the templates \eqref{eq:P_of_k_sharp} and \eqref{eq:P_of_k_resonant} is the envelope $\Pbar(k)$ of the scalar power spectrum, whose functional form is generally different for every feature model. To generate a contribution to the SGWB that can be detected with the upcoming generation of GW observatories the scalar power spectrum will have to be enhanced by several orders of magnitude compared to the amplitude $\PCMB \sim 10^{-9}$ at CMB scales. Thus, in this work we will focus on features associated with an envelope $\Pbar$ that is enhanced at small scales compared to $\PCMB$.

Interestingly, for sharp features there exists a link between the enhancement of $\Pbar$ and the amplitude of oscillations $A_\textrm{lin}$, as was explained in \cite{Fumagalli:2020nvq}. The sharp feature can be understood as effectively replacing the Bunch-Davies vacuum by an excited state, which can be described with the help of the Bogoliubov coefficients $\alpha_k$ and $\beta_k$. The maximal value of the power spectrum is controlled by $|\alpha_k|$ while the amplitude of oscillations is set by the ratio $|\beta_k| / |\alpha_k|$. Most importantly, the quantisation conditions require that $|\alpha_k|^2-|\beta_k|^2=1$, so that neither one of the two coefficients can become large without the other growing at the same rate. This implies that a sharp feature with a strongly enhanced envelope $\Pbar$ will necessarily also exhibit $\mathcal{O}(1)$ oscillations.

\begin{figure}[t]
\centering
\begin{overpic}[width=0.90\textwidth]{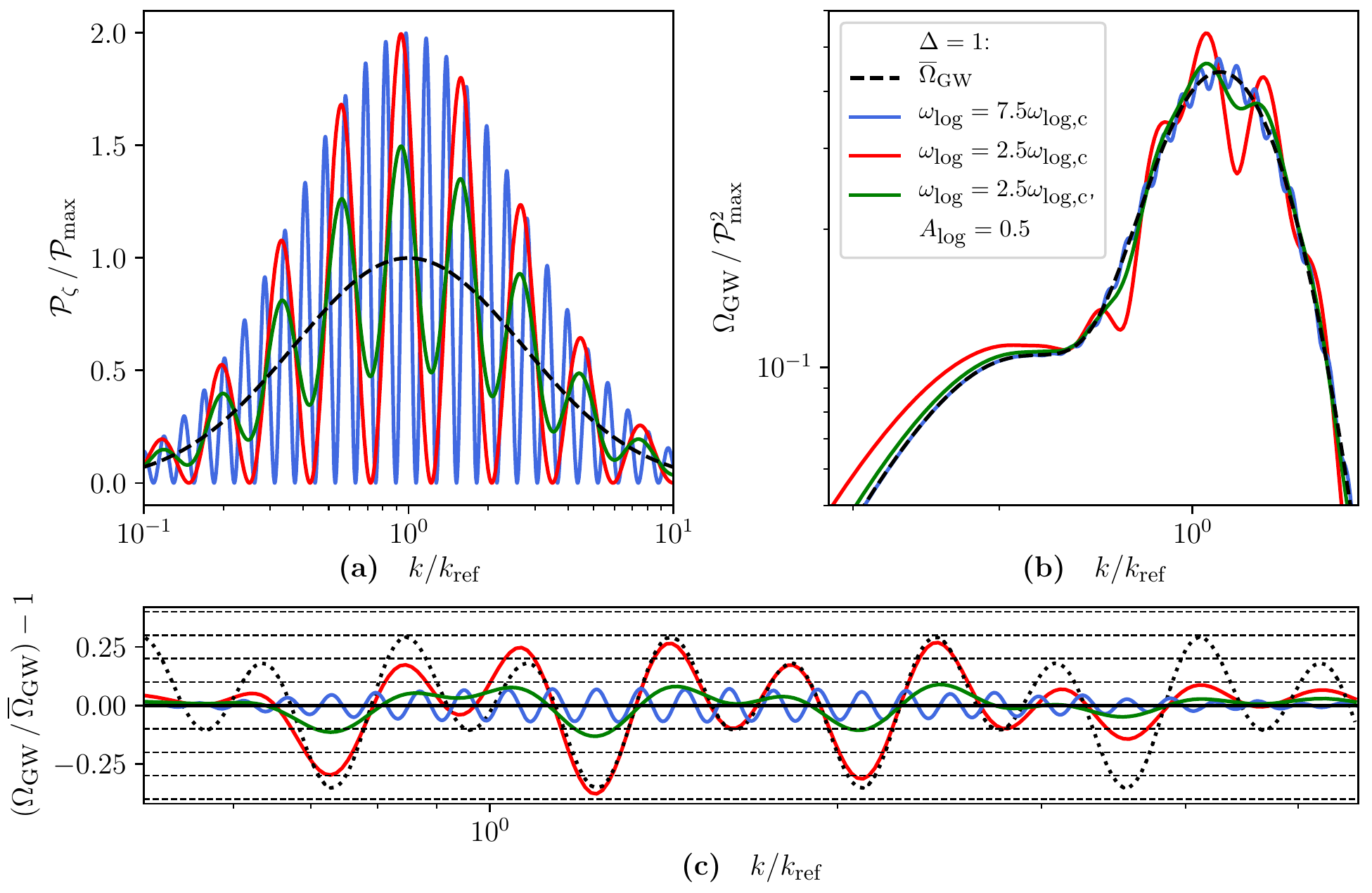}
\end{overpic}
\caption{\textit{Scalar power spectrum $\Pzeta$ (panel \textbf{(a)}) and GW energy density fraction $\OGW$ (panel \textbf{(b)}) vs.~$k / \kref$ for three examples of a resonant feature of type \eqref{eq:P_of_k_resonant} with  envelope $\Pbar$ given by a lognormal peak as in \eqref{eq:Pbar-LN} with standard deviation $\Delta=1$. The red and blue curves are for $A_\textrm{log}=1$ and $\omegalog = 2.5 \omegalogc$ or $\omegalog = 7.5 \omegalogc$, respectively, with $\omegalogc \simeq 4.77$ defined in \eqref{eq:omegalogc-def}. The green curve is for $\omegalog = 2.5 \omegalogc$, but with reduced amplitude $A_\textrm{log}=0.5$. The black dashed curve corresponds to the envelope $\Pbar$ and the corresponding GW spectrum $\overline{\Omega}_\textrm{GW}$. In panel \textbf{(c)} we isolated the oscillatory part of $\OGW$ by plotting the ratio $\OGW / \overline{\Omega}_\textrm{GW}$ vs.~$k / \kref$ for the three examples. Over the range of scales of the principal peak in $\OGW$ the modulations can be approximated by a superposition of two sinusoidal oscillations with frequencies $\omegalog$ and $2 \omegalog$ as recorded in \eqref{eq:resonant_template_intro}. For the red curve this is shown as the dotted black curve, where the values of $\mathcal{A}_{\textrm{log},1/2}$ and $\phi_{\textrm{log},1/2}$ in \eqref{eq:resonant_template_intro} were chosen to best fit the numerical result. Overall, we find that for $\omegalog = 2.5 \omegalogc$ (red and green curves) the two sinusoidal terms have comparable amplitude, leading to a complicated oscillatory pattern in $\OGW$. For larger values of $\omegalog$ the oscillatory contribution with frequency $\omegalog$ becomes suppressed compared to that with $2 \omegalog$. Thus, for $\omegalog = 7.5 \omegalogc$ (blue curve) we find that the oscillatory part is well-matched by a single sinusoidal term with frequency $2 \omegalog$.}}
\label{fig:P_OGW_ratios_res_LN1p0_2p5wc_7p5wc_a0p5}
\end{figure}

In \cite{Fumagalli:2020nvq} we initiated a dedicated study of the GW spectrum linked to a small-scale feature in the scalar power spectrum and assessed its detectability. There we analysed $\OGW(k)$ for the case of a sharp feature, using an explicit realisation in form of a sharp turn in the inflationary trajectory as an example. The goal of the present work is to extend this analysis to the case of a resonant features and to study how this type of feature manifests itself in the SGWB.\footnote{Some initial results on resonant features were already presented in \cite{Fumagalli:2020nvq}.} The analysis presented here will be done in a model-independent way for any resonant feature that follows the schematic form \eqref{eq:P_of_k_resonant}. We will then describe in detail how the frequency $\omegalog$, the amplitude of oscillations $A_\textrm{log}$ and the choice of envelope $\Pbar(k)$ affect $\OGW(k)$. The scalar-induced GW spectrum due to a small-scale feature was also analysed in \cite{Braglia:2020taf} for a two-field model of inflation, which, depending on the choice of parameters, exhibits a sharp or resonant feature or a combination of the two. Both in \cite{Fumagalli:2020nvq} and here we exclusively focus on contributions to the SGWB induced by scalar fluctuations during the post-inflationary era. GWs sourced during inflation will be the subject of a dedicated publication \cite{GWinfSharp} with focus on sharp features (see also \cite{An:2020fff,Dalianis:2020gup,daCunha:2021wyy} for other cosmological sources leading to oscillations in the GW spectrum). \vspace{0.1cm}

\noindent \textbf{Summary of results --} Our results can be summarised as follows, where we also briefly review our findings for sharp features from \cite{Fumagalli:2020nvq}. Most importantly, we find that the oscillations in the scalar power spectrum translate into corresponding modulations in $\OGW$. The periodicity in $k$ of $\Pzeta$ for a sharp feature is inherited by $\OGW$, albeit with a frequency $\omegagwlin = \sqrt{3} \omegalin$, where the numerical factor $\sqrt{3}$ is for GWs sourced during a period of radiation-domination. For a resonant feature $\OGW$ exhibits a superposition of two $\log(k)$-periodic oscillations with frequencies $\omegalog$ and $2 \omegalog$, respectively. For power spectra of the form \eqref{eq:P_of_k_sharp} and \eqref{eq:P_of_k_resonant} we show that the GW energy fraction $\OGW$ can be well-matched by the following templates:
\begin{align}
    \label{eq:sharp_template_intro} 
    \textrm{Sharp feature:} \quad & \Omega_{\textrm{GW}}(k) = \overline{\Omega}_{\textrm{GW}}(k) \Big[1+ \mathcal{A}_\textrm{lin} \cos \big(\omega_\textrm{lin}^\textsc{gw} k + \phi_\textrm{lin}  \big) \Big] \, , \\
    \label{eq:resonant_template_intro}
    \textrm{Resonant feature:} \quad & \Omega_{\textrm{GW}}(k) = \overline{\Omega}_{\textrm{GW}}(k) \Big[1+ \mathcal{A}_{\textrm{log},1} \cos \big(\omegalog \log (k/k_\textrm{ref}) + \phi_{\textrm{log},1} \big) \\
    \nonumber & \hphantom{\Omega_{\textrm{GW}}(f) = \overline{\Omega}_{\textrm{GW}}(k) \Big[1} + \mathcal{A}_{\textrm{log},2} \cos \big(2 \omegalog \log (k/k_\textrm{ref}) + \phi_{\textrm{log},2} \big) \Big] \, ,
\end{align}
which hold for the scales where $\OGW$ is maximally enhanced. The envelope $\overline{\Omega}_\textrm{GW}$ is the GW spectrum associated with the envelope of the scalar power spectrum $\Pbar$, which thus determines the overall shape of $\OGW$. The amplitudes of the oscillatory pieces in \eqref{eq:sharp_template_intro} and \eqref{eq:resonant_template_intro} can be shown to depend not only on the amplitude of the oscillation in $\Pzeta$, but also on its frequency. For a fixed envelope $\Pbar$ we observe that the amplitudes of oscillation decrease both for the case of a sharp and a resonant feature as $\omegalin$ and $\omegalog$ are increased. For a resonant feature the amplitude $\mathcal{A}_{\textrm{log},1}$ decreases faster than $\mathcal{A}_{\textrm{log},2}$. This can e.g.~be seen in fig.~\ref{fig:P_OGW_ratios_res_LN1p0_2p5wc_7p5wc_a0p5}, where we plot $\Pzeta$ and $\OGW$ for three examples of a resonant feature with common envelope $\Pbar$, but different choices for $\omegalog$ and $A_\textrm{log}$. Remarkably, we find a significant change in the oscillatory structure of $\OGW$ tied to a universal frequency $\omegalogc \simeq 4.77$, defined in \eqref{eq:omegalogc-def}. For $\omegalog < \omegalogc$ the relative amplitudes and phases of the two oscillatory pieces in \eqref{eq:resonant_template_intro} adjust one another in such a way as to give rise to a single series of peaks with frequency $\omegalog$. Overall, we demonstrate that for small values of $\omegalog$ the oscillation with frequency $\omegalog$ dominates over that with frequency $2 \omegalog$, while for large values of $\omegalog$ the situation is reversed. The precise value of $\omegalog$ where the cross-over occurs also depends on the value of $A_\textrm{log}$ in \eqref{eq:P_of_k_resonant}, but is generically of the order $\omegalog \gtrsim \mathcal{O}(1) \omegalogc$. One key result of this work are semi-analytic expressions for the oscillatory part of $\OGW$, given in (\ref{analyticalO1}, \ref{analyticalO2b}) for a constant or broad peak in $\Pbar$ and generalised to more general peak profiles in (\ref{eq:OmegaGW1-fit}, \ref{eq:OmegaGW2-fit}), from which one can read off $\mathcal{A}_{\textrm{log},1/2}$ and $\phi_{\textrm{log},1/2}$ in \eqref{eq:resonant_template_intro}.    

If the scalar power spectrum is sufficiently enhanced, the SGWB contribution $\OGW$ will have an amplitude that will permit a detection in upcoming GW wave observatories such as LISA. Thus, our results lead to the exciting prospect that by measuring an oscillation in the primordial contribution to the SGWB one may extract crucial physical information about the physical mechanism active in the late stage of inflation, probing the `dark ages' of inflation in a very direct manner. Yet, there exist open questions regarding the detectability of the GW signal due to primordial features motivating further research. One important line of work concerns developments in data analysis techniques for future GW observatories dedicated to reconstructing GW signals with oscillations.\footnote{Some initial work on reconstructing a wiggly GW spectrum with LISA has been presented in \cite{Caprini:2019pxz, Flauger:2020qyi}.} For this task, the templates in \eqref{eq:sharp_template_intro} and \eqref{eq:resonant_template_intro} can be used as test signals. On the theory side there remain questions regarding theoretical control over models with an enhancement of scalar fluctuations. If a loss of perturbativity is to be avoided, scalar fluctuations cannot be arbitrarily large, which in turn bounds the maximal amplitude of $\OGW$. Constraints from perturbativity (and also backreaction) on the fluctuations have been addressed in \cite{Fumagalli:2020nvq} for an example of a sharp feature, but a more general investigation of this issue is desirable. Another important task is to determine to what extent the scalar power spectrum can be reconstructed from a measurement of $\OGW$. What is exciting is that our results suggest that such a reconstruction may indeed be possible for inflation models with features: The frequency of modulations in $\OGW$ is directly related to that of $\Pzeta$. The amplitude of the modulations in $\OGW$ is also causally linked to various properties of $\Pzeta$. Our analytical expressions for $\mathcal{A}_{\textrm{log}, 1/2}$ and $\phi_{\textrm{log}, 1/2}$ will be highly valuable for accomplishing this task which we leave for future work. If this was successful, small-scale features would offer direct access to the early-universe through their GW signal. This would only strengthen feature models as a rewarding target for future GW experiments.

The paper is structured as follows. In sec.~\ref{sec:features} we collect some general results regarding the post-inflationary GW spectrum due to features in $\Pzeta$. Most of the facts recorded in this section are not new, but will be useful later and are sometimes presented in a novel fashion. In sec.~\ref{sec:general-observations} we analyse the effect of a peak in $\Pzeta$ on $\OGW$ and how an oscillation in $\Pzeta$ can be modelled as a series of peaks. Sec.~\ref{sec:sharp-review} contains a brief review of the GW spectrum due to a sharp feature in the scalar power spectrum. Sec.~\ref{sec:resonant} is dedicated to the analysis of the effect of a resonant feature in $\Pzeta$ on $\OGW$ and contains the bulk of the new results of this work. In sec.~\ref{sec:resonant-peak-structure} we derive the periodicity of modulations in $\OGW$ from a study of resonance peaks. In sec.~\ref{sec:resonant-templates} we present an analytical method for computing the oscillatory part of $\OGW$ for a resonant feature. Sec.~\ref{sec:conclusion} contains a more detailed summary of our results and a discussion of open questions. Some more technical results are relegated to the appendix.

\section{Small-scale features from inflation and scalar-induced GWs}
\label{sec:features}
Scalar fluctuations source GWs at second order, \cite{Acquaviva:2002ud, Mollerach:2003nq}, resulting in a post-inflationary contribution to the GW energy density fraction given by:\footnote{We  do  not  discuss  here the impact on the  GW  spectrum of the  connected four-point  function  of the  curvature  perturbation,  i.e.~the  primordial  trispectrum, see e.g.~\cite{Garcia-Bellido:2017aan,Unal:2018yaa,Cai:2018dig,Atal:2021jyo,Adshead:2021hnm}.} 
\cite{Ananda:2006af, Baumann:2007zm}
\begin{align}
\label{eq:OmegaGW-i}
    \Omega_{\textrm{GW}}(k) = \int_0^{\frac{1}{\sqrt{3}}} \textrm{d} d \int_{\frac{1}{\sqrt{3}}}^\infty \textrm{d} s \, \TRD(d,s) \, \mathcal{P}_\zeta \bigg(\frac{\sqrt{3}k}{2}(s+d)\bigg) \mathcal{P}_\zeta \bigg(\frac{\sqrt{3}k}{2}(s-d)\bigg) \, .
\end{align}
The integration kernel $\TRD$ depends on the equation of state of the universe when the relevant scalar fluctuations re-enter the horizon. Here we consider a standard cosmological history so that for all modes of interest, this occurs during a period of radiation-domination, in which case the kernel $\TRD$ is given by:\footnote{See \cite{Kohri:2018awv} for the corresponding expression for GWs produced during a period of matter-domination.}
\begin{align}
\label{eq:TRDds}
\TRD(d,s) = 36 \frac{{\big(d^2-\tfrac{1}{3}\big)}^2{\big(s^2-\tfrac{1}{3}\big)}^2{\big(d^2+s^2-2\big)}^4}{{\big(s^2-d^2\big)}^8} \bigg[ \bigg(\ln \frac{1-d^2}{|s^2-1|} + \frac{2 \big(s^2-d^2\big)}{d^2+s^2-2}\bigg)^2 + \pi^2 \Theta(s-1) \bigg] .
\end{align}
To relate \eqref{eq:OmegaGW-i} to the energy density fraction in GWs today, we need to multiply the result by a factor $c_g \Omega_{\textrm{r},0}$, where $\Omega_{\textrm{r},0}$ is the fraction of energy density in all of radiation today, and the factor $c_g$ depends on the number of effective species during radiation domination (RD) and today (0), see \cite{Espinosa:2018eve}:
\begin{align}
    c_g \equiv \frac{g_{*,\textsc{rd}}}{g_{*,0}} {\bigg( \frac{g_{*S,0}}{g_{*S,\textsc{rd}}} \bigg)}^{4/3} \, .
\end{align}
In this work we will be mainly concerned with describing the spectral shape of $\OGW(k)$ and its absolute numerical value will be less important. Thus we will frequently omit the factor $c_g \Omega_{\textrm{r},0}$ and only include it when making predictions for observations today.

\subsection{Frequency profile of $\OGW(k)$ from peaks in $\Pzeta(k)$}
\label{sec:general-observations}
For reasons of detectability of the corresponding GW signal, we are mainly interested in scalar power spectra that are enhanced at small scales compared to $\PCMB$. This enhancement can take the form of a peak, implying a violation of scale-invariance throughout the enhanced region, or take a form of a plateau, where $\Pzeta$ has returned to near-scale-invariance, albeit at a higher amplitude than $\PCMB$. For brevity, we will henceforth always refer to this enhanced region as a peak in $\Pzeta$, with the latter case of a plateau corresponding to an example of a broad peak.

In this section we will analyse the spectral shape of the GW signal due to a single and/or many peaks in $\Pzeta(k)$. The results for a single peak will be important for understanding the effect of a peak in the envelope $\Pbar(k)$ on the GW spectrum. The analysis for multiple peaks will be employed to explain the effect of the oscillatory feature on $\Omega_\textrm{GW}$. 

We begin with the case of a single peak in $\Pzeta(k)$, i.e.~an interval $k_\textrm{min} \leq k \leq k_\textrm{max}$ over which $\Pzeta(k)$ is significantly enhanced compared to its value outside this interval.
To compute the dominant part of $\Omega_\textrm{GW}$ we can then, to a good approximation, ignore the contribution coming from $\Pzeta(k)$ away from the peak, and take the scalar power spectrum as
\begin{align}
\label{Pzeta_peak_via_heaviside} \Pzeta(k) = \mathcal{P}_\textrm{peak}(k) \, \Theta (k-k_{\textrm{min}}) \, \Theta (k_{\textrm{max}}-k) \, .
\end{align}
Inserting this into \eqref{eq:OmegaGW-i} one obtains 
\begin{align}
\label{eq:OmegaGW-i-for-peak}
\nonumber    \Omega_{\textrm{GW}}(k) = \int_0^{\frac{1}{\sqrt{3}}} \textrm{d} d \int_{\frac{1}{\sqrt{3}}}^\infty \textrm{d} s \, & \TRD(d,s) \, \mathcal{P}_\textrm{peak} \bigg(\frac{\sqrt{3}k}{2}(s+d)\bigg) \mathcal{P}_\textrm{peak} \bigg(\frac{\sqrt{3}k}{2}(s-d)\bigg) \times \\
\nonumber    & \times \Theta \bigg(\frac{\sqrt{3}k}{2}(s+d) -k_\textrm{min} \bigg) \Theta \bigg(\frac{\sqrt{3}k}{2}(s-d) -k_\textrm{min} \bigg) \\
& \times \Theta \bigg(k_\textrm{max}- \frac{\sqrt{3}k}{2}(s+d) \bigg) \Theta \bigg(k_\textrm{max}-\frac{\sqrt{3}k}{2}(s-d) \bigg)  \, .
\end{align}
The finite extension of the peak in $\Pzeta(k)$ then has the effect of restricting the range of integration over the variables $d$ and $s$, as follows from the four Heaviside theta factors. This comes in addition to the fact that the integration is already over a restricted area in the $d$-$s$-plane:
\begin{align}
\label{eq:d-s-int-range}
    d \in \Big[0, \frac{1}{\sqrt{3}} \Big] \, , \quad \textrm{and} \quad s \in \Big[\frac{1}{\sqrt{3}}, \infty \Big] \, .
\end{align} 
The Heaviside theta factors in \eqref{eq:OmegaGW-i-for-peak} enforce the constraints:
\begin{align}
\nonumber    s+d &> \frac{2 k_\textrm{min}}{\sqrt{3} k} \, , \qquad s+d < \frac{2 k_\textrm{max}}{\sqrt{3} k} \, , \\
\label{eq:s-d-constraints}    s-d &> \frac{2 k_\textrm{min}}{\sqrt{3} k} \, , \qquad s-d < \frac{2 k_\textrm{max}}{\sqrt{3} k} \, .
\end{align}
Note that these define a square in the $(s-d)$-$(s+d)$-plane. The extension of the square depends on the value of $k$, but also on $k_\textrm{min}$ and $k_\textrm{max}$ and thus on the extension of the peak.

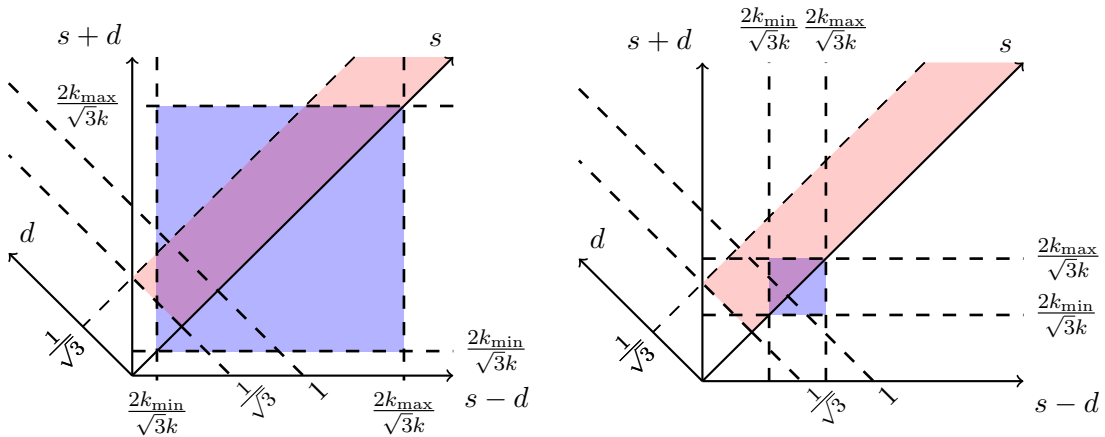
\begin{figure}[t]
		\centering
		\begin{tikzpicture}[scale=0.65]
		%\fill[yellow!20!white] (2, 0) -- (-2.5, 2+2.5) -- (-2.5, 6.5) -- (6.5, 6.5) -- (6.5,0) -- cycle;
		\fill[blue!30!white] (0.5,0.5) rectangle (5.5,5.5);
		%\fill[red!100!white, opacity=0.2] (0, 0) -- (-1.0, 1.0) -- (6.5-2, 6.5) -- (6.5, 6.5) -- cycle;
		\fill[red!100!white, opacity=0.2] (1, 1) -- (0, 2.0) -- (6.5-2, 6.5) -- (6.5, 6.5) -- cycle;
		\draw[thick,->] (0,0) -- (6.5,0) node[anchor=north west] {$s-d$};
        \draw[thick,->] (0,0) -- (0,6.5) node[anchor=south east] {$s+d$};
        \draw[thick,->] (0,0) -- (-2.5,2.5) node[anchor=south west] {$d$};
        \draw[thick,->] (0,0) -- (6.5,6.5) node[anchor=south east] {$s$};
        \draw [line width=1.0pt,dash pattern=on 5pt off 5pt] (0.5,6.5)-- (0.5,-0.1) node[anchor=north] {$\frac{2 k_\textrm{min}}{\sqrt{3} k}$};
        \draw [line width=1.0pt,dash pattern=on 5pt off 5pt] (5.5,6.5)-- (5.5,-0.1) node[anchor=north] {$\frac{2 k_\textrm{max}}{\sqrt{3} k}$};
        \draw [line width=1.0pt,dash pattern=on 5pt off 5pt] (0,0.5)-- (6.5,0.5) node[anchor=west] {$\frac{2 k_\textrm{min}}{\sqrt{3} k}$};
        \draw [line width=1.0pt,dash pattern=on 5pt off 5pt] (6.5,5.5)-- (0,5.5) node[anchor=east] {$\frac{2 k_\textrm{max}}{\sqrt{3} k}$};
        \draw [line width=0.6pt,dash pattern=on 5pt off 5pt] (6.5-2, 6.5) -- (-1.0,1.0) node[anchor=east, rotate=45] {$\frac{1}{\sqrt{3}}$};
        \draw [line width=0.6pt,dash pattern=on 5pt off 5pt] (6.5-2, 6.5) -- (-1.0,1.0) node[anchor=east, rotate=45] {$\frac{1}{\sqrt{3}}$};
        \draw [line width=1.0pt,dash pattern=on 5pt off 5pt] (1,1)-- (2,0) node[anchor=north, rotate=45] {$\frac{1}{\sqrt{3}}$};
        \draw [line width=1.0pt,dash pattern=on 5pt off 5pt] (1.0,1.0)-- (-2.5,4.5);
        \draw [line width=1.0pt,dash pattern=on 5pt off 5pt] (1/0.57735,1/0.57735)-- (2/0.57735,0) node[anchor=north, rotate=45] {$1$};
        \draw [line width=1.0pt,dash pattern=on 5pt off 5pt] (-2.5,2/0.57735+2.5)-- (1/0.57735,1/0.57735);
        %\draw [thick, dotted] (5.5,5.5)-- (4.5,6.5);
		\end{tikzpicture} 
		\quad
		\begin{tikzpicture}[scale=0.65]
		%\fill[yellow!20!white] (2, 0) -- (-2.5, 2+2.5) -- (-2.5, 6.5) -- (6.5, 6.5) -- (6.5,0) -- cycle;
		\fill[blue!30!white] (1.35,1.35) rectangle (2.5,2.5);
		%\fill[red!100!white, opacity=0.2] (0, 0) -- (-1.0, 1.0) -- (6.5-2, 6.5) -- (6.5, 6.5) -- cycle;
		\fill[red!100!white, opacity=0.2] (1, 1) -- (0, 2.0) -- (6.5-2, 6.5) -- (6.5, 6.5) -- cycle;
		\draw[thick,->] (0,0) -- (6.5,0) node[anchor=north west] {$s-d$};
        \draw[thick,->] (0,0) -- (0,6.5) node[anchor=south east] {$s+d$};
        \draw[thick,->] (0,0) -- (-2.5,2.5) node[anchor=south west] {$d$};
        \draw[thick,->] (0,0) -- (6.5,6.5) node[anchor=south east] {$s$};
        \draw [line width=1.0pt,dash pattern=on 5pt off 5pt] (1.35,0)-- (1.35,6.5) node[anchor=south] {$\frac{2 k_\textrm{min}}{\sqrt{3} k}$};
        \draw [line width=1.0pt,dash pattern=on 5pt off 5pt] (2.5,0)-- (2.5,6.5) node[anchor=south] {$\ \ \frac{2 k_\textrm{max}}{\sqrt{3} k}$};
        \draw [line width=1.0pt,dash pattern=on 5pt off 5pt] (0,1.35)-- (6.5,1.35) node[anchor=west] {$\frac{2 k_\textrm{min}}{\sqrt{3} k}$};
        \draw [line width=1.0pt,dash pattern=on 5pt off 5pt] (0,2.5)-- (6.5,2.5) node[anchor=west] {$\frac{2 k_\textrm{max}}{\sqrt{3} k}$};
        \draw [line width=0.6pt,dash pattern=on 5pt off 5pt] (6.5-2, 6.5) -- (-1.0,1.0) node[anchor=east, rotate=45] {$\frac{1}{\sqrt{3}}$};
        \draw [line width=0.6pt,dash pattern=on 5pt off 5pt] (6.5-2, 6.5) -- (-1.0,1.0) node[anchor=east, rotate=45] {$\frac{1}{\sqrt{3}}$};
        \draw [line width=1.0pt,dash pattern=on 5pt off 5pt] (1,1)-- (2,0) node[anchor=north, rotate=45] {$\frac{1}{\sqrt{3}}$};
        \draw [line width=1.0pt,dash pattern=on 5pt off 5pt] (1.0,1.0)-- (-2.5,4.5);
        \draw [line width=1.0pt,dash pattern=on 5pt off 5pt] (1/0.57735,1/0.57735)-- (2/0.57735,0) node[anchor=north, rotate=45] {$1$};
        \draw [line width=1.0pt,dash pattern=on 5pt off 5pt] (-2.5,2/0.57735+2.5)-- (1/0.57735,1/0.57735);
		\end{tikzpicture}
		\caption{\textit{Red: Region of integration over the parameters $(d,s)$ in the computation of $\Omega_\textrm{GW}$ via \eqref{eq:OmegaGW-i}. Blue: Region in $(d,s)$-space with non-zero contributions to $\Omega_\textrm{GW}$ from $\Pzeta(k)$ with finite support in $k_\textrm{min} \leq k \leq k_\textrm{max}$. Left panel: Example of a broad peak in $\Pzeta(k)$ with $k_\textrm{min} \ll k_\textrm{max}$. Right panel: Narrow peak example with $k_\textrm{min} \lesssim k_\textrm{max}$.}}
		\label{fig:int-ranges}
	\end{figure}

In the following we will focus on the two cases of broad and narrow peaks in $\Pzeta$, as in both instances we will be able to give approximate analytic expressions for $\OGW$ near its maximum. 

\subsubsection{Broad and narrow peaks in $\Pzeta$}
\label{sec:broad-narrow}
We will refer to a peak as broad if its centre region is both far from $k_\textrm{min}$ and $k_\textrm{max}$. That is, there is a centre region with values of $k$ in $k_\textrm{min} \leq k \leq k_\textrm{max}$ that satisfy:
\begin{align}
\label{eq:k-broad}
  \textrm{broad:} \quad  \frac{k_\textrm{min}}{k} \ll 1 \, , \qquad \frac{k_\textrm{max}}{k} \gg 1 \, . 
\end{align}
In contrast, a peak will be designated as narrow if its width $\Delta k$ is small compared to the location $k_\star$ of the peak, i.e.
\begin{align}
\label{eq:narrowdef}
  \textrm{narrow:} \quad \frac{\Delta k}{k_\star} \ll 1 \, , \quad \textrm{with} \quad \Delta k \equiv k_\textrm{max}-k_\textrm{min} \, , \quad \textrm{and} \quad \Pzeta (k_\star) = \textrm{max} \Big( \Pzeta (k) \Big) \, .
\end{align}
In both cases we will be mainly interested in values of $k$ close to the maximum of $\Pzeta$, as the corresponding GW spectrum will also be enhanced in this range. That is, for a broad peak we focus on values of $k$ satisfying \eqref{eq:k-broad} while for a narrow peak we consider $k \sim k_\star$.

From \eqref{eq:k-broad} it then follows that for a broad peak the constraints \eqref{eq:s-d-constraints} define a comparatively `large' square, while for a narrow peak satisfying \eqref{eq:narrowdef} this square has a `small' area. This is illustrated in figure \ref{fig:int-ranges} where we plot the area defined by \eqref{eq:s-d-constraints} as a blue square on the $(s-d)$-$(s+d)$-plane together with the integration domain \eqref{eq:d-s-int-range} shown as a red strip. Here, the left panel is for a broad peak and the left panel for a narrow peak as defined above. Most importantly, we observe that for a broad peak the blue square overlaps with the integration range (red strip) over a significant area, while for a narrow peak there is only a small area of intersection between the two. This observation will be central for deriving approximate expressions for $\OGW$ for both the broad and narrow peak case.

In the following we will outline the main arguments and give the results while details and computations can be found in appendix \ref{app:peaks}. For the broad peak case the crucial point is that the blue square defined by the constraints \eqref{eq:s-d-constraints} is sufficiently large to contain the part of the integration domain \eqref{eq:d-s-int-range} that is responsible for the dominant contribution to $\OGW$. This corresponds to the vicinity of $s=1$ where the integration kernel $\TRD$ in \eqref{eq:TRDds} diverges. The constraints \eqref{eq:s-d-constraints} are effectively irrelevant for computing $\OGW$ for values of $k$ satisfying \eqref{eq:k-broad} and we can integrate over the whole range \eqref{eq:d-s-int-range}.\footnote{This is explained with some more detail in appendix \ref{app:broad}.} For a sufficiently broad peak $\mathcal{P}_\textrm{peak}(k)$ in \eqref{Pzeta_peak_via_heaviside} that is to a good approximation constant over its central region, i.e.~$\mathcal{P}_\textrm{peak}(k) \sim \mathcal{P}_\textrm{max} = \textrm{const}$, we can factor out a term $\mathcal{P}_\textrm{max}^2$ from the integral in \eqref{eq:OmegaGW-i-for-peak} to obtain\footnote{Recall that to relate this to the GW energy density fraction today, we need to multiply by a factor $c_g \Omega_{\textrm{r},0}$.}
\begin{align}
\label{eq:OmegaGW-i-broad}
    \Omega_{\textrm{GW}, \textrm{broad}} \approx \mathcal{P}_\textrm{max}^2 \int_0^{\frac{1}{\sqrt{3}}} \textrm{d} d \int_{\frac{1}{\sqrt{3}}}^\infty \textrm{d} s \, \TRD(d,s)  \simeq 0.823 \, \mathcal{P}_\textrm{max}^2 \, ,
\end{align}
which we identify as the result for a constant scalar power spectrum with amplitude $\mathcal{P}_\textrm{max}$ \cite{Kohri:2018awv}. That is, if the peak in $\Pzeta(k)$ is broad and smooth enough, the value of $\Omega_\textrm{GW}(k)$ for $k_\textrm{min} \ll k \ll k_\textrm{max}$ can be approximated by the result \eqref{eq:OmegaGW-i-broad} for a constant scalar power spectrum.

In the narrow peak case we exploit the fact that the blue area is small as follows. As the integration over $d$ is now bounded as $0 < d < d_\epsilon$ with $d_\epsilon = \Delta k / (\sqrt{3} k) \ll 1$, one can trivialise the integral over $d$ by expanding in powers of $d_\epsilon$.\footnote{The quantity $d_\epsilon$ is the value of $d$ at the top left corner of the blue square defined by the constraints \eqref{eq:s-d-constraints}.} The remaining integral over $s$ can then be written as an average over an interval $\Delta k$:\footnote{See appendix \ref{app:narrow} for more details.}  
\begin{align}
    \label{eq:OmegaGW-narrow-average}
    \Omega_{\textrm{GW}, \textrm{narrow}}(k) = \frac{\mathcal{A}^2}{3 k^2} \frac{1}{\Delta k} \int_{k_\textrm{min}}^{k_\textrm{max}} \textrm{d}q \, \TRD \Big(0,\tfrac{2 q}{\sqrt{3} k} \Big) \, \Theta \Big(2- \tfrac{k}{q} \Big) \, ,
\end{align}
where we identify the integrand as the GW spectrum $\OGW^{(\delta)}$ due to a $\delta$-peak in $\Pzeta$ at location $k_\star$, \cite{Ananda:2006af, Kohri:2018awv}:
\begin{align}
\label{eq:OmegaGW-delta}
    \Omega_{\textrm{GW}}^{(\delta)}(k) = \frac{\mathcal{A}^2}{3 k^2} \TRD \Big(0,\tfrac{2 k_\star}{\sqrt{3} k}\Big) \Theta \Big(2- \tfrac{k}{k_\star} \Big) \, .
\end{align}
Similar proposals for computing $\OGW$ for a narrow peak by smoothing the $\delta$-peak result have appeared in \cite{Byrnes:2018txb, Pi:2020otn, Fumagalli:2020nvq}, but here we arrive at this result through a controlled approximation. As a result, the GW spectrum for a narrow peak inherits many properties from the expression $\Omega_{\textrm{GW}}^{(\delta)}$. This exhibits a broad peak at $k \sim k_\star / 2$ and a sharp principal peak at $k=2 k_\star / \sqrt{3}$, where $\Omega_{\textrm{GW}}^{(\delta)}$ diverges due to resonant amplification. Most importantly, for a narrow peak of finite width this peak-structure of $\OGW$ is preserved, albeit the resonance peak at $k=2 k_\star / \sqrt{3}$ is now of finite amplitude.

\begin{figure}[t]
\centering
\begin{overpic}[width=0.7\textwidth]{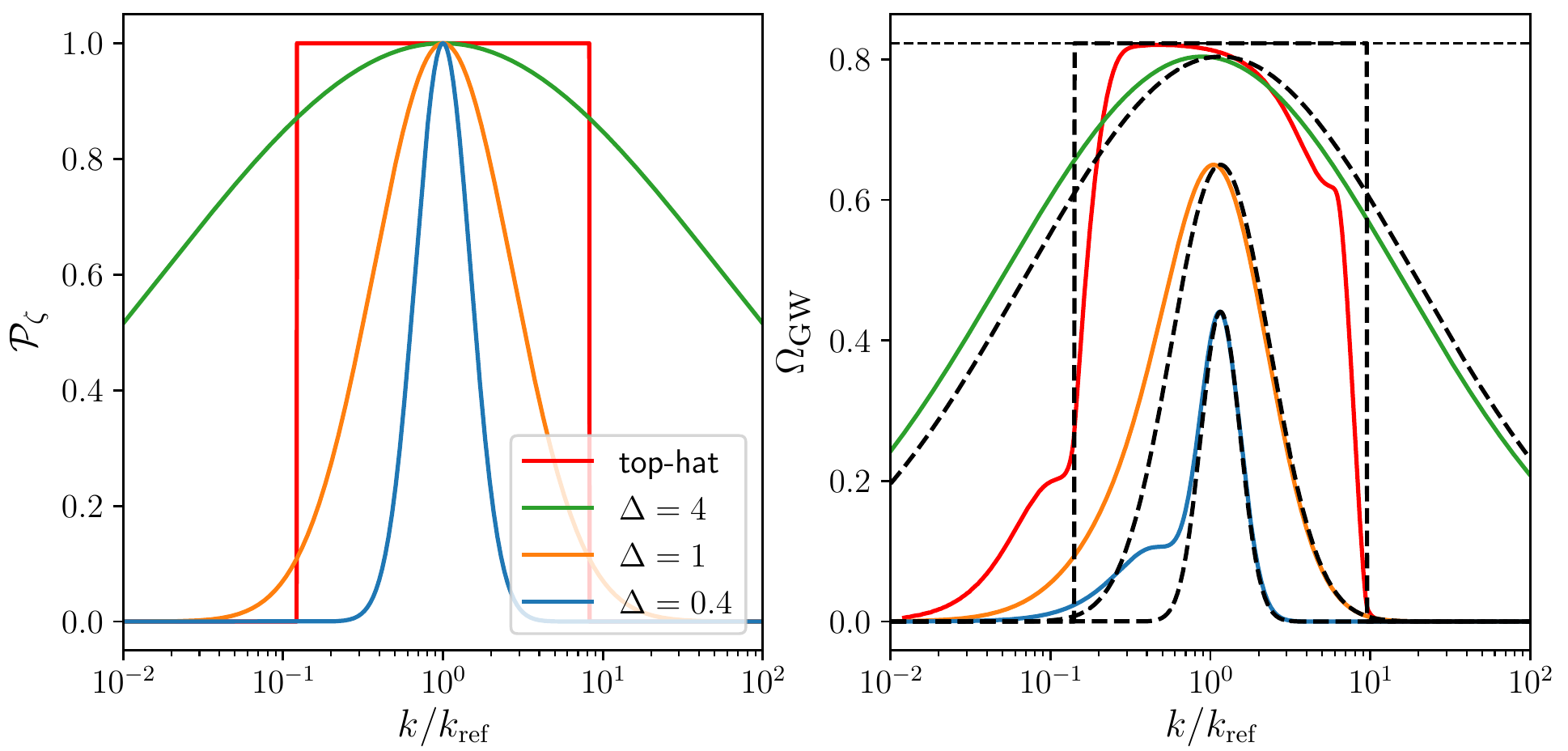}
\end{overpic}
\caption{\textit{$\Pzeta(k)$ (left panel) and the corresponding GW spectrum $\OGW(k)$ (right panel) vs.~$k/\kref$ for four examples with $\mathcal{P}_\textrm{max}=1$. The red curve is for $\Pzeta$ with a top-hat profile with $\sqrt{k_\textrm{min} k_\textrm{max}}=\kref$ and $k_\textrm{max} \simeq 68 k_\textrm{min}$. The remaining three examples are for $\Pzeta$ given by a lognormal peak, \eqref{eq:lognormal-def}, with $\Delta=0.4,1,4$. The dashed horizontal line in the right panel is the predicted amplitude of $\OGW$ for a broad-peaked $\Pzeta$ with $\mathcal{P}_\textrm{max}=1$, see \eqref{eq:OmegaGW-i-broad}. The dashed black curves are proportional to $\Pzeta^2 (\sqrt{3} k /2)$ for the four examples, with the amplitude fitted to that of $\OGW$ at its maximum.}}
\label{fig:Broad_Narrow_top_LN}
\end{figure}

For peaks in $\Pzeta$ that are moderately narrow, we can arrive at a different approximate expression for $\OGW$ that will be especially useful later. In particular, for $\Delta k / k_\star \gtrsim 0.3$ (see appendix \ref{app:narrow} for the origin of this) we expect the principal peak of $\OGW$ to be well-approximated by
\begin{align}
    \label{eq:OmegaGW-as-Psquared-fit}
    \Omega_{\textrm{GW}, \textrm{fit}}(k)= 0.823 \,
    \tilde{\gamma} \, \mathcal{P}_\zeta^2 \bigg(\frac{\sqrt{3}k}{2}\bigg) \, , \quad \textrm{for} \quad k \sim k_\star \, ,
\end{align}
where $\tilde{\gamma}$ is a numerical factor to be determined by e.g.~fitting to the numerical result and the factor $0.823$ was included for later convenience and to allow for easier comparison with the expression \eqref{eq:OmegaGW-i-broad}. The upshot is that in the moderately narrow case the shape of the peak in $\Pzeta$ determines the shape of the principal peak in $\OGW$.

We find support for this and our results regarding a broad peak by considering explicit examples. In fig.~\ref{fig:Broad_Narrow_top_LN} we plot $\Pzeta(k)$ (left panel) and the corresponding GW spectrum $\OGW(k)$ (right panel) for four examples (with $\mathcal{P}_\textrm{max}=1$): a top-hat with $\sqrt{k_\textrm{min} k_\textrm{max}}=\kref$ and $k_\textrm{max} \simeq 68 k_\textrm{min}$, and three examples of a lognormal peak, i.e.
\begin{align}
\label{eq:lognormal-def}
    \Pzeta(k) = \mathcal{P}_\textrm{max} \exp \bigg( -\frac{1}{2 \Delta^2} \bigg(\log \frac{k}{\kref} \bigg)^2 \bigg) \, ,
\end{align}
with $\Delta=0.4,1,4$. We make the following observations. For the top-hat example the broad-peak result \eqref{eq:OmegaGW-i-broad} gives a very good estimate of $\OGW$ over the central region of its peak, even for this moderately broad example with $k_\textrm{max} \simeq 68 k_\textrm{min}$. The expression \eqref{eq:OmegaGW-i-broad} also gives a good estimate for the maximal value of $\OGW$ for the broadest lognormal peak with $\Delta =4$, predicting the maximum up to an error of $\sim 2 \%$, but not for the narrower examples with $\Delta=0.4,1$. In the right panel of  fig.~\ref{fig:Broad_Narrow_top_LN} we also plot fits to the GW spectrum based on \eqref{eq:OmegaGW-as-Psquared-fit} for all four examples. This gives the best result for the example with the narrowest peak, i.e.~the lognormal example with $\Delta=0.4$, closely reproducing the shape of the principal peak.\footnote{Defining the width $\Delta k$ of the peak as the full width at half-maximum, for a lognormal peak with $\Delta=0.4$ we find $\Delta k / k_\star \sim 1$ and hence \eqref{eq:OmegaGW-as-Psquared-fit} is expected to apply.} For the examples with a broader peak in $\Pzeta$ (the lognormal cases with $\Delta =1$ and $\Delta =4$ and the broad top-hat with $k_\textrm{max} \simeq 68 k_\textrm{min}$) the fit \eqref{eq:OmegaGW-as-Psquared-fit} works less well as expected. Still, the expression in \eqref{eq:OmegaGW-as-Psquared-fit} fares better in approximating $\OGW(k)$ than may have been anticipated, even for the broad lognormal example with $\Delta=4$, with only a $\%$-level error near the centre of the peak. For the top-hat case \eqref{eq:OmegaGW-as-Psquared-fit} accurately predicts the extent in $k$ of the principal peak. We will use these observations later in sec.~\ref{sec:resonant-templates} when proposing templates for $\OGW$.

\subsubsection{Oscillatory features as multiple peaks}
\label{sec:multiple}
We now turn to the subject of principal interest, that is scalar power spectra with a sharp or resonant oscillatory feature, as given in (\ref{eq:P_of_k_sharp}, \ref{eq:P_of_k_resonant}). In this work we will put special emphasis on the case where the oscillatory feature is pronounced in the sense that the oscillations are $\mathcal{O}(1)$, which corresponds to $A_\textrm{lin}, A_\textrm{log}=1$ or at most $A_\textrm{lin}, A_\textrm{log} \lesssim 1$. In addition, we also assume that the envelope $\Pbar (k)$ exhibits a peak, which may be be broad or narrow.\footnote{For $A_\textrm{lin}, A_\textrm{log} \ll 1$ the oscillations only provide a small correction to the scalar power spectrum which is effectively given by $\Pbar(k)$. The corresponding GW spectrum $\OGW(k)$ is then given at leading order by that obtained for $\Pbar(k)$, which can be understood using the results of sections \ref{sec:broad-narrow} depending on whether the peak in $\Pbar(k)$ is broad or narrow. However, in section \ref{sec:resonant-templates} we analyse $\OGW(k)$ for any value of $A_\textrm{log}$ for the case of a resonant feature.}

The reasons for giving increased attention to $\mathcal{O}(1)$ oscillations are two-fold. The first reason is phenomenological: pronounced oscillations in $\Pzeta(k)$ lead to a characteristic signal in $\OGW(k)$ whose description is one of the main subjects of this paper. Secondly, if the oscillations are sufficiently pronounced, the scalar power spectrum can be modelled as a comb of individual peaks. This picture will be instrumental for the analytical understanding of the corresponding GW spectrum in sections \ref{sec:sharp-review} and \ref{sec:resonant-peak-structure}. 

An important observation is that the individual peaks in $\Pzeta(k)$ due to an oscillatory feature will mostly fall into the category of narrow peaks in the sense of eq.~\eqref{eq:narrowdef}. This is automatically the case if the envelope $\Pbar(k)$ is already narrow-peaked, as a modulation of a narrow bump will exhibit even narrower peaks by default. However, even when the envelope $\Pbar(k)$ is broadly-peaked, the peaks due to the oscillation will be narrow as long as the frequency of oscillation $\omega_{\textrm{lin/log}}$ is sufficiently large.

One key observation from sec.~\ref{sec:broad-narrow} for a single narrow peak at $k=k_\star$ in $\Pzeta(k)$ is the existence of a principal peak in $\OGW(k)$ at $k = 2 k_\star / \sqrt{3}$ resulting from resonant amplification. One can then show that for multiple narrow peaks in $\Pzeta(k)$, with loci denoted by $k_{\star i}$, one finds multiple resonance peaks in $\OGW(k)$, see \cite{Cai:2019amo}, where this has been used to study the contribution to the SGWB due to a series of $\delta$-function and Gaussian peaks in $\Pzeta(k)$. More precisely, every peak $k_{\star i}$ will produce a resonance peak in $\OGW(k)$ at $k=2 k_{\star i}/ \sqrt{3}$. In addition, due to the two factors of $\Pzeta$ in \eqref{eq:OmegaGW-i} there will also be resonant amplification in $\OGW(k)$ due to interactions between different peaks in $\Pzeta$. Together, these effects lead to a series of peaks in $\OGW(k)$ at the loci $k_{\textrm{max}, ij}$ given by \cite{Cai:2019amo}:
\begin{align}
\label{eq:kmaxij-def}
k_{\textrm{max},ij} = \frac{1}{\sqrt{3}}(k_{\star i}+ k_{\star j}) \, , \quad \textrm{with} \quad k_{\textrm{max},ij} > |k_{\star i}- k_{\star j}| \, ,
\end{align}
The constraint arises due to the restricted integration range over $(d,s)$ as given in \eqref{eq:d-s-int-range}. In \cite{Fumagalli:2020nvq} we used \eqref{eq:kmaxij-def} to analyse the peak structure of $\OGW(k)$ due to a sharp feature in detail and also presented initial results for a resonant feature. For the analysis of the sharp feature case in \cite{Fumagalli:2020nvq} we used an explicit realisation in terms of a strong sharp turn in the inflationary trajectory. In the following section \ref{sec:sharp-review} we will review our results, but phrase them with respect to a general sharp feature of the form \eqref{eq:P_of_k_sharp}. The case of a resonant feature will be addressed in detail in section \ref{sec:resonant}. Background on the applicability of \eqref{eq:kmaxij-def} in the context of a sharp and resonant feature is collected in appendix \ref{app:multiple}.

\subsection{Probing sharp features with the SGWB: brief review}
\label{sec:sharp-review}
Here we will briefly review the results of \cite{Fumagalli:2020nvq} concerning the spectral shape of the scalar-induced contribution to the SGWB due to a sharp feature in the scalar power spectrum. We also assume that the scalar fluctuations are enhanced over a finite range of scales, i.e.~we consider a scalar power spectrum of the form \eqref{eq:P_of_k_sharp} with a peaked envelope $\Pbar(k)$.

\begin{figure}[t]
\centering
\begin{overpic}[width=0.90\textwidth]{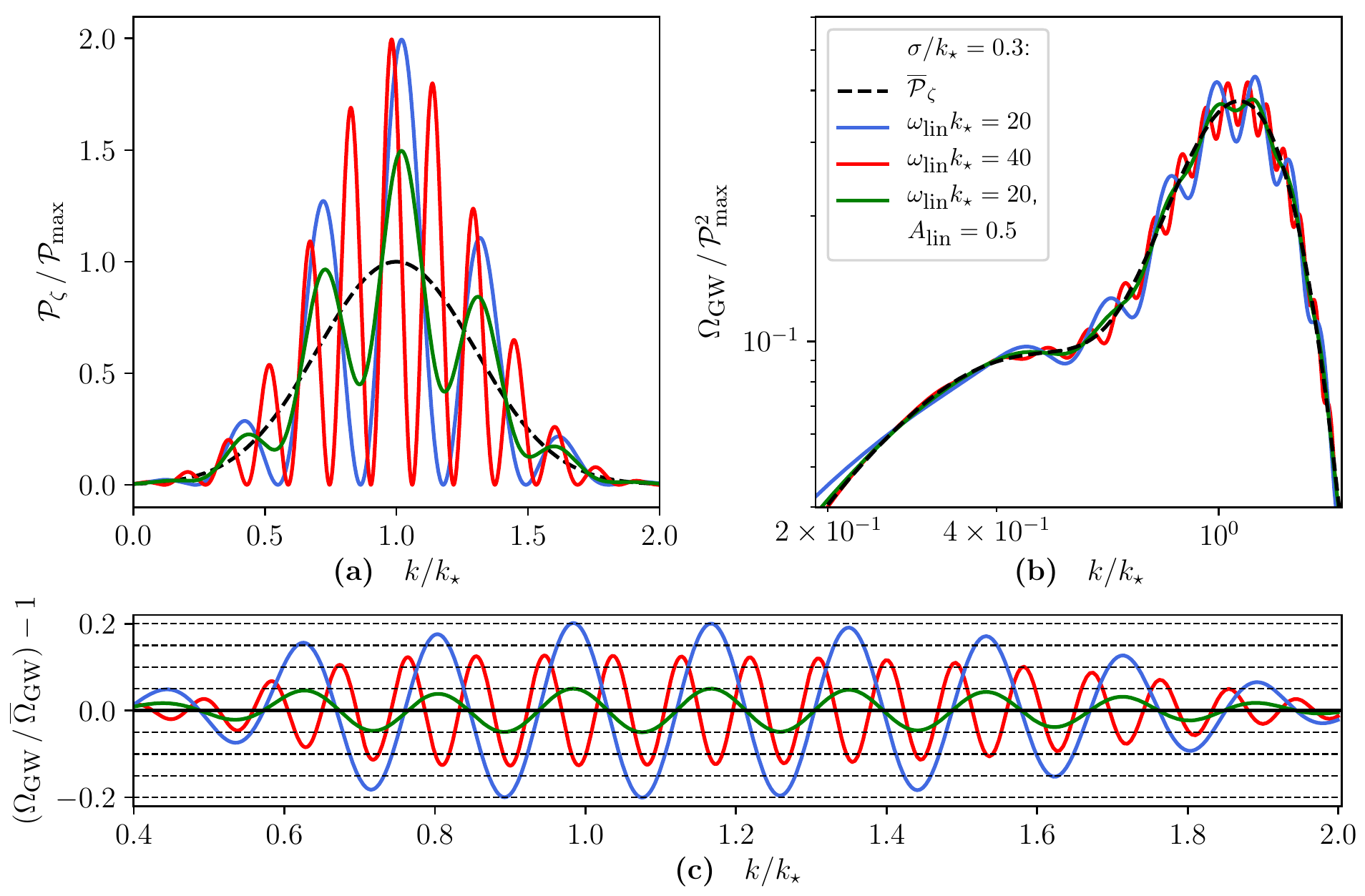}
\end{overpic}
\caption{\textit{$\Pzeta$ (panel \textbf{(a)}) and $\OGW$ (panel \textbf{(b)}) vs.~$k / k_\star$ for three example power spectra given by \eqref{eq:P-sharp-turn-analytic} (parameters recorded in legend). The green curve is for an example with $A_\textrm{lin}=0.5$, in which case we re-instated a factor of $A_\textrm{lin}$ multiplying the $\sin$-term in \eqref{eq:P-sharp-turn-analytic}. The black dashed line denotes $\Pbar$ and the corresponding GW spectrum $\overline{\Omega}_\textrm{GW}$. In panel \textbf{(c)} we plot the ratio $\OGW / \overline{\Omega}_\textrm{GW}$ vs.~$k / k_\star$ for the three examples.}}
\label{fig:P_OGW_ratios_sharp_s0p3_w20_w40_a0p5}
\end{figure}

We will illustrate this with the help of numerical results obtained for a sharp feature due to a strong sharp turn in the inflationary trajectory. For a large and constant rate of turn during a short interval of time, the scalar power spectrum can be computed analytically \cite{Palma:2020ejf, Fumagalli:2020nvq}. Over the enhanced scales this result can be well-approximated by:
\begin{align}
\label{eq:P-sharp-turn-analytic}
    \Pzeta(k) = \mathcal{P}_\textrm{max} \, e^{-\frac{(k - k_\star)^2}{2 \sigma^2}} \Big[1 + \sin \big( \omegalin k \big) \Big] \, ,
\end{align}
i.e.~the scalar power spectrum takes the form of a sharp feature with $A_\textrm{lin}=1$ modulating an envelope $\Pbar$ that is given by a Gaussian peak centred at $k_\star$. This approximation is applicable for $\sigma/k_\star \ll 1$ and $\omegalin k_\star \gg 1$. The parameters $\mathcal{P}_\textrm{max}$, $k_\star$, $\sigma$ and $\omegalin$ are related to the rate of turn $\eta_\perp$ and its duration $\delta$ (in units of $e$-folds) as
\begin{align}
\label{eq:P-sharp-turn-parameters}
\mathcal{P}_\textrm{max} = \Pzero e^{2 \eta_\perp \delta} \, , \quad k_\star= \eta_\perp k_\textrm{f} \, , \quad \sigma = \frac{k_\star}{\sqrt{2(\eta_\perp \delta -1)}} \, , \quad \omegalin = \frac{2 e^{-\delta / 2} \eta_\perp}{k_\star} \, , 
\end{align}
where $\Pzero$ is the amplitude of the scalar power spectrum in absence of the turn and $k_\textrm{f}$ is the scale associated with the turn, corresponding to the wavenumber of modes that leave the horizon halfway during the turn. In fig.~\ref{fig:P_OGW_ratios_sharp_s0p3_w20_w40_a0p5} we plot $\Pzeta$ as given in \eqref{eq:P-sharp-turn-analytic} for two example models with $(\sigma/k_\star, \, \omegalin k_\star)=(0.3, 20)$ and $(0.3,40)$ together with the corresponding results for $\OGW$.\footnote{This is equivalent to the parameter choices $(\delta, \, \eta_\perp)=(0.508, \, 12.89)$ and $(0.284, \, 23.06)$.} The enhancement factor of fluctuations is $\mathcal{P}_\textrm{max}/ \Pzero \approx 5 \cdot 10^5$ in both cases. We also show the results for an example with $(\sigma/k_\star, \, \omegalin k_\star)=(0.3, 20)$ and $A_\textrm{lin}=0.5$, for which we re-instate a factor of $A_\textrm{lin}$ multiplying the $\sin$-term in \eqref{eq:P-sharp-turn-analytic}. Note that all three examples share the same envelope $\Pbar$.

The results of \cite{Fumagalli:2020nvq} regarding the GW spectrum due to a sharp feature associated with a peak can then be summarised as follows.
\begin{itemize}
    \item The overall spectral shape of $\OGW(k)$ is determined by the functional form of the envelope $\Pbar(k)$ of the scalar power spectrum. We can see this explicitly in panel (b) of fig.~\ref{fig:P_OGW_ratios_sharp_s0p3_w20_w40_a0p5} where the GW spectrum for all three examples (blue, red and green curves) traces the corresponding result for $\Pbar$ (black dashed curve) once the short-wavelength modulations are ignored.
    
    If $\Pbar(k)$ is sufficiently narrowly peaked, the corresponding GW spectrum will exhibit a broad lower peak at smaller values of $k$ and a narrower principal peak at $k \simeq 2 k_\star / \sqrt{3}$, where $k_\star$ denotes the maximum of $\Pbar(k)$, cf.~sec.~\ref{sec:broad-narrow}. When the peak in $\Pbar(k)$ is broadened, the two peaks in $\OGW(k)$ begin to merge into one, until for sufficiently broad $\Pbar(k)$ the GW spectrum $\OGW(k)$ only exhibits one peak. The overall shape of $\OGW$ in fig.~\ref{fig:P_OGW_ratios_sharp_s0p3_w20_w40_a0p5} is that for a narrow peak in $\Pzeta$, exhibiting a broad peak at $k \sim 0.5 k_\star$ and the principal peak at $k \simeq 2 k_\star / \sqrt{3}$.
    
    \item One characteristic property of GW spectrum due to a sharp feature is a modulation of $\OGW(k)$ on its principal peak (or, equivalently, its single peak in the case of a broad envelope). Most importantly, this oscillation is periodic in $k$ with frequency $\omegagwlin=\sqrt{3} \omegalin$. This can be e.g.~understood by modelling the oscillation in $\Pzeta(k)$ as a series of peaks and then predicting the peak pattern of $\OGW(k)$ using \eqref{eq:kmaxij-def}: The periodic peak-structure in $k$ of $\Pzeta(k)$ is translated into a periodic peak-structure in $k$ of $\OGW(k)$ with the factor $1/\sqrt{3}$ in \eqref{eq:kmaxij-def} responsible for the $\sqrt{3}$ enhancement of the frequency. As a result, over the range of scales covering the principal peak the GW spectrum is well-approximated by the template 
    \begin{align}
    \label{eq:OGW_sharp_template} 
    \Omega_{\textrm{GW}}(k) = \overline{\Omega}_{\textrm{GW}}(k) \Big[1+ \mathcal{A}_\textrm{lin} \cos \big(\omega_\textrm{lin}^\textsc{gw} k + \phi_\textrm{lin}  \big) \Big] \, , \quad \textrm{with} \quad \omegagwlin=\sqrt{3} \omegalin \, ,
    \end{align}
    where $\overline{\Omega}_{\textrm{GW}}(k)$ denotes the GW spectrum with the modulations smoothed out.\footnote{In practice, while $\mathcal{A}_\textrm{lin}$, $\omegagwlin$ and $\phi_\textrm{lin}$ are taken as constant here, they can exhibit a mild $k$-dependence that becomes more important away from the center of the principal peak.} This envelope $\overline{\Omega}_{\textrm{GW}}(k)$ can be identified with the scalar-induced GW spectrum due to the envelope $\Pbar$, i.e.~\eqref{eq:OmegaGW-i} with $\Pzeta$ replaced by $\Pbar$, which follows from the findings listed in the first bullet point above.
    
    The oscillations are visible in panel (b) of fig.~\ref{fig:P_OGW_ratios_sharp_s0p3_w20_w40_a0p5}  as the modulations of the principal peak of $\OGW$. In panel (c) we plot $\OGW / \overline{\Omega}_\textrm{GW}-1$, which we observe to follow the template \eqref{eq:OGW_sharp_template} over the range of the principal peak, exhibiting a sinusoidal modulation with an approximately constant amplitude and frequency $\omega_\textrm{lin}^\textsc{gw}$.

    \item The amplitude of oscillations $\mathcal{A}_\textrm{lin}$ in \eqref{eq:OGW_sharp_template} depends on the amplitude of oscillations $A_\textrm{lin}$ in \eqref{eq:P_of_k_sharp}, but also on the ratio of the period of oscillations in $\Pzeta$ and the width $\Delta k$ of the peak in $\Pbar$. This ratio affects the magnitude of $\mathcal{A}_\textrm{lin}$ as follows. For the time being, we assume that $A_\textrm{lin} \sim \mathcal{O}(1)$ so that we can model $\Pzeta$ as a comb of individual peaks as described in sec.~\ref{sec:multiple}. The modulation in $\OGW(k)$ then arises from a superposition of the contributions due to the individual peaks in $\Pzeta$ and their interactions. Now imagine increasing $\omegalin$ while keeping the envelope $\Pbar$ fixed. In this case the overall envelope of $\OGW(k)$ is unchanged, as explained above, but the number of maxima of $\OGW(k)$ in a given $k$-interval is increased. The superposition of a larger number of peaks over the same interval in $k$ then leads to an `averaging-out' effect, resulting in a smaller amplitude $\mathcal{A}_\textrm{lin}$ of the modulation. The upshot is that scalar power spectra with larger values of $\omegalin \Delta k$ lead to smaller values for the amplitude of oscillations $\mathcal{A}_\textrm{lin}$.
    
    This averaging-out effect also leads to the observation that the amplitude of oscillations in $\Omega_\textrm{GW}(k)$ is at most $\mathcal{A}_\textrm{lin} \lesssim \mathcal{O}(10 \%)$ for any example of a sharp feature with a peak. This follows from the fact that to have visible oscillations in the range where $\Pzeta$ is enhanced, the width of the peak in the envelope $\Delta k$ needs to be at least as large as the period of oscillation $2 \pi / \omegalin$, i.e.~in all cases of interest $\omegalin \Delta k > 2 \pi$. The largest modulations are thus expected for $\omegalin \Delta k \sim 2 \pi$ and $A_\textrm{lin}=1$, in which case numerical analyses of various examples give $\mathcal{A}_\textrm{lin} \sim \mathcal{O}(10 \%)$. 
    
    This is what we observe for the model with $(\sigma/k_\star, \, \omegalin k_\star)=(0.3, 20)$ and $A_\textrm{lin}=1$ (blue curve) in fig.~\ref{fig:P_OGW_ratios_sharp_s0p3_w20_w40_a0p5}. This exhibits $\sim 3$ periods of oscillation over the range where the scalar power spectrum is enhanced. The resulting amplitude of oscillations in $\OGW$ can be read off from panel (c) as $\mathcal{A}_\textrm{lin} \sim 20 \%$. Doubling the frequency of oscillations (red curve) this amplitude is reduced to $\mathcal{A}_\textrm{lin} \sim 12 \%$.
    
    \item Decreasing $A_\textrm{lin}$ in \eqref{eq:P_of_k_sharp} has the predictable effect of decreasing the amplitude of the modulations $\mathcal{A}_\textrm{lin}$ in $\OGW(k)$. The attenuation can be quite drastic. For the example with $A_\textrm{lin}=0.5$ in fig.~\ref{fig:P_OGW_ratios_sharp_s0p3_w20_w40_a0p5} (green curve) the amplitude $\mathcal{A}_\textrm{lin}$ is reduced by a factor of $\sim 4$ compared to the example with $A_\textrm{lin}=1$ but otherwise identical model parameters (blue curve). In \cite{Fumagalli:2020nvq}, an example with $A_\textrm{lin}=10 \%$ was shown to give $\mathcal{A}_\textrm{lin} \sim 0.1 \%$, which raises doubt over the detectability of such a modulation with GW observatories.
\end{itemize}

\section{Probing resonant features with the SGWB}
\label{sec:resonant}
In this section we will analyse the scalar-induced contribution to the SGWB due to a resonant feature in the scalar power spectrum, i.e.~a $\log(k)$-periodic modulation of $\Pzeta$. Here we also assume that the resonant feature is associated with an enhancement of scalar fluctuations, i.e.~we consider a scalar power spectrum \eqref{eq:P_of_k_resonant} with a peak in the envelope $\Pbar$. 

Our goal is a general analysis of the spectral shape of $\OGW$ due to a resonant feature, without reference to a specific model. We do this in two complementary ways. Firstly, by approximating the power spectrum as a series of individual peaks displaced from one another by the period set by the frequency of oscillation. This allows for a first understanding of the possible periodic structures that could emerge in $\OGW$ as well to extract a universal frequency $\omegalogc\simeq 4.77$ that signals a qualitative change in the behaviour of $\OGW$. Secondly, we expand the power spectrum in terms of a modulation over an otherwise smooth background and derive semi-analytical templates for $\OGW$. 

There are limitations on how model-independent these analyses can be. The reason is that the shape of the peak in $\Pbar$ will vary between different realisations of a resonant feature and hence one cannot be entirely model-independent when describing the induced GW spectrum. To account for this, we will consider several shapes of peaks in $\Pbar$.
To focus on the effect of the periodic structure only, we will initially consider the case of a top-hat envelope. We will then also present examples with a nontrivial profile for $\Pbar$ such as a lognormal peak, which we will employ in the main text. Our results will however also be valid for other peak profiles as we demonstrate in appendix \ref{app:Gausslike}, where we consider a Gaussian-like peak.

\subsection{Periodic frequency profile of the SGWB}
\label{sec:resonant-peak-structure}
Here we will show that the periodic structure of peaks in $\Pzeta$ leads to periodic structures of peaks in $\OGW$. To demonstrate this we will model the oscillations in $\Pzeta$ as a series of individual peaks, as described in sec.~\ref{sec:multiple}. This can be done as long as the amplitude of oscillations is $A_\textrm{log} \sim \mathcal{O}(1)$, which is also the most interesting case phenomenologically. An analysis valid for any value $A_\textrm{log}$ using a different method will be presented in sec.~\ref{sec:resonant-templates}. 

\subsubsection{Analytical considerations}
\label{sec:resonant-peak-structure-analytical}
We begin by identifying the loci $k_{\star j}$ of the peaks in $\Pzeta$ as given by \eqref{eq:P_of_k_resonant}. Here we assume that the envelope $\Pbar$ is varying more slowly compared to the variation due to the oscillation. As a result, the maxima of $\Pzeta$ in \eqref{eq:P_of_k_resonant} are given to a good approximation by the maxima of the $\cos$-term, corresponding to a $\log(k)$-periodic series of peaks. To remove clutter, note that one can always absorb $\vartheta_\textrm{log}$ by redefining $\kref$, which we will do in the following. The maxima $k_{\star j}$ are thus given by
\begin{align}
\label{eq:k-star-j-def}
    k_{\star j} \equiv \kref \, e^{\frac{2 \pi j}{\omegalog}} \, ,
\end{align}
We can then investigate the peak structure of the GW spectrum by analysing the resonance peaks in $\OGW$ due to the peaks \eqref{eq:k-star-j-def} as described in sec.~\ref{sec:multiple}. However, this approach presupposes that the individual peaks in $\Pzeta$ are narrow in the sense \eqref{eq:narrowdef}. That is, the width of the peak $\Delta k$ has to be small compared to its central value $k_{\star j}$. Here one can confirm that this will always be the case for any resonant feature of phenomenological interest. As the width of a peak we take the interval between two intersections of $\Pzeta$ with $\Pbar$. This gives
\begin{align}
\label{eq:resonant-narrow-condition}
 \frac{\Delta k}{k_{\star j}} = \frac{e^{\frac{\pi}{\omegalog}\big(2j +\frac{1}{2} \big)}-e^{\frac{\pi}{\omegalog}\big(2j -\frac{1}{2} \big)}}{e^{\frac{2 \pi j}{\omegalog}}} = 2 \sinh \Big( \frac{\pi}{2 \omegalog} \Big) \ll 1 \, , \quad \Rightarrow \quad \omegalog \gg \frac{\pi}{2} \, .
\end{align}
Explicit realisations of a resonant feature typically exhibit $\omegalog \gg 1$, so that the condition \eqref{eq:resonant-narrow-condition} is generically satisfied.

This implies that the resonance peak analysis of sec.~\ref{sec:multiple} is applicable in the resonant feature case and we can use \eqref{eq:kmaxij-def} to predict the peak structure of $\OGW$. From \eqref{eq:kmaxij-def} it then follows that every peak $k_{\star j}$ in $\Pzeta$ gives rise to a peak in $\OGW$ at
\begin{align}
\label{eq:kmaxij-jj}
\frac{k_{\textrm{max},jj}}{k_\textrm{ref}} = \frac{2}{\sqrt{3}} e^{\frac{2 \pi j}{\omegalog}} \, .    
\end{align}
Note that this set of peaks is $\log(k)$-periodic with frequency $\omegalog$, i.e.~the periodic peak-structure of $\Pzeta$ leads to a periodic peak-structure in $\OGW$ with the same frequency. This is to be contrasted with the sharp feature case, reviewed in sec.~\ref{sec:sharp-review}, where the frequency of the modulations in $\OGW$ is different from that in $\Pzeta$.

In addition to the peaks given by \eqref{eq:kmaxij-jj}, the GW spectrum exhibits further maxima due to interactions between different peaks in $\Pzeta$. In this case, the constraint in \eqref{eq:kmaxij-def} becomes important as it limits which peaks can interact with one another. For example, consider two neighbouring peaks at $k_{\star j}$ and $k_{\star (j+1)}$ in $\Pzeta$. According to the constraint in \eqref{eq:kmaxij-jj} these can only give rise to a resonance peak in $\OGW$ if the frequency $\omegalog$ is sufficiently large. In particular, one finds that for nearest-neighbour interactions between peaks in $\Pzeta$ to be possible one requires $\omegalog > \omegalogc$ with
\begin{align}
    \label{eq:omegalogc-def}
    \omegalogc= 2 \pi / \log \bigg( \frac{\sqrt{3}+1}{\sqrt{3}-1} \bigg) \simeq 4.77 \, .
\end{align}
If this is satisfied, nearest-neighbour interactions lead to resonance peaks in $\OGW$ at
\begin{align}
\label{eq:kmaxij-jjplus1}
\frac{k_{\textrm{max},j(j+1)}}{k_\textrm{ref}} = \frac{2}{\sqrt{3}} e^{\frac{\pi (2j+1)}{\omegalog}} \, \cosh \Big( \tfrac{\pi}{\omegalog} \Big) \, .    
\end{align}
This again corresponds to a $\log(k)$-periodic structure with frequency $\omegalog$, but the peaks due to nearest-neighbour interactions are shifted in $\log(k)$ compared to the peaks \eqref{eq:kmaxij-jj} due to self-interactions by an amount $\pi/\omegalog + \log \big( \cosh (\pi / \omegalog) \big)$.

This pattern repeats if we take into account interactions between more distant peaks in $\Pzeta$. From the constraint in \eqref{eq:kmaxij-def} it follows that for peaks separated by $n$ periods one requires $\omegalog > n \, \omegalogc$ for interactions to be possible. If this is a case one again finds a $\log(k)$-periodic structure of resonance peaks in $\OGW$ with frequency $\omegalog$.

This suggests that for large $\omegalog \gg \omegalogc$ and a power spectrum with a large number of oscillations the resulting GW spectrum can be quite complicated with multiple $\log(k)$-periodic structures that are shifted w.r.t.~one another. However, one can show that there are additional patterns in $\OGW$ that will ensure that its peak-structure is much simpler. 

To show this, consider two peaks in $\Pzeta$ that are separated from one another by an even number of periods. Without loss of generality we can choose $k_{\star (j-n)}$ and $k_{\star (j+n)}$. The corresponding resonance peak in $\OGW$ due to their interaction is given by
\begin{align}
    \label{eq:kmaxij-even}
    \frac{k_{\textrm{max},(j-n)(j+n)}}{k_\textrm{ref}} = \frac{2}{\sqrt{3}} e^{\frac{2 \pi j}{\omegalog}} \, \cosh \Big( \tfrac{2\pi n}{\omegalog} \Big) \, .  
\end{align}
However, note that for these two peaks to be able to interact one requires $\omegalog > 2n \, \omegalogc$. This implies that the argument of the $\cosh$-term in \eqref{eq:kmaxij-even} is small, $2 \pi n / \omegalog < \pi / \omegalogc < 1$, and decreases further if $\omegalog$ is increased. As a result, for large $\omegalog \gg 2n \, \omegalogc$ we can treat the argument of the $\cosh$-term as a small parameter and expand, finding
\begin{align}
    \label{eq:kmaxij-even-large-omega}
    \frac{k_{\textrm{max},(j-n)(j+n)}}{k_\textrm{ref}} \underset{\omegalog \gg 2n \, \omegalogc}{=} \frac{2}{\sqrt{3}} e^{\frac{2 \pi j}{\omegalog}} \, \Big( 1 + \mathcal{O}\big( \tfrac{2\pi n}{\omegalog} \big)^2  \Big) \, .  
\end{align}
This coincides with the resonance peak in $\OGW$ due to the self-interaction of the peak at $k_{\star j}$, see \eqref{eq:kmaxij-jj}, up to the small quadratic correction. The result is that for sufficiently large $\omegalog$ the resonance peaks due to interactions of peaks that are separated by an even number of periods effectively coincide with the resonance peaks due to self-interactions. That is, these interactions do not give rise to a new $\log(k)$-periodic structure in $\OGW$, but rather reinforce the one already present from self-interactions.  

We can perform a similar analysis of the resonance peaks in $\OGW$ due to peaks in $\Pzeta$ that are separated by an odd number of periods. Here we take these to be at $k_{\star (j-n)}$ and $k_{\star (j+n+1)}$ in which case the resonance peak due to their interaction is given by
\begin{align}
    \label{eq:kmaxij-odd}
    \frac{k_{\textrm{max},(j-n)(j+n+1)}}{k_\textrm{ref}} = \frac{2}{\sqrt{3}} e^{\frac{\pi (2j+1)}{\omegalog}} \, \cosh \Big( \tfrac{\pi (2n+1)}{\omegalog} \Big) \, .  
\end{align} 
This resonance peak is produced as long as $\omegalog > (2n+1) \, \omegalogc$. For sufficiently large $\omegalog \gg (2n+1) \, \omegalogc$ we can again expand the $\cosh$-term in \eqref{eq:kmaxij-odd}. The observation now is that up to quadratic corrections, the peak \eqref{eq:kmaxij-odd} coincides with that due to the nearest-neighbour interaction between the peaks $k_{\star j}$ and $k_{\star (j+1)}$ in $\Pzeta$, see \eqref{eq:kmaxij-jjplus1}. The upshot is that for large $\omegalog$ the interactions due to peaks in $\Pzeta$ that are separated by an odd number of periods do not give rise to new $\log(k)$-periodic structures in $\OGW$, but they instead effectively coincide with the peaks due to nearest-neighbour interactions.

This shows that as $\omegalog$ is increased we do not find new $\log(k)$-periodic peak-structures in $\OGW$, but instead the additional resonance peaks align with either the peaks due to self-interactions, \eqref{eq:kmaxij-jj}, or the ones due to nearest-neighbour interactions, \eqref{eq:kmaxij-jjplus1}. For large $\omegalog$ we can also show that the relation between these two sets of peaks simplifies. As mentioned above, these two sets of peaks are separated w.r.t.~one another by a shift $\pi/\omegalog + \log \big( \cosh (\pi / \omegalog) \big)$ in $\log(k)$. For large $\omegalog \gg \pi$ we can once more expand the $\cosh$-term to find
\begin{align}
    \tfrac{\pi}{\omegalog} + \log \Big( \cosh \tfrac{\pi}{\omegalog} \Big) = \tfrac{\pi}{\omegalog} + \mathcal{O} \Big( \big( \tfrac{\pi}{\omegalog} \big)^2 \Big) \, ,
\end{align}
i.e.~the shift is given by $\pi / \omegalog$ at leading order. Interestingly, a shift by $\pi / \omegalog$ is exactly what one would expect for a $\log(k)$-periodic structure with frequency $2 \omegalog$. That is, for large $\omegalog$ one finds that the two sets of peaks corresponding to self-interactions and nearest-neighbour interactions arrange one another in such a way as to give rise to a single $\log(k)$-periodic structure with frequency $2 \omegalog$.

\begin{figure}[t]
\centering
\begin{overpic}[width=0.8\textwidth]{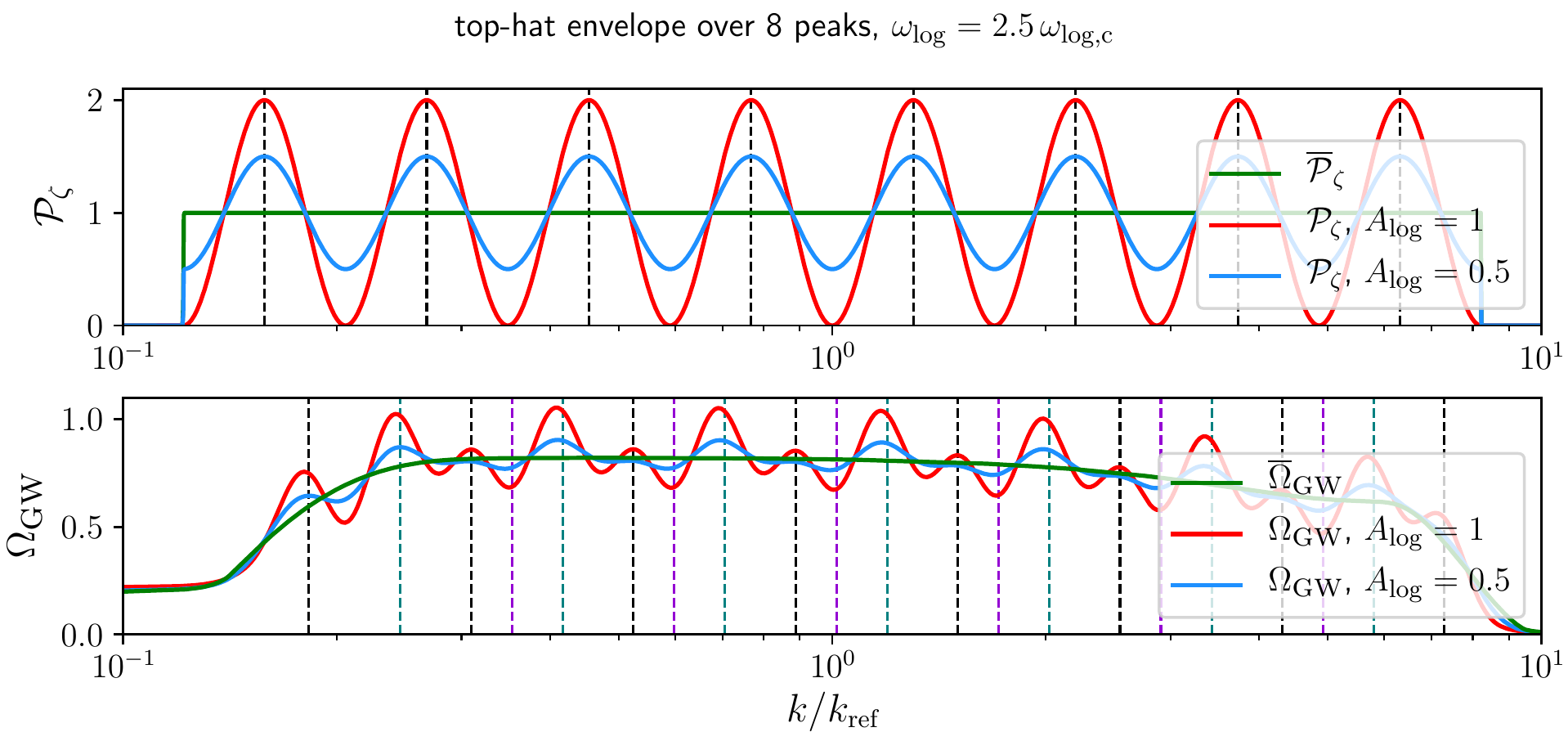}
\end{overpic}
\caption{\textit{$\Pzeta$ (upper panel) and $\OGW$ (lower panel) vs.~$k/\kref$ for a scalar power spectrum consisting of 8 peaks of equal amplitude with $\omegalog= 2.5 \omegalogc$ and $A_\textrm{log}=1$ (red curves) or $A_\textrm{log}=0.5$ (blue curves). The green curves denote $\Pbar$ and the corresponding GW spectrum $\overline{\Omega}_\textrm{GW}$. The vertical black dashed lines in the upper panel denote the loci of the maxima of $\Pzeta$. The vertical dashed lines in the lower panel denote the expected locations of resonance peaks in $\OGW$ from the interactions of peaks in $\Pzeta$ with themselves (black), their nearest neighbours (blue) and next-to-nearest-neighbours (violet).}}
\label{fig:P_OGW_2p5wc_8}
\end{figure}

\subsubsection{Numerical confirmations}
\label{sec:resonant-peak-structure-numerical}
To illustrate the above findings we turn to explicit examples. Firstly, to solely focus on the effect of the oscillation in $\Pzeta$ on $\OGW$, we consider the case of a scalar power spectrum with a top-hat envelope. This can be seen as a toy model for a more realistic example with a finite number of oscillations of comparable amplitude. Here we will consider \eqref{eq:P_of_k_resonant} with
\begin{align}
    \label{eq:P-top-hat-envelope}
   \Pbar(k) = \begin{cases} \mathcal{P}_\textrm{\textrm{max}}=1 \, , \quad & e^{-\pi n_\textrm{p} / \omegalog} \leq k/k_\textrm{ref} \leq e^{+\pi n_\textrm{p} / \omegalog} \, , \\
    0 \, , \quad & \textrm{otherwise} \, , \end{cases}
\end{align}
and $\vartheta_\textrm{log}=\pi (n_\textrm{p}+1)$, 
corresponding to a resonant feature over $n_\textrm{p}$ peaks (i.e.~periods of oscillation) of equal height.

In fig.~\ref{fig:P_OGW_2p5wc_8} we plot the scalar power spectrum and the corresponding GW energy density fraction for $\omegalog = 2.5 \omegalogc$ and $n_\textrm{p}=8$. In the upper panel we plot $\Pbar$ (green) together with $\Pzeta$ for the choices $A_\textrm{log}=1$ (red) and $A_\textrm{log}=0.5$ (blue). The loci of the peaks are marked by the vertical black dashed lines. In the lower panel we display the corresponding plots for the GW spectrum, with $\overline{\Omega}_\textrm{GW}$ denoting the GW spectrum for a power spectrum given by $\Pbar$.\footnote{We have encountered this result before: The top-hat scalar power spectrum employed in fig.~\ref{fig:Broad_Narrow_top_LN} is given by \eqref{eq:P-top-hat-envelope} with $\omegalog = 2.5 \omegalogc$ and $n_\textrm{p}=8$, which together determine the range of the top hat.} The vertical dashed lines indicate the predicted loci of the resonance peaks. For $\omegalog = 2.5 \omegalogc$ we expect resonance from self-interactions of the peaks in $\Pzeta$, from nearest-neighbour-interactions and next-to-nearest-neighbour-interactions. The expected peak locations are shown as the black, blue and violet dashed lines, respectively. 

\begin{figure}[t]
\centering
\begin{overpic}[width=0.8\textwidth]{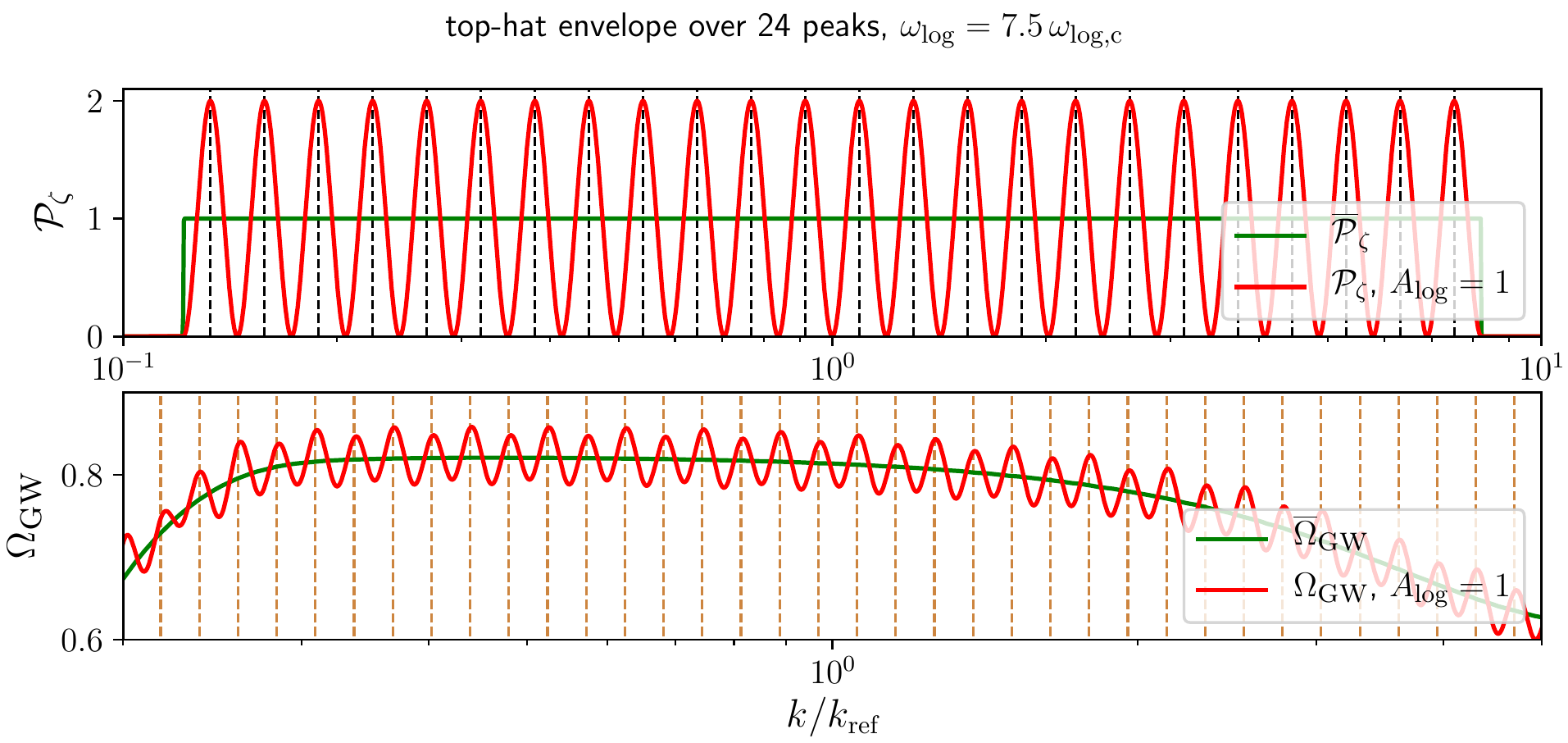}
\end{overpic}
\caption{\textit{$\Pzeta$ (upper panel) and $\OGW$ (lower panel) vs.~$k/\kref$ for a scalar power spectrum consisting of 24 peaks of equal amplitude with $\omegalog= 7.5 \omegalogc$ and $A_\textrm{log}=1$. The green curves denote $\Pbar$ and the corresponding GW spectrum $\overline{\Omega}_\textrm{GW}$. The vertical black dashed lines in the upper panel denote the loci of the maxima of $\Pzeta$. The brown vertical dashed lines in the lower panel are for a $\log(k)$-periodic structure with frequency $2 \omegalog$.}}
\label{fig:P_OGW_7p5wc_24}
\end{figure}

We make the following observations: For $A_\textrm{log}=1$ the result for $\OGW$ exhibits two series of peaks, whose maxima coincide with the predicted locations of peaks due to self-interaction and nearest-neighbour interactions (black and blue dashed lines). For $A_\textrm{log}=0.5$ the overall amplitude of the modulations is suppressed and as a result the maxima due to self-interactions have effectively disappeared. The remaining visible peaks are however still consistent with the expected loci from the resonance analysis. The peaks expected from next-to-nearest-neighbour-interactions (violet dashed lines) are however absent for both values of $A_\textrm{log}$. This can be explained as follows. For $\omegalog = 2.5 \omegalogc$ the frequency is just above the threshold for next-to-nearest-neighbour interactions to be possible, which require $\omegalog > 2 \omegalogc$. In this case the resulting resonance peaks are suppressed in amplitude compared to the contribution from self- or nearest-neighbour-interactions and remain invisible in the full result.\footnote{This can e.g.~be seen by writing $\Pzeta$ as a sum of individual peaks and computing the contribution to $\OGW$ from the various terms individually. The contribution from self- or nearest-neighbour-interactions is then observed to dominate over that from next-to-nearest-neighbour interactions.}

For comparison, in fig.~\ref{fig:P_OGW_7p5wc_24} we consider an example with the same envelope as in fig.~\ref{fig:P_OGW_2p5wc_8}, but with a larger value for the frequency $\omegalog = 7.5 \omegalogc$ and $n_\textrm{p}=24$ peaks. In the upper panel we again display $\Pbar$ (green) and $\Pzeta$ (red), now only considering the case with $A_\textrm{log}=1$. In the lower panel we once more plot $\overline{\Omega}_\textrm{GW}$ and $\OGW$, focussing on a smaller interval in $k / \kref$ for better visibility. Compared to the example with $\omegalog = 2.5 \omegalogc$, the amplitude of the modulations is now more homogeneous across the various peaks. In addition, the spacing of peaks is now more regular, effectively coinciding with a $\log(k)$-periodic structure with frequency $2 \omegalog$ as denoted by the brown vertical dashed lines. The two examples in figs.~\ref{fig:P_OGW_2p5wc_8} and \ref{fig:P_OGW_7p5wc_24} thus illustrate the findings recorded in subsection \ref{sec:resonant-peak-structure-analytical}. For smaller values of $\omegalog$ the GW spectrum $\OGW$ displays several $\log(k)$ periodic structures with frequency $\omegalog$, while for larger values of $\omegalog$ a single $\log(k)$-periodic structure with frequency $2 \omegalog$ emerges. We will confirm this quantitatively with another method in section \ref{sec:resonant-templates}.\\

\begin{figure}[t]
\centering
\begin{overpic}[width=0.8\textwidth]{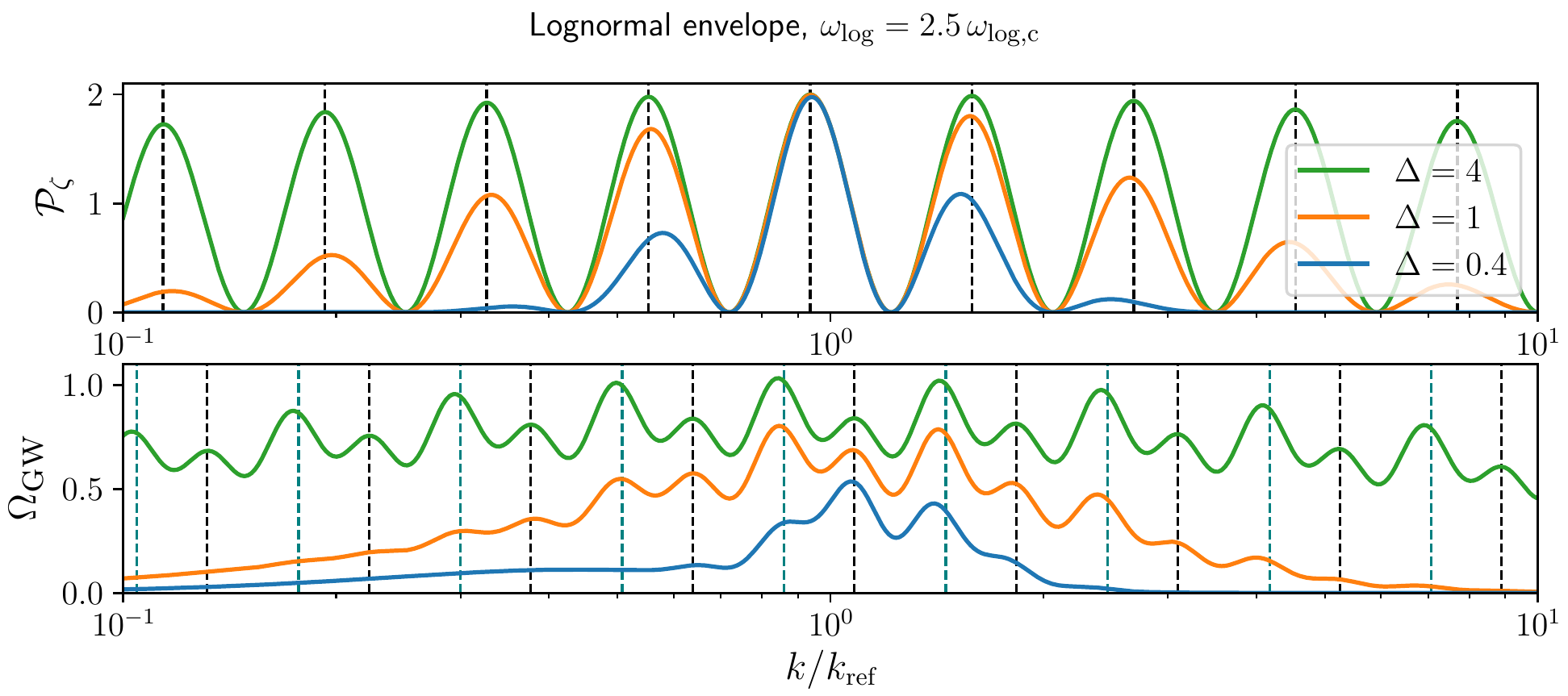}
\end{overpic}
\caption{\textit{$\Pzeta$ (upper panel) and $\OGW$ (lower panel) vs.~$k/\kref$ for a scalar power spectrum \eqref{eq:P_of_k_resonant} with $\omegalog = 2.5 \omegalogc$, $\vartheta_\textrm{log} = \pi /4$ and $A_\textrm{log}=1$ and envelope $\Pbar$ given by lognormal peaks as in \eqref{eq:Pbar-LN} with different choices for the `width' $\Delta$. The curves shown are for $\Delta=4$ (green), $\Delta=1$ (orange) and  $\Delta=0.4$ (blue). The black dashed lines in the upper panel correspond to the maxima of the $\cos$-term in \eqref{eq:P_of_k_resonant} which coincide with the maxima of $\Pzeta$ to a good approximation. The vertical dashed lines in the lower panel indicate the expected locations of resonance peaks from the self-interaction of peaks in $\Pzeta$ (black) and from nearest-neighbour interactions (blue).}}
\label{fig:P_OGW_LN_2p5wc}
\end{figure}

The resonance peak analysis is also successful in predicting the loci of maxima in $\OGW$ if the envelope $\Pbar$ is not constant as in \eqref{eq:P-top-hat-envelope}. This can be e.g.~seen in fig.~\ref{fig:P_OGW_LN_2p5wc} where we plot $\Pzeta$ and $\OGW$ for a scalar power spectrum of resonant feature type \eqref{eq:P_of_k_resonant} with 
\begin{align}
    \label{eq:Pbar-LN}
   \Pbar(k) = \mathcal{P}_\textrm{max} \, \exp \bigg(- \frac{1}{2 \Delta^2} \bigg( \log \frac{k}{\kref} \bigg)^2 \bigg) \, , \quad \textrm{with} \quad \mathcal{P}_\textrm{max}=1 \, .
\end{align}
Here we have further set $A_\textrm{log}=1$ for better visibility of the modulations and chosen $\vartheta_\textrm{log} = \pi /4$ so that we avoid the tuned situation where a maximum, minimum or zero of the cos-term coincides with the maximum of $\Pbar$. In fig.~\ref{fig:P_OGW_LN_2p5wc} we plot $\Pzeta$ (upper panel) and $\OGW$ (lower panel) for $\omegalog = 2.5 \omegalogc$ and $\Delta =4,1,0.4$. The black dashed lines in the upper panel denote the approximate loci of the peaks $\Pzeta$ given as the maxima of the $\cos$-term in \eqref{eq:P_of_k_resonant}.\footnote{The exact positions of the peaks in $\Pzeta$ are somewhat shifted from the maxima of the $\cos$-term due to the non-constant envelope, but this only results in a small shift, even for the narrowest envelope with $\Delta=0.4$.} The vertical dashed lines in the lower panel correspond to the expected positions of resonance peaks due to the self-interaction of peaks in $\Pzeta$ (black) and nearest-neighbour-interactions (blue). We observe a visible modulation of $\OGW$ in all three cases and the maxima are predicted to a high degree of accuracy by the resonance-peak analysis. 

We can also make observations in the numerical results in figs.~\ref{fig:P_OGW_2p5wc_8}, \ref{fig:P_OGW_7p5wc_24} and \ref{fig:P_OGW_LN_2p5wc} which cannot be explained simply with the resonance-peak analysis alone:
\begin{itemize}
    \item  For example, the modulation in $\OGW$ can be observed to take the form of an oscillation about the GW spectrum $\overline{\Omega}_\textrm{GW}$ obtained for the envelope of the scalar power spectrum. This behaviour had also been observed in \cite{Fumagalli:2020nvq} for the case of a sharp feature in $\Pzeta$. 
    \item Comparing the GW spectra for $\omegalog= 2.5 \omegalogc$ (and $A_\textrm{log}=1$) with that for $\omegalog= 7.5 \omegalogc$, one finds that the increase in $\omegalog$ results in a decrease of the amplitude of the modulations in $\OGW$.
    \item Also, for $\omegalog= 2.5 \omegalogc$ the amplitude of alternating peaks in $\OGW$ is fairly different while for $\omegalog= 7.5 \omegalogc$ this amplitude of oscillations is much more homogeneous.
\end{itemize}
In the following, we will present a different method for analysing the GW spectrum for a resonant feature which will permit us to explain these findings.

\subsection{Theoretically-motivated templates for $\OGW(k)$}
\label{sec:resonant-templates}
In this section we will present a complementary approach for analysing the spectral shape of $\OGW(k)$ due to a resonant feature, which is at the same time more powerful than the resonance peak analysis described in sec.~\ref{sec:resonant-peak-structure}. In particular, given a scalar power spectrum of the type \eqref{eq:P_of_k_resonant} we will be able to give a semi-analytical template for the corresponding GW energy density spectrum $\OGW(k)$.

\subsubsection{Analytical considerations}
\label{sec:resonant-templates-analytical}
Here, we will exploit the fact that the scalar power spectrum can be written in terms of a modulation about a smooth background to consider the contributions to $\OGW(k)$ from the smooth background and the modulations separately. That is, we insert the scalar power spectrum \eqref{eq:P_of_k_resonant} into the expression \eqref{eq:OmegaGW-i} for $\OGW(k)$ and expand in powers of $A_\textrm{log}$:
\begin{align}
\label{eq:OGW-analyt-expansion}
        \Omega_{\textrm{GW}} (k)= \Omega_{\textrm{GW},0} (k) + A_\textrm{log} \Omega_{\textrm{GW},1} (k) + A_\textrm{log}^2 \Omega_{\textrm{GW},2} (k) \, ,
\end{align}
with
\begin{align}
\label{eq:OmegaGW-i-0}
    \Omega_{\textrm{GW},0}(k) &= \int_0^{\frac{1}{\sqrt{3}}} \textrm{d} d \int_{\frac{1}{\sqrt{3}}}^\infty \textrm{d} s \, \TRD (d,s) \, \Pbar \bigg(\frac{\sqrt{3}k}{2}(s+d)\bigg) \Pbar \bigg(\frac{\sqrt{3}k}{2}(s-d)\bigg) \, , \\
    \label{eq:OmegaGW-i-1}
    \Omega_{\textrm{GW},1}(k) &= \int_0^{\frac{1}{\sqrt{3}}} \textrm{d} d \int_{\frac{1}{\sqrt{3}}}^\infty \textrm{d} s \, \TRD (d,s) \, \times \\
    \nonumber &\hphantom{=} \times \bigg[ \hphantom{+} \Pbar \bigg(\frac{\sqrt{3}k}{2}(s+d)\bigg) \Pbar \bigg(\frac{\sqrt{3}k}{2}(s-d)\bigg) \, \cos \bigg(\omegalog \log \bigg( \frac{\sqrt{3}k}{2 \kref}(s+d) \bigg) \bigg) \\
    \nonumber &\hphantom{= \ \times } + \Pbar \bigg(\frac{\sqrt{3}k}{2}(s+d)\bigg) \Pbar \bigg(\frac{\sqrt{3}k}{2}(s-d)\bigg) \, \cos \bigg(\omegalog \log \bigg( \frac{\sqrt{3}k}{2 \kref}(s-d) \bigg) \bigg) \bigg] \, , \\
    \label{eq:OmegaGW-i-2}
    \Omega_{\textrm{GW},2}(k) &= \int_0^{\frac{1}{\sqrt{3}}} \textrm{d} d \int_{\frac{1}{\sqrt{3}}}^\infty \textrm{d} s \, \TRD (d,s) \, \Pbar \bigg(\frac{\sqrt{3}k}{2}(s+d)\bigg) \Pbar \bigg(\frac{\sqrt{3}k}{2}(s-d)\bigg) \ \times \, \\
    \nonumber &\hphantom{=} \times \cos \bigg(\omegalog \log \bigg( \frac{\sqrt{3}k}{2 \kref}(s+d) \bigg) \bigg) \cos \bigg(\omegalog \log \bigg( \frac{\sqrt{3}k}{2 \kref}(s-d) \bigg) \bigg) \, ,
\end{align}
where we have also absorbed $\vartheta_\textrm{log}$ by a redefinition of $\kref$.
In the following, we will analyse the three contributions $\Omega_{\textrm{GW},0}$, $\Omega_{\textrm{GW},1}$ and $\Omega_{\textrm{GW},2}$ in turn. Interestingly, it is the $\log(k)$-periodic structure that will allow us to make analytical progress.

The quantity $\Omega_{\textrm{GW},0}(k)$ can be identified as the GW energy density spectrum induced by scalar fluctuations with power spectrum $\Pbar(k)$. In this work we consider a resonant feature associated with an enhancement of fluctuations which implies that $\Pbar(k)$ exhibits a peak. While the shape of this peak is in general model-dependent and will depend on the explicit mechanism of enhancement, some general observations can still be made: The contribution $\Omega_{\textrm{GW},0}(k)$ will exhibit either one or two peaks (one lower broad one and a taller narrower one) depending on the width of the peak in $\Pbar(k)$. We can also be more precise by using the quantitative results from section \ref{sec:broad-narrow} which apply if the peak in $\Pbar(k)$ is broad or narrow, respectively. In particular, for broad peaks we can use \eqref{eq:OmegaGW-i-broad}, while for moderately narrow peaks we will employ the heuristic result from
\eqref{eq:OmegaGW-as-Psquared-fit}, i.e.:
\begin{align}
\label{eq:OGW0-broad} \textrm{broad peak in } \Pbar\textrm{:} \quad \Omega_{\textrm{GW},0}(k) &\simeq 0.823 \, \bigg( \textrm{max} \Big( \Pbar(k) \Big) \bigg)^2 \, , \\
\label{eq:OGW0-narrow} \textrm{narrow peak in } \Pbar\textrm{:} \quad \Omega_{\textrm{GW},0}(k) &\simeq 0.823 \, \tilde{\gamma} \, \Pbar^2 \bigg(\frac{\sqrt{3}k}{2}\bigg) \, ,
\end{align}
where these expressions are expected to apply in the vicinity of the maximum of $\Omega_{\textrm{GW},0}$ and $\tilde{\gamma}$ is a numerical factor that needs to be determined by fitting to e.g.~the numerical result.

Next, we turn to the contributions $\Omega_{\textrm{GW},1}(k)$ and $\Omega_{\textrm{GW},2}(k)$. Both expressions contain an integration over two factors of $\Pbar$ and an oscillatory function. We will now continue the analysis assuming that the peak in $\Pbar$ is broad as defined in \eqref{eq:k-broad}, however we will later state how our findings can be generalised. In this case, when evaluated at values of $k$ near the central part of the peak in $\Pbar$, the two factors of $\Pbar$ can be treated as effectively constant and factored out from the integral.\footnote{The reason is that the dominant contribution to the integral comes from a finite part of the integration domain over $(d,s)$ where the kernel $\TRD(d,s)$ is unsuppressed. A sufficiently broad peak can be seen as effectively constant over this integration domain, cf.~the discussion in sec.~\ref{app:broad}.} Thus, \eqref{eq:OmegaGW-i-1} and \eqref{eq:OmegaGW-i-2} become 
\begin{align}
\label{eq:OmegaGW-i-1-factored}
    \Omega_{\textrm{GW},1}(k) &= \Pbar^2 \int_0^{\frac{1}{\sqrt{3}}} \textrm{d} d \int_{\frac{1}{\sqrt{3}}}^\infty \textrm{d} s \, \TRD (d,s) \bigg[ \cos \Big(X(d,s) + \varphi_k \Big) + \cos \Big(Y(d,s) + \varphi_k \Big) \bigg] , \\
    \label{eq:OmegaGW-i-2-factored}
    \Omega_{\textrm{GW},2}(k) &= \Pbar^2 \int_0^{\frac{1}{\sqrt{3}}} \textrm{d} d \int_{\frac{1}{\sqrt{3}}}^\infty \textrm{d} s \, \TRD (d,s) \cos \Big(X(d,s) + \varphi_k \Big) \cos \Big(Y(d,s) + \varphi_k \Big) \, ,
\end{align}
where the presence of the $\log$ allowed us to write the arguments of the $\cos$-terms as a sum and introduced the quantities
\begin{align}
\label{eq:XYphidef}
    X(d,s) \equiv \omegalog \log(s+d) , \quad Y(d,s) \equiv \omegalog \log(s-d) \, ,\quad \varphi_k \equiv \omegalog \log \bigg( \frac{\sqrt{3}k}{2 \kref} \bigg) \, .
\end{align}
We can then use trigonometric identities to factor out the $k$-dependent parts from the integrands. In particular, in \eqref{eq:OmegaGW-i-1-factored} we write:
\begin{align}
\cos(X+\varphi_k) + \cos(Y+\varphi_k) = \Big[\cos X + \cos Y \Big] \cos \varphi_k - \Big[\sin X + \sin Y \Big] \sin \varphi_k \, ,
\end{align}
while in \eqref{eq:OmegaGW-i-2-factored} we employ:
\begin{align}
\cos(X+\varphi_k)\cos(Y+\varphi_k) = \frac{1}{2}\Big[\cos(X+Y)\cos(2 \varphi_k)-\sin(X+Y)\sin(2 \varphi_k )+\cos(X-Y)\Big].
\end{align}
Inserting this into \eqref{eq:OmegaGW-i-1-factored}, and \eqref{eq:OmegaGW-i-2-factored} and rearranging, we arrive at the following expressions for $\Omega_{\textrm{GW},1}(k)$ and $\Omega_{\textrm{GW},2}(k)$:\footnote{To simplify, we have used the following identity: $$ a\cos z -b\sin z = (a^2+b^2)^{1/2}\cos(z+\theta) \, , \quad \textrm{with} \quad \tan \theta = b/a \, .$$}
\begin{align}
 \label{analyticalO1}
 \Omega_{\textrm{GW},1}(k) &= \Pbar^2 \, \Big(A_1(\omegalog)^2+B_1(\omegalog)^2\Big)^{1/2}\cos \bigg[ \omegalog \log \bigg( \frac{\sqrt{3}k}{2 \kref} \bigg) + \theta_1(\omegalog) \bigg] \, , \\
 \label{analyticalO2b} 
 \Omega_{\textrm{GW},2}(k) &= \Pbar^2 \Bigg\{  \Big(A_2(\omegalog)^2+B_2(\omegalog)^2 \Big)^{1/2} \cos\bigg[2 \omegalog \log \bigg( \frac{\sqrt{3}k}{2 \kref} \bigg) + \theta_2(\omegalog) \bigg] + C_2(\omegalog) \bigg\} \, ,
\end{align}
with 
\begin{align}
\label{eq:tan1_and_tan2_def}
\tan \big( \theta_1(\omegalog) \big)= \frac{B_1(\omegalog)}{A_1(\omegalog)} \, , \qquad \tan \big( \theta_2(\omegalog) \big)= \frac{B_2(\omegalog)}{A_2(\omegalog)} \, , 
\end{align}
and
\begin{align}\label{eq:A1}
    A_1(\omegalog) &= \int_0^{\frac{1}{\sqrt{3}}} \textrm{d} d \int_{\frac{1}{\sqrt{3}}}^\infty \textrm{d} s \, \TRD (d,s) \, \Big[ \cos \big( \omegalog \log(s+d) \big) + \cos \big( \omegalog \log(s-d) \big) \Big],\\
    \label{eq:B1}
    B_1(\omegalog) &= \int_0^{\frac{1}{\sqrt{3}}} \textrm{d} d \int_{\frac{1}{\sqrt{3}}}^\infty \textrm{d} s \, \TRD (d,s) \, \Big[\sin \big( \omegalog \log(s+d) \big) + \sin \big( \omegalog \log(s-d) \big) \Big],
\end{align}
and
\begin{align}\label{eq:A2}
    A_2(\omegalog) &= \frac{1}{2} \int_0^{\frac{1}{\sqrt{3}}} \textrm{d} d \int_{\frac{1}{\sqrt{3}}}^\infty \textrm{d} s \, \TRD (d,s) \, \cos\big(\omegalog \log (s+d)(s-d)\big),\\
    \label{eq:B2}
    B_2(\omegalog) &= \frac{1}{2} \int_0^{\frac{1}{\sqrt{3}}} \textrm{d} d \int_{\frac{1}{\sqrt{3}}}^\infty \textrm{d} s \, \TRD (d,s) \, \sin\big(\omegalog \log (s+d)(s-d)\big),\\
    \label{eq:C2}
    C_2(\omegalog) &= \frac{1}{2} \int_0^{\frac{1}{\sqrt{3}}} \textrm{d} d \int_{\frac{1}{\sqrt{3}}}^\infty \textrm{d} s \, \TRD (d,s) \, \cos\bigg(\omegalog \log \frac{s+d}{s-d}\bigg) .
\end{align}
What \eqref{analyticalO1} and \eqref{analyticalO2b} show is that both $\Omega_{\textrm{GW},1}(k)$ and $\Omega_{\textrm{GW},2}(k)$ are sinusoidal functions of $\log(k)$. However, while $\Omega_{\textrm{GW},1}(k)$ exhibits a modulation with frequency $\omegalog$, the contribution $\Omega_{\textrm{GW},2}(k)$ oscillates with frequency $2\omegalog$. This is consistent with our analysis of the peak-structure in sec.~\ref{sec:resonant-peak-structure}, where we also found $\OGW(k)$ to be a superposition of periodic structures with frequencies $\omegalog$ and $2\omegalog$. The relative amplitude of the two structures and their relative phase depend on the value of $\omegalog$ through the parameters $A_{1,2}(\omegalog)$ and $B_{1,2}(\omegalog)$. Also note that $\Omega_{\textrm{GW},2}(k)$ contains a constant offset $C_2(\omegalog)$, which will however only play a negligible role for $\omegalog \gtrsim \omegalogc$. The phases $\theta_1(\omegalog)$ and $\theta_2(\omegalog)$ are defined implicitly in \eqref{eq:tan1_and_tan2_def} and some care is needed to ensure the correct quadrant is chosen when inverting the $\tan$.

\begin{figure}[t]
\centering
\begin{overpic}[width=0.7\textwidth]{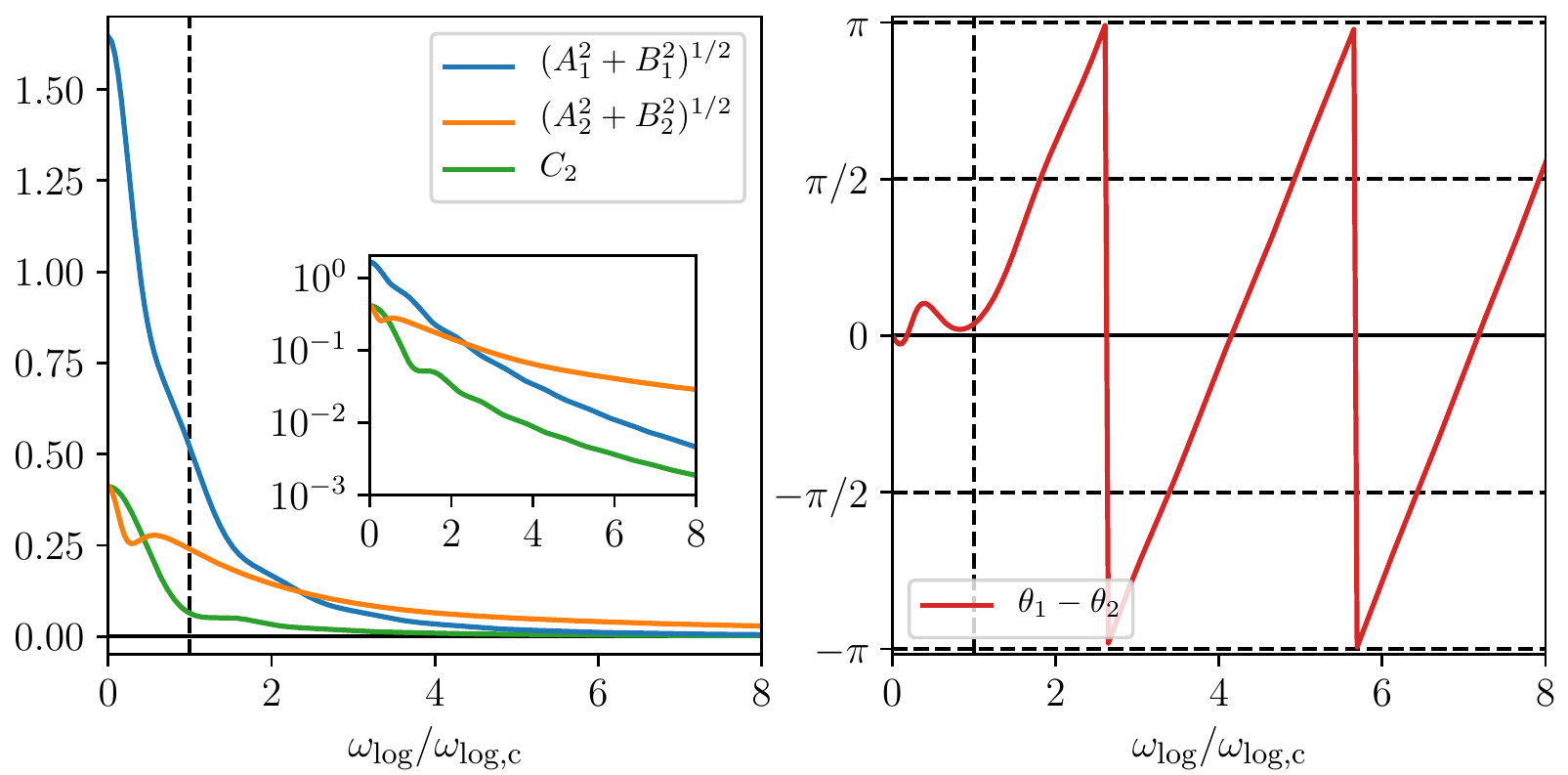}
\end{overpic}
\caption{\textit{Amplitudes $(A_1^2+B_1^2)^{1/2}$, $(A_2^2+B_2^2)^{1/2}$ of the cosine-terms in \eqref{analyticalO1}, \eqref{analyticalO2b} and the offset $C_2$ (left panel), and the phase difference $\theta_1-\theta_2$ (right panel) plotted vs.~$\omegalog / \omegalogc$, with $A_1$, $B_1$ defined in \eqref{eq:A1}, \eqref{eq:B1}, $A_2$, $B_2$, $C_2$ given in \eqref{eq:A2}-\eqref{eq:C2} and $\theta_1$, $\theta_2$ defined via \eqref{eq:tan1_and_tan2_def}. The dashed vertical line in the left panel denotes $\omegalog=\omegalogc$. The inset shows the same data as a log-plot for better visibility of the drop-off for larger values of $\omegalog$.}}
\label{fig:ABC}
\end{figure}

In fig.~\ref{fig:ABC} we plot the amplitudes of the two cosine terms in \eqref{analyticalO1} and \eqref{analyticalO2b} as a function of $\omegalog / \omegalogc$ (LHS) and also their phase difference $\theta_1-\theta_2$ (RHS). We make the following observations: 
\begin{itemize}
    \item Both amplitudes $(A_1^2+B_1^2)^{1/2}$ and $(A_2^2+B_2^2)^{1/2}$ of the $\cos$-terms in \eqref{analyticalO1} and \eqref{analyticalO2b} decrease as $\omegalog$ is increased. Thus, the size of the modulations on $\OGW(k)$ is diminished for higher oscillation frequencies, independently of the value of $A_\textrm{log}$. This is consistent with the behaviour of the examples corresponding to the red curves in figs.~\ref{fig:P_OGW_2p5wc_8} and \ref{fig:P_OGW_7p5wc_24} where the amplitude of the modulation is smaller for the example with the larger value of $\omegalog$. We also found a similar effect for the case of a sharp feature, where this attenuation of the modulation was understood due to an `averaging-out' effect, see sec.~\ref{sec:sharp-review} and \cite{Fumagalli:2020nvq}. Here we find that this also occurs for resonant features.
    \item Now assume that $A_\textrm{log}=1$, so that the relative importance of $\Omega_{\textrm{GW},1}$ and $\Omega_{\textrm{GW},2}$ is just determined by the amplitudes $(A_1^2+B_1^2)^{1/2}$ and $(A_2^2+B_2^2)^{1/2}$ (and the offset $C_2$). From fig.~\ref{fig:ABC} it then follows that for small values $\omegalog < \omegalogc$ the contribution $\Omega_{\textrm{GW},1}$ (with frequency $\omegalog$) dominates over $\Omega_{\textrm{GW},2}$ (with frequency $2\omegalog$). The phase difference is also not very different from 0 in this regime, so that the maxima of $\Omega_{\textrm{GW},2}$ coincide with the extrema (both maxima and minima) of $\Omega_{\textrm{GW},1}$. The effect is that the maxima of $\Omega_{\textrm{GW},1}$ are increased by a small amount, while the minima are slightly boosted upwards. Overall, one finds a periodic structure with frequency $\omegalog$, consistent with the resonance-peak analysis from sec.~\ref{sec:resonant-peak-structure}.
    \item As $\omegalog$ is increased, the amplitude of $\Omega_{\textrm{GW},1}$ drops off faster than that of $\Omega_{\textrm{GW},2}$ so that for sufficiently large $\omegalog$ the contribution $\Omega_{\textrm{GW},2}$ dominates over $\Omega_{\textrm{GW},1}$ as predicted by the analysis in  sec.~\ref{sec:resonant-peak-structure}. The cross-over, i.e.~the point where both contributions are equal, happens at $\omegalog \approx 2.3 \omegalogc$ and e.g.~for $\omegalog = 6 \omegalogc$ one has that $(A_1^2+B_1^2)^{1/2} \approx \tfrac{1}{4} (A_2^2+B_2^2)^{1/2}$. Thus, for $\omegalog \gtrsim 6 \omegalogc$ the oscillatory part of $\OGW$ can be very well described by the periodic structure with frequency $2 \omegalog$ alone.\footnote{The precise value for this threshold is however a matter of choice, depending on how well one wishes to approximate the oscillatory part of $\OGW$.} This is for example what we observed in fig.~\ref{fig:P_OGW_7p5wc_24} for the example with $\omegalog=7.5 \omegalogc$.
    
\begin{figure}[t]
\centering
\begin{overpic}[width=0.8\textwidth]{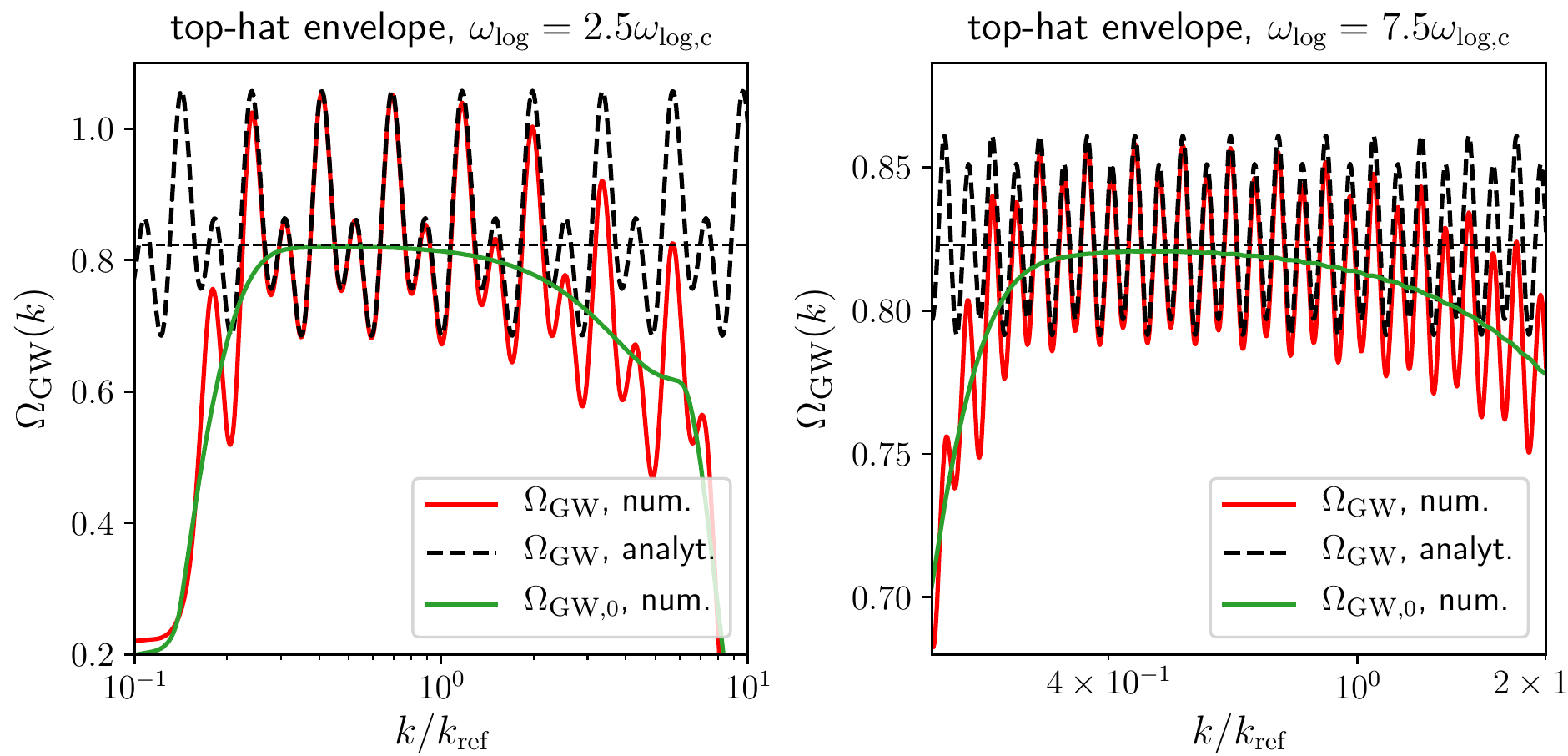}
\end{overpic}
\caption{\textit{Numerical result for $\OGW$ (red) and $\Omega_{\textrm{GW},0}$ (green) vs.~$k/\kref$ for a scalar power spectrum \eqref{eq:P_of_k_resonant} with $A_\textrm{log}=1$ and a broad top-hat envelope and $\omegalog =2.5 \omegalogc$ (left panel) and $\omegalog =7.5 \omegalogc$ (right panel). The envelope is identical in both cases and covers 8 or 24 periods of oscillation, respectively. We compare the numerical result for $\OGW$ with our analytical approximation for $\OGW =\Omega_{\textrm{GW},0}+ \Omega_{\textrm{GW},1}+\Omega_{\textrm{GW},2}$. Here we use the broad-peak result \eqref{eq:OmegaGW-i-broad} for $\Omega_{\textrm{GW},0}$ and \eqref{analyticalO1}, \eqref{analyticalO2b} for $\Omega_{\textrm{GW},1}$, $\Omega_{\textrm{GW},2}$. As expected, the analytical result matches the full numerical solution very well near the centre of the peak of $\OGW$. The horizontal dashed line denotes 0.823, i.e.~the maximal amplitude of $\OGW$ for a broad peak in $\Pzeta$ with unit amplitude.}}
\label{fig:OGW_2p5_7p5_vs_analyt}
\end{figure}
    
    \item There is then an intermediate regime $\omegalogc < \omegalog \lesssim 6 \omegalogc$ where the two contributions $\Omega_{\textrm{GW},1}$ and $\Omega_{\textrm{GW},2}$ have comparable magnitude.\footnote{Again, the upper limit $6 \omegalogc$ is to some extent a matter of choice which we included for definiteness, but readers may choose a different value $\lesssim \mathcal{O}(10) \omegalogc$.} Also, for $\omegalog > \omegalogc$ their phase difference evolves linearly with $\omegalog / \omegalogc$ with slope $\simeq 2$. In the regime $\omegalogc < \omegalog \lesssim 6 \omegalogc$ we hence expect a more complicated peak structure in $\OGW(k)$ with one taller and one lower set of maxima (and minima). This is what was obtained numerically for the example with $\omegalog= 2.5 \omegalogc$ shown in fig.~\ref{fig:P_OGW_2p5wc_8}.
    \item If $A_\textrm{log}$ is decreased, the overall size of modulations is reduced, but also the relative importance of $\Omega_{\textrm{GW},2}$ compared to $\Omega_{\textrm{GW},1}$ is diminished, as follows from \eqref{eq:OGW-analyt-expansion}. The effect is that the various regimes described above are shifted to larger values of $\omegalog$, i.e.~the periodic structure with frequency $\omegalog$ dominates over a larger range of frequencies and the double-frequency structure becomes dominant only at even larger values of $\omegalog$. This effect is however limited, as by reducing $A_\textrm{log}$ too much the modulations in $\OGW$ become unobservably small. See appendix \ref{app:reduce-A-log} for details.
    \item For $\omegalog \rightarrow 0$ the results in fig.~\ref{fig:ABC} are consistent with a constant scalar power spectrum with no oscillations, i.e.~where the cosine term in $\Pzeta$ is replaced by unity. In particular, for $\omegalog \rightarrow 0$ we find $\OGW = 4 \cdot 0.823 \cdot \Pbar^2$ which we identify as the broad-peak result in \eqref{eq:OmegaGW-i-broad} for a constant scalar power spectrum $\Pzeta=2 \Pbar$.
\end{itemize}

\begin{figure}[t]
\centering
\begin{overpic}[width=0.8\textwidth]{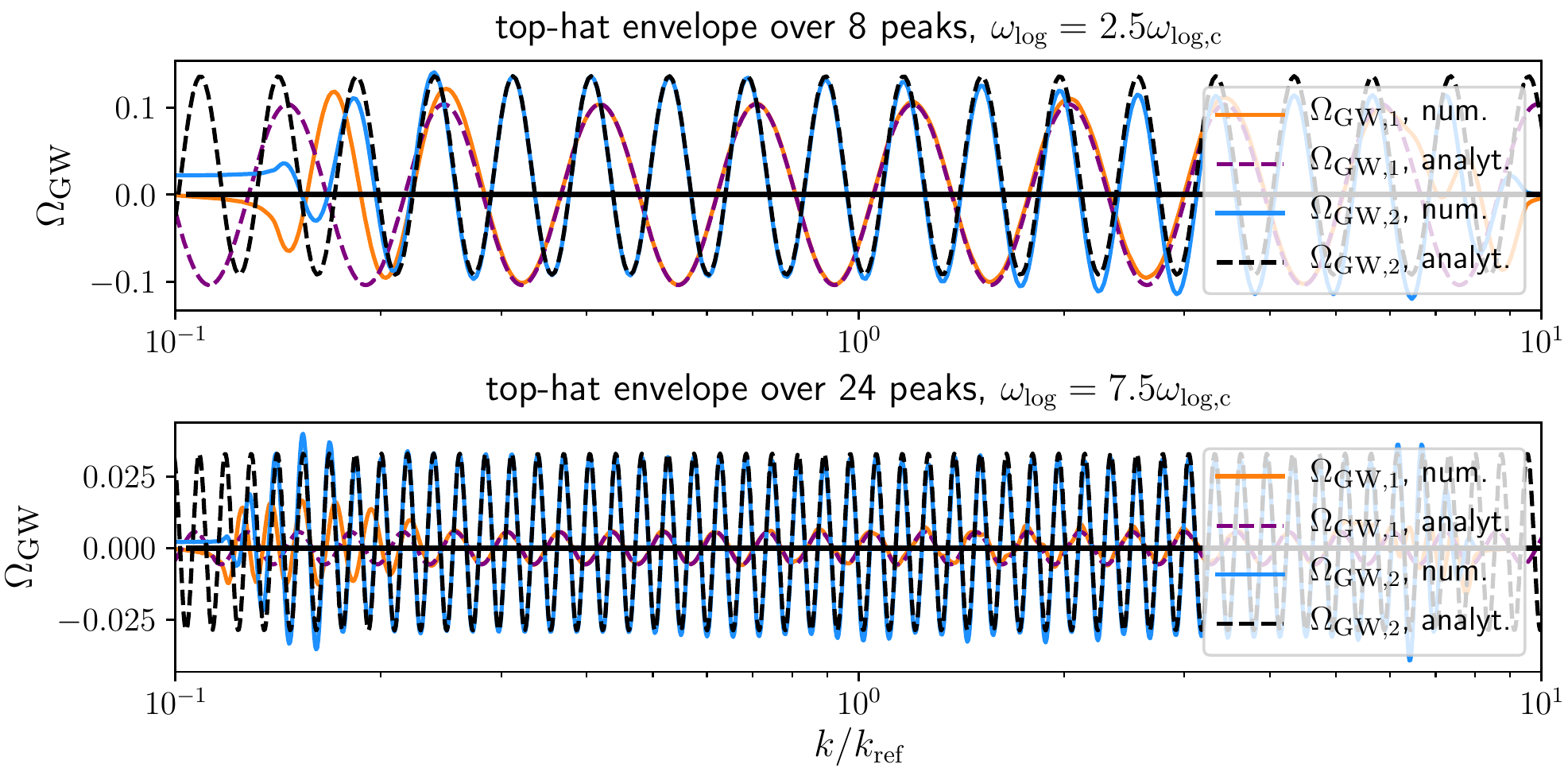}
\end{overpic}
\caption{\textit{Numerical and analytical results \eqref{analyticalO1}, \eqref{analyticalO2b} for $\Omega_{\textrm{GW},1}$ and $\Omega_{\textrm{GW},2}$ for a scalar power spectrum \eqref{eq:P_of_k_resonant} with a broad top-hat envelope over 8 periods of oscillation and $\omegalog =2.5 \omegalogc$ (upper panel) and 24 periods of oscillation and $\omegalog =7.5 \omegalogc$ (lower panel). The analytical results are a very good fit to the numerical results over most of the central region.}}
\label{fig:OGW_1_2_2p5_vs_analyt}
\end{figure}

\subsubsection{Numerical confirmations}
\label{sec:resonant-templates-numerical}
We now compare our analytical expression for $\OGW$ with numerical results. As remarked previously, we expect our above analysis to be applicable for scalar power spectra with a broad envelope, i.e.~with $\Pbar$ that is nearly constant over a sufficiently broad interval. We thus return to the examples from sec.~\ref{sec:resonant-peak-structure} with $\omegalog= 2.5 \omegalogc$ and $\omegalog= 7.5 \omegalogc$ and a top-hat profile for the envelope given by \eqref{eq:P-top-hat-envelope}. The envelope is shared by the two examples and was chosen to cover 8 and 24 periods of oscillation, respectively, resulting in a broad peak that stretches over an interval $[k_\textrm{min}, k_\textrm{max}]$ with $k_\textrm{max} \simeq 68 k_\textrm{min}$. For these two examples, in fig.~\ref{fig:OGW_2p5_7p5_vs_analyt} we plot the numerical results for $\OGW$ and $\Omega_{\textrm{GW},0}$ (red and green curves), as well as the analytical result for $\OGW$ (black dashed curve) for $A_\textrm{log}=1$. The analytical result is computed from \eqref{eq:OGW-analyt-expansion}, where for $\Omega_{\textrm{GW},0}$ we take the approximate analytical result \eqref{eq:OGW0-broad} for a broad peak in $\Pbar$, and $\Omega_{\textrm{GW},1}$ and $\Omega_{\textrm{GW},2}$ are given by the expressions \eqref{analyticalO1} and \eqref{analyticalO2b} with the factor $\Pbar^2$ replaced by $\big( \textrm{max}(\Pbar) \big)^2=1$. From fig.~\ref{fig:OGW_2p5_7p5_vs_analyt} we observe that our analytical expression gives a very good approximation to the numerical result near the maximum of $\OGW$, faithfully reproducing the modulations, both for $\omegalog= 2.5 \omegalogc$ and $\omegalog= 7.5 \omegalogc$. 

In fig.~\ref{fig:OGW_2p5_7p5_vs_analyt} we only considered $A_\textrm{log}=1$ for better visibility of the modulations, but we can now confirm that our analytical expression will be equally successful for any value of $A_\textrm{log}$. To this end, in fig.~\ref{fig:OGW_1_2_2p5_vs_analyt} we show numerical results for $\Omega_{\textrm{GW},1}$ (orange curve) and $\Omega_{\textrm{GW},2}$ (blue curve), together with our analytical expressions \eqref{analyticalO1} and \eqref{analyticalO2b} (purple and black dashed lines), again with $\Pbar^2$ set to $\big( \textrm{max}(\Pbar) \big)^2=1$. Over the central region $0.3 \lesssim k / \kref \lesssim 5$ we observe that the analytic expression provides a nearly perfect match to the numerical result.\\

\begin{figure}[t]
\centering
\begin{overpic}[width=0.8\textwidth]{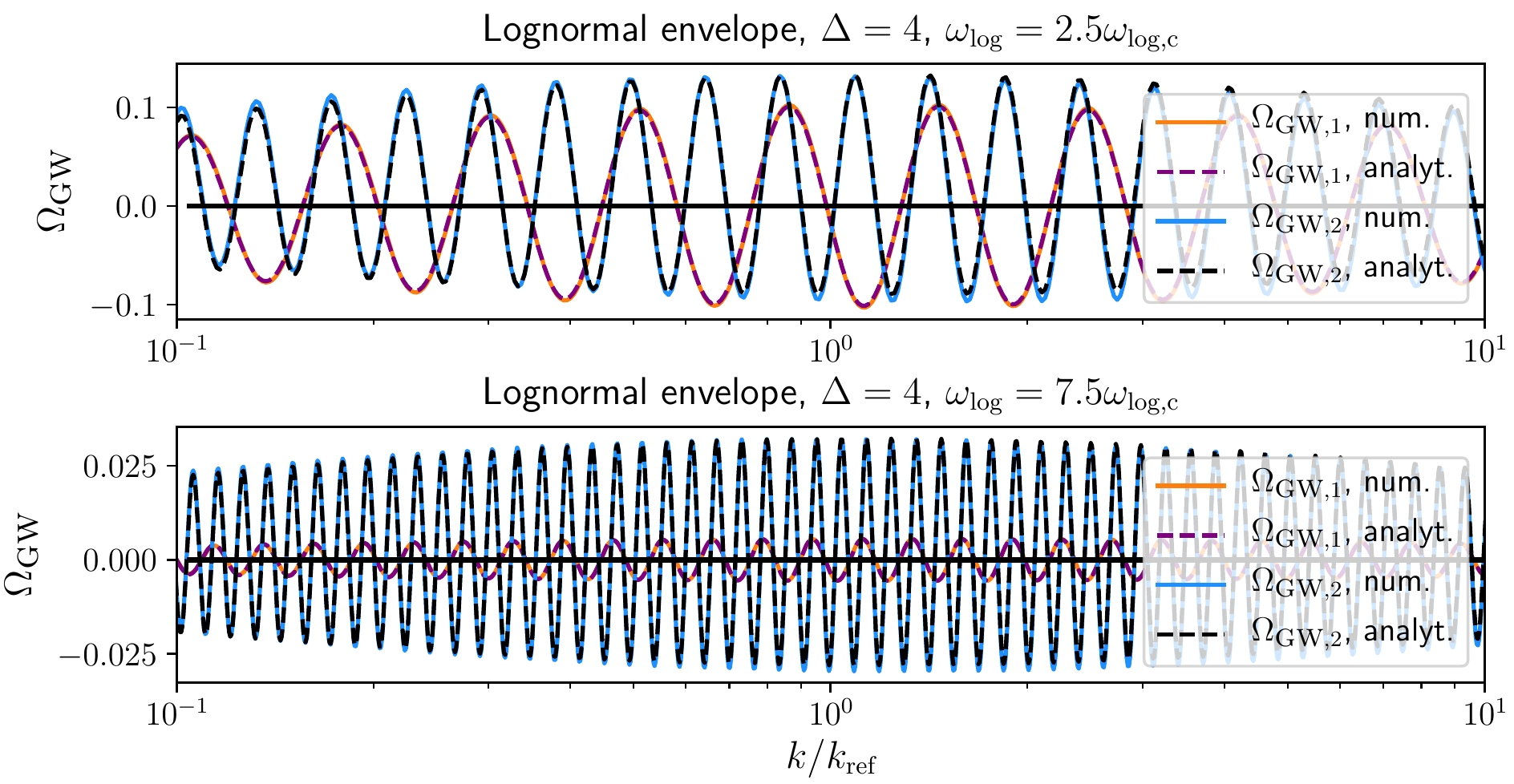}
\end{overpic}
\caption{\textit{Numerical and analytical results \eqref{analyticalO1}, \eqref{analyticalO2b} for $\Omega_{\textrm{GW},1}$ and $\Omega_{\textrm{GW},2}$ for a scalar power spectrum \eqref{eq:P_of_k_resonant} with a lognormal envelope given in \eqref{eq:Pbar-LN} with $\Delta=4$ and $\omegalog =2.5 \omegalogc$ (upper panel) and $\omegalog =7.5 \omegalogc$ (lower panel). The corresponding analytical results are given by \eqref{analyticalO1} and \eqref{analyticalO2b} with the factor $\Pbar^2$ replaced by $0.977 \, \Pbar^2 (\sqrt{3} k /2)$ and are observed to closely follow the numerical plots over the whole plotted range.}}
\label{fig:OGW_1_2_LN4p0_2p5_7p5_vs_analyt}
\end{figure}

\begin{figure}[t]
\centering
\begin{overpic}[width=0.8\textwidth]{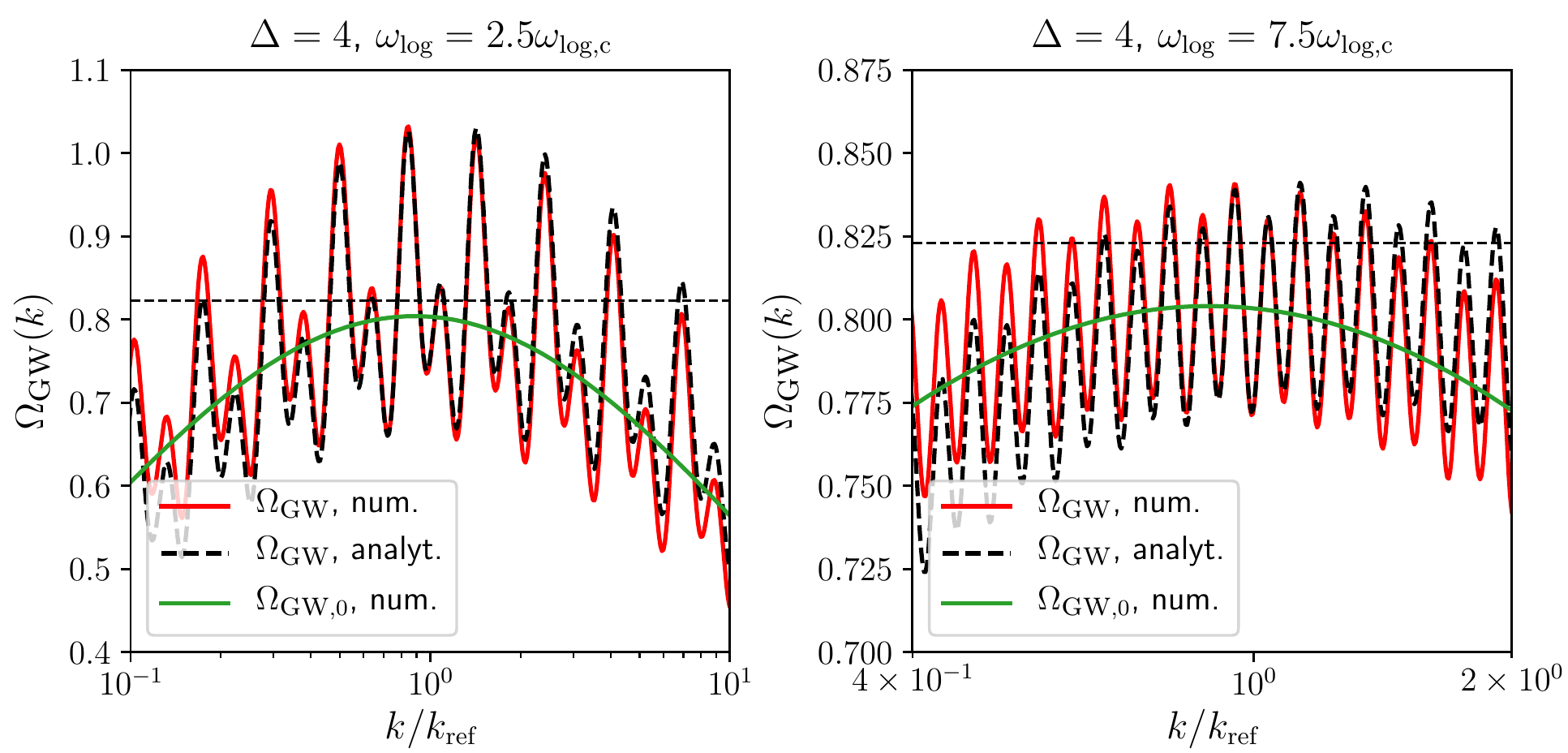}
\end{overpic}
\caption{\textit{Numerical result for $\OGW$ (red) and $\Omega_{\textrm{GW},0}$ (green) vs.~$k/\kref$ for a scalar power spectrum \eqref{eq:P_of_k_resonant} with $A_\textrm{log}=1$ and envelope $\Pbar$ given by a broad lognormal peak \eqref{eq:Pbar-LN} with $\Delta=4$. The left panel is for $\omegalog =2.5 \omegalogc$ and the right panel for $\omegalog =7.5 \omegalogc$. The black dashed curve corresponds to the analytical approximation which now accounts for the non-constant envelope as described in the main text near eq.~\eqref{eq:envelope-guess}. We find a very good agreement between the numerical and analytical result over the central part of the peak of $\OGW$. The horizontal dashed line denotes 0.823, i.e.~the maximal amplitude of $\OGW$ for a broad peak in $\Pzeta$ with unit amplitude.}}
\label{fig:OGW_LN4p0_2p5_7p5_vs_analyt}
\end{figure}

We now turn to examples with a non-constant envelope. To be specific, we once again consider the case of a lognormal peak in $\Pbar$ as given in \eqref{eq:Pbar-LN}. The `width' of the peak is controlled by the parameter $\Delta$ and we begin with the case of a broadly-peaked envelope with $\Delta=4$. The question now is how to account for the non-constancy of $\Pbar$ in our analytic expressions. To model $\Omega_{\textrm{GW},0}$ we will use an observation from fig.~\ref{fig:Broad_Narrow_top_LN}. There we found that the narrow-peak approximation of sec.~\ref{sec:broad-narrow}, when applied to a broadly peaked example, still reproduces $\OGW$ up to only $\%$-level errors. Thus for the analytical result for $\Omega_{\textrm{GW},0}$ we will use \eqref{eq:OGW0-narrow} and \emph{not} \eqref{eq:OGW0-broad}. Here we choose $\tilde{\gamma} =0.977$ so that the value of $\Omega_{\textrm{GW},0}$ matches the numerical result at the peak. That is, compared to the top-hat case, we replace
\begin{align}
\label{eq:envelope-guess}
    \bigg( \textrm{max} \Big( \Pbar(k) \Big) \bigg)^2 \longrightarrow \tilde{\gamma} \, \Pbar^2 \bigg(\frac{\sqrt{3}k}{2}\bigg)
\end{align}
in the analytic expression for $\Omega_{\textrm{GW},0}$.

\begin{figure}[t]
\centering
\begin{overpic}[width=0.8\textwidth]{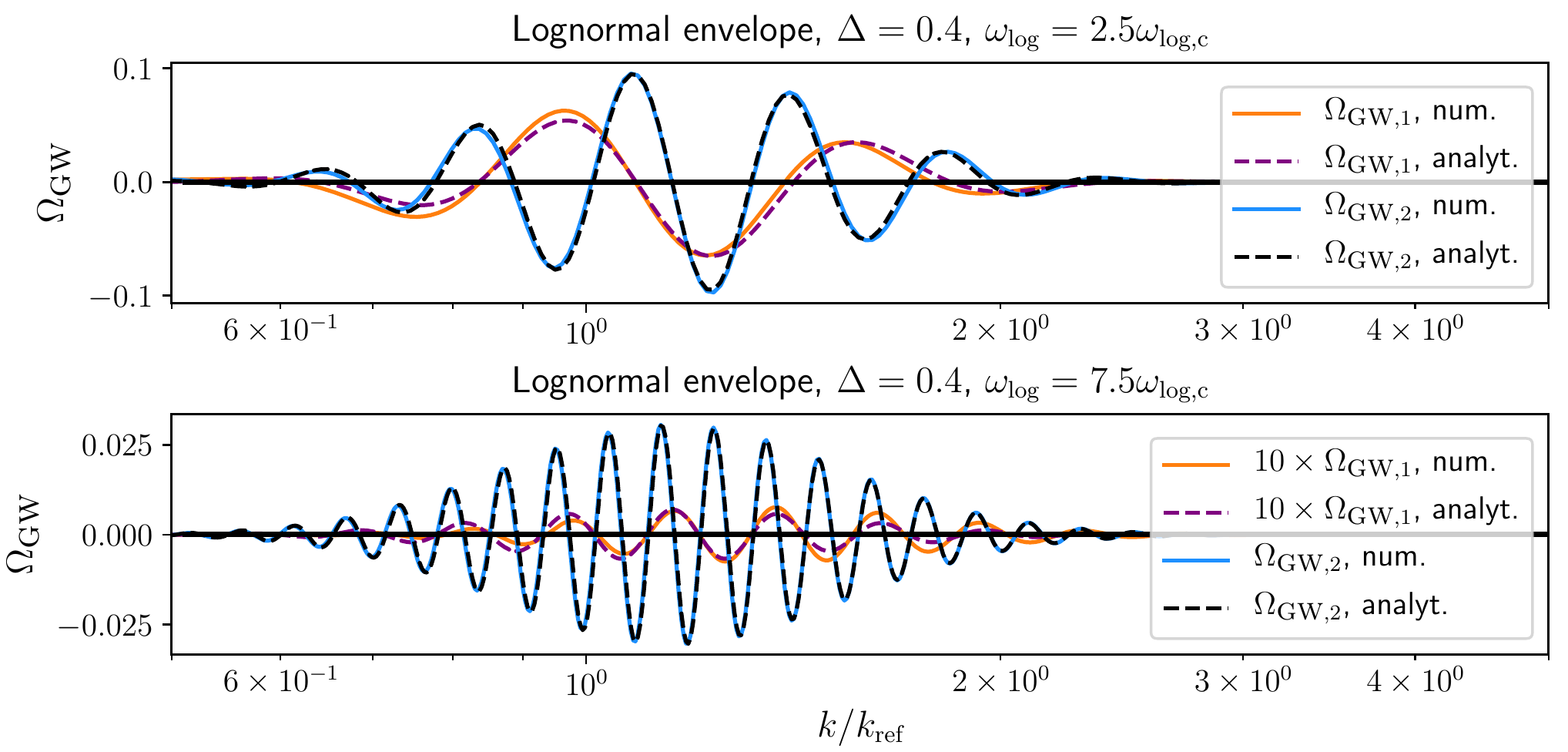}
\end{overpic}
\caption{\textit{Numerical result and analytical templates \eqref{eq:OmegaGW1-fit}, \eqref{eq:OmegaGW2-fit} for $\Omega_{\textrm{GW},1}$ and $\Omega_{\textrm{GW},2}$ for a scalar power spectrum \eqref{eq:P_of_k_resonant} with a lognormal envelope given in \eqref{eq:Pbar-LN} with $\Delta=0.4$ and $\omegalog =2.5 \omegalogc$ (upper panel) and $\omegalog =7.5 \omegalogc$ (lower panel). The parameters $\tilde{\gamma}_{1,2}$ and $\phi_{1,2}$ in the templates \eqref{eq:OmegaGW1-fit} and \eqref{eq:OmegaGW2-fit} are chosen to best match the numerical curves.}}
\label{fig:OGW_1_2_LN0p4_2p5_7p5_vs_analyt}
\end{figure}

\begin{figure}[t]
\centering
\begin{overpic}[width=0.8\textwidth]{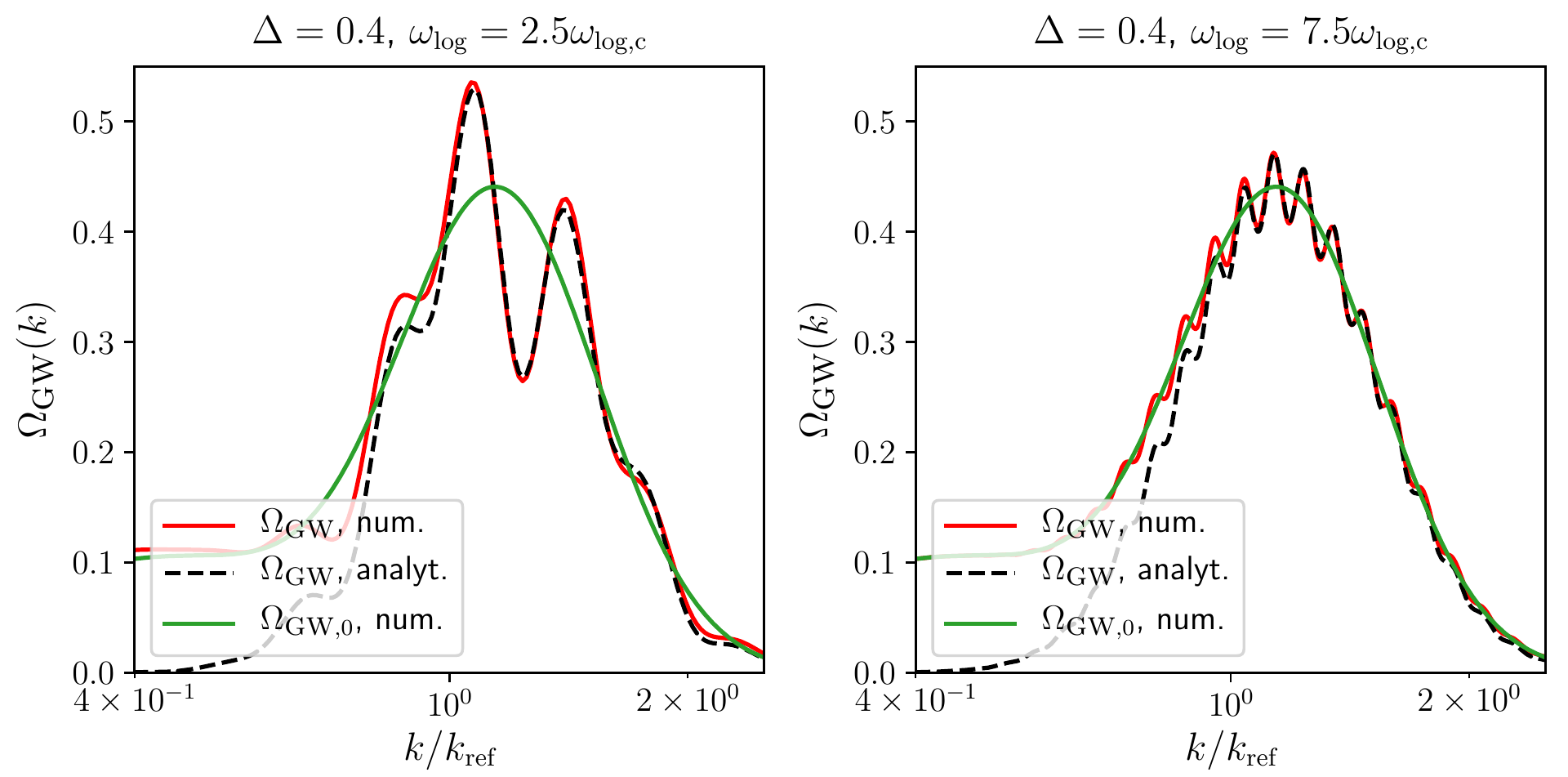}
\end{overpic}
\caption{\textit{Numerical result for $\OGW$ (red) and $\Omega_{\textrm{GW},0}$ (green) vs.~$k/\kref$ for a scalar power spectrum \eqref{eq:P_of_k_resonant} with $A_\textrm{log}=1$ and envelope $\Pbar$ given by a broad lognormal peak \eqref{eq:Pbar-LN} with $\Delta=0.4$. The left panel is for $\omegalog =2.5 \omegalogc$ and the right panel for $\omegalog =7.5 \omegalogc$. The black dashed curve corresponds to the analytical approximation as described in the main text in the vicinity of eqs.~\eqref{eq:OmegaGW1-fit}, \eqref{eq:OmegaGW2-fit}. We find a very good agreement between the numerical and analytical result over the central part of the peak of $\OGW$.}}
\label{fig:OGW_LN0p4_2p5_7p5_vs_analyt}
\end{figure}

We then make an educated guess for how to account for the non-constant envelope in the analytic expressions for $\Omega_{\textrm{GW},1}$ and $\Omega_{\textrm{GW},2}$. Here, we will follow what we have done for $\Omega_{\textrm{GW},0}$ and take our result for a top-hat envelope and again perform the replacement in \eqref{eq:envelope-guess}. This is equivalent to swapping the factor $\Pbar^2$ in \eqref{analyticalO1} and \eqref{analyticalO2b} with $\tilde{\gamma} \, \Pbar^2 (\sqrt{3} k /2)$ where we again take $\tilde{\gamma} =0.977$. Interestingly, this appears to remarkably account for the changing envelope in $\Omega_{\textrm{GW},1}$ and $\Omega_{\textrm{GW},2}$, as can be seen in fig.~\ref{fig:OGW_1_2_LN4p0_2p5_7p5_vs_analyt}. There we plot the numerical results for $\Omega_{\textrm{GW},1}$ (orange) and $\Omega_{\textrm{GW},2}$ (blue) for a lognormal envelope with $\Delta =4$ and $\omegalog= 2.5 \omegalogc$ (upper panel) and $\omegalog= 7.5 \omegalogc$ (lower panel). The corresponding analytical results are plotted as the purple and black dashed lines and can be observed to closely follow the numerical result over the whole plotted range. In fig.~\ref{fig:OGW_LN4p0_2p5_7p5_vs_analyt} we then plot both the numerical and the analytical result for $\OGW$ for $A_\textrm{log}=1$, given by the red curves and the black dashed curves, respectively. We find good agreement in the vicinity of the maximum. The discrepancies mainly arise from our choice for the analytical expression for $\Omega_{\textrm{GW},0}$. \\

For examples with a narrow-peaked envelope $\Pbar$ we do not expect our analytic expressions \eqref{analyticalO1} and \eqref{analyticalO2b} for $\Omega_{\textrm{GW},1}$ and $\Omega_{\textrm{GW},2}$ to reproduce the modulations in the GW correctly. Still, we will find that several qualitative but also quantitative properties of \eqref{analyticalO1} and \eqref{analyticalO2b} hold even in the narrow-peak case. In fig.~\ref{fig:OGW_1_2_LN0p4_2p5_7p5_vs_analyt} we plot numerical results for $\Omega_{\textrm{GW},1}$ (orange curve) and $\Omega_{\textrm{GW},2}$ (blue curve) for an example scalar power spectrum with lognormal envelope \eqref{eq:Pbar-LN} with $\Delta=0.4$. The upper panel is for $\omegalog=2.5 \omegalogc$ and the lower one for $\omegalog=7.5 \omegalogc$.\footnote{The value of $\Omega_{\textrm{GW},1}$ is boosted by a factor of 10 in the lower panel for better visibility.} While we observe that the relative amplitudes of $\Omega_{\textrm{GW},1}$ and $\Omega_{\textrm{GW},2}$ are not given by the factors in \eqref{analyticalO1} and \eqref{analyticalO2b} to a very good accuracy anymore, we still find that $\Omega_{\textrm{GW},1}$ is suppressed compared to $\Omega_{\textrm{GW},2}$ when $\omegalog$ is increased. Most importantly, $\Omega_{\textrm{GW},1}$ and $\Omega_{\textrm{GW},2}$ still exhibit a sinusoidal modulation with frequency $\omegalog$ and $2 \omegalog$, respectively. However, the phase is not reproduced correctly anymore by the expression in \eqref{analyticalO1} and \eqref{analyticalO2b}. Regarding the envelope of the oscillation, we find that the observation from the broad-peaked case still holds: The envelope of both $\Omega_{\textrm{GW},1}$ and $\Omega_{\textrm{GW},2}$ is very-well approximated by $\Pbar^2(\sqrt{3} k/2)$. Putting everything together, this suggests that we can fit $\Omega_{\textrm{GW},1}$ and $\Omega_{\textrm{GW},2}$ with
\begin{align}
\label{eq:OmegaGW1-fit}  
\Omega_{\textrm{GW},1} &= \tilde{\gamma}_1 \, \Pbar^2 \big( \sqrt{3} k / 2\big) \, \cos \Big(\omegalog \log \big(k / \kref \big) + \phi_1 \Big) \, , \\
\label{eq:OmegaGW2-fit}  
\Omega_{\textrm{GW},2} &= \tilde{\gamma}_2 \, \Pbar^2 \big( \sqrt{3} k / 2\big) \, \cos \Big(2\omegalog \log \big(k / \kref \big) + \phi_2 \Big) \, ,
\end{align}
without a constant offset in $\Omega_{\textrm{GW},2}$. These are shown as the purple and black dashed lines in fig.~\ref{fig:OGW_1_2_LN0p4_2p5_7p5_vs_analyt} where we have chosen the values of $\tilde{\gamma}_{1,2}$ and $\phi_{1,2}$ to best match the numerical result.\footnote{For $\omegalog=2.5 \omegalogc$ we choose $\tilde{\gamma}_1=0.067$, $\tilde{\gamma}_2=0.098$, $\phi_1=2.14$, $\phi_2=1.42$ and for $\omegalog=7.5 \omegalogc$ we use $\tilde{\gamma}_1=0.0007$, $\tilde{\gamma}_2=0.0305$, $\phi_1=-0.18$, $\phi_2=1.12$.} We observe that the templates \eqref{eq:OmegaGW1-fit} and \eqref{eq:OmegaGW2-fit} provide an excellent fit to the numerical results, especially $\Omega_{\textrm{GW},2}$ is reproduced effectively exactly by \eqref{eq:OmegaGW2-fit}.

In fig.~\ref{fig:OGW_LN0p4_2p5_7p5_vs_analyt} we then show the numerical result for $\OGW$ together with an analytical fit for this example with lognormal envelope \eqref{eq:Pbar-LN} with $\Delta=0.4$ and $A_\textrm{log}=1$. The analytic expression is given by $\OGW = \Omega_{\textrm{GW},0} + \Omega_{\textrm{GW},1} + \Omega_{\textrm{GW},2}$ where for $\Omega_{\textrm{GW},0}$ we use the narrow-peak approximation \eqref{eq:OGW0-narrow} with $\tilde{\gamma}=0.54$ and for $\Omega_{\textrm{GW},1}$ and $\Omega_{\textrm{GW},2}$ we use \eqref{eq:OmegaGW1-fit} and \eqref{eq:OmegaGW2-fit} with the parameter choices as in fig.~\ref{fig:OGW_1_2_LN0p4_2p5_7p5_vs_analyt}. Again, the analytic result gives a good approximation to the numerical result in the vicinity of the principal peak.

While we so far only presented numerical examples with a top-hat or a lognormal peak for the envelope $\Pbar$, our results can also be shown to apply for different envelope profiles. As an example, in appendix \ref{app:Gausslike} we display numerical results for a resonant feature with a Gaussian-like peak in $\Pbar$ and confirm that the templates \eqref{eq:OmegaGW1-fit} and \eqref{eq:OmegaGW2-fit} again provide an excellent fit to the oscillatory part. This gives evidence that the templates \eqref{eq:OmegaGW1-fit} and \eqref{eq:OmegaGW2-fit} are indeed valid more generally for a resonant feature, and do not rely on the envelope $\Pbar$ being broad or having a particular functional form or symmetry. 

\vspace{0.1cm}
\noindent \textbf{Theoretically motivated templates for $\OGW$ --} We can draw the following lessons from the results in this section and our findings in section \ref{sec:resonant-peak-structure}. The main result is that for a resonant feature in $\Pzeta$ the scalar-induced GW spectrum $\OGW$ exhibits a modulation that can be written as a superposition of an oscillation in $\log(k)$ with frequency $\omegalog$ and one with $2\omegalog$. This was observed in the resonance-peak analysis of sec.~\ref{sec:resonant-peak-structure} and in the expansion in powers of $A_\textrm{log}$ above. Moreover, in the comparison with numerical results we found that both these oscillatory pieces can be understood as a sinusoidal modulation of a common envelope, which also provides a good approximation for $\Omega_{\textrm{GW},0}$, at least close to its maximum. Putting everything together we arrive at a template for $\OGW$ for a resonant feature in $\Pzeta$, given by   
\begin{align}
    \label{eq:resonant_template}
    \Omega_{\textrm{GW}}(k) = \overline{\Omega}_{\textrm{GW}}(k) \Big[1 &+ \mathcal{A}_{\textrm{log},1} \cos \big(\omega_\textrm{log} \log (k/k_\textrm{ref}) + \phi_{\textrm{log},1} \big) \\
    \nonumber & + \mathcal{A}_{\textrm{log},2} \cos \big(2 \omega_\textrm{log} \log (k/k_\textrm{ref}) + \phi_{\textrm{log},2} \big) \Big] \, ,
\end{align}
where $\overline{\Omega}_{\textrm{GW}}$ is the GW spectrum for the envelope $\Pbar$ of the scalar power spectrum. This piece is model-dependent, as the functional form of $\overline{\Omega}_{\textrm{GW}}$ depends on the shape of $\Pbar$.

One use of \eqref{eq:resonant_template} is as a template that can be employed as a target in the analysis of GW data, with free parameters $\omegalog$, $\mathcal{A}_{\textrm{log},1/2}$ and $\phi_{\textrm{log},1/2}$ to be fitted to data. However, our analysis above implies that there is further structure in the GW spectrum resulting in interdependencies between the parameters in \eqref{eq:resonant_template}. That is, the ratio of the amplitudes $\mathcal{A}_{\textrm{log},1}$ and $\mathcal{A}_{\textrm{log},2}$ depends on the frequency $\omegalog$ and the value of the amplitude of oscillation $A_\textrm{log}$ in $\Pzeta$.\footnote{A more detailed summary of these findings will be presented in the following conclusion section \ref{sec:conclusion} to make it visible to the casual reader.} For scalar power spectra with a broad envelope $\Pbar$ we even have exact analytical predictions for the parameters $\mathcal{A}_{\textrm{log},1/2}$ and $\phi_{\textrm{log},1/2}$ for any value of $\omegalog$. This opens the possibility that, given a GW spectrum with a $\log(k)$-periodic structure, one may determine whether this is due to a resonant feature in $\Pzeta$ and further reconstruct the scalar power spectrum. We leave this exciting prospect for future work.\\

Eventually, note that the presence of the sinusoidal modulation with frequency $2 \omegalog$ could have been predicted by requiring consistency with our results for a sharp feature in \cite{Fumagalli:2020nvq}. The observation is that if the frequency of oscillation $\omegalog$ is sufficiently high and $\Pbar$ exhibits a narrow peak, a resonant feature in $\Pzeta$ is difficult to distinguish from a sharp feature, and should give a similar effect in $\OGW$. That is, consider an envelope $\Pbar$ with a peak at $k=\kref$ and width $\Delta k$ with $\Delta k / \kref \ll 1$. To exhibit visible modulations across this peak the frequency needs to be sufficiently large, i.e.~$\omegalog > 2 \pi \kref / \Delta k$. Close to $\kref$ the oscillations due to a resonant feature then resemble that of a sharp feature:
\begin{align}
    \nonumber \Pzeta (\kref + \delta k) \supset &\cos \bigg(\omegalog \log \frac{\kref + \delta k}{\kref} + \thetalog \bigg) = \cos \bigg( \frac{\omegalog}{\kref} \delta k + \ldots \bigg) \\
    & = \cos \big(\omegalin \, \delta k + \ldots \big) \, , \qquad \textrm{with} \qquad \omegalin = \frac{\omegalog}{\kref} \, ,
    \label{resonant-sharp-matching}
\end{align}
up to correction that are small for $\delta k < \Delta k$. For such examples we expect a GW spectrum whose envelope peaks near $k = 2 \kref / \sqrt{3}$ (see sec.~\ref{sec:broad-narrow}) modulated by sinusoidal oscillations. Modelling the example as a sharp feature, these oscillations are expected to take the form
\begin{align}
\label{eq:OGW-sharp-osc-near-peak}
    \OGW \Big(\tfrac{2}{\sqrt{3}}\kref + \delta k \Big) \supset \cos \big( \sqrt{3} \omegalin \, \delta k + \ldots \big) \, ,
\end{align}
see \eqref{eq:OGW_sharp_template}. Understanding the example as a resonant feature our results imply the following. The situation described can only occur for sufficiently large values of $\omegalog$ and hence in general in the regime where the oscillatory contribution with frequency $2 \omegalog$ dominates over that with frequency $\omegalog$, see e.g.~the discussion at the end of sec.~\ref{sec:resonant-templates-analytical} and the numerical examples in sec.~\ref{sec:resonant-peak-structure-numerical}. Thus, from \eqref{eq:resonant_template} we expect the oscillations in $\OGW$ to be given by 
\begin{align}
    \OGW \Big(\tfrac{2}{\sqrt{3}}\kref + \delta k \Big) \supset \cos \bigg( 2 \omegalog \log \frac{2 \kref / \sqrt{3} + \delta k}{\kref} + \ldots \bigg) = \cos \bigg( \sqrt{3} \frac{\omegalog}{\kref} \,  \delta k + \ldots \bigg) \, ,
\end{align}
which is consistent with \eqref{eq:OGW-sharp-osc-near-peak} when taking into account the identification \eqref{resonant-sharp-matching} $\omegalin = \omegalog / \kref$. Thus, the term with frequency $2 \omegalog$ in \eqref{eq:resonant_template} is crucial for consistency with the sharp feature results.

\section{Conclusions and outlook}
\label{sec:conclusion}
Primordial features in the scalar power spectrum signal a departure of inflation from the single-field slow-roll regime and are motivated by embeddings of inflation in high energy physics. Features manifest themselves as oscillations in the scalar power spectrum that can be periodic in $k$ (sharp feature) or $\log(k)$ (resonant feature) or exhibit a combination of the two. Features are also frequently accompanied by an enhancement of fluctuations, resulting in a peak in the scalar power spectrum modulated by oscillations.  

In this work we analysed the spectral shape of the scalar-induced contribution to the SGWB due to a resonant feature in the scalar power spectrum $\Pzeta$, complementing the previous work \cite{Fumagalli:2020nvq} dedicated to sharp features. As in \cite{Fumagalli:2020nvq}, here we focused exclusively on GWs produced in the post-inflationary period, leaving the model-dependent and often $\epsilon$-suppressed contribution from the inflationary era for future work. Our analysis does not rely on a particular realisation of a resonant feature. Instead, we consider scalar power spectra as given in \eqref{eq:P_of_k_resonant} and investigate the corresponding GW spectrum for all possible values of the parameters $\omegalog$ and $A_\textrm{log}$ setting the frequency and amplitude of the oscillation in $\Pzeta$, respectively. To account for the model-dependent envelope $\Pbar$ we consider several examples of various shapes (broad and narrow) and functional forms (top-hat, lognormal peak, Gaussian-like). 

We show that the oscillations in $\Pzeta$ are processed into corresponding modulations of the GW energy density fraction $\OGW$. To analyse these, we model the oscillations in $\Pzeta$ as a series of individual peaks and study the structure of the resulting resonance peaks in $\OGW$ (sec.~\ref{sec:resonant-peak-structure}). In addition, we exploit a mathematical property of $\log(k)$-periodic oscillations to separate the $\omegalog$-dependence of the amplitude of modulations in $\OGW$ from the $k$-dependence of the GW spectrum, and hence derive analytic expressions for the oscillatory contribution to $\OGW$ (sec.~\ref{sec:resonant-templates}). These latter results are obtained for example power spectra with sufficiently broad envelope $\Pbar$, but we observe in numerical examples that their main  qualitative properties still apply in examples with a narrowly-peaked envelope. Altogether, we find that the contribution to the SGWB due to a resonant feature can be modelled by the template \eqref{eq:resonant_template}. That is, the oscillation in $\OGW$ can be expressed as a superposition of two sinusoidal oscillations with frequencies $\omegalog$ and $2 \omegalog$ modulating a common envelope $\overline{\Omega}_\textrm{GW}$.

\begin{figure}[t]
\centering
\begin{overpic}[width=0.95\textwidth]{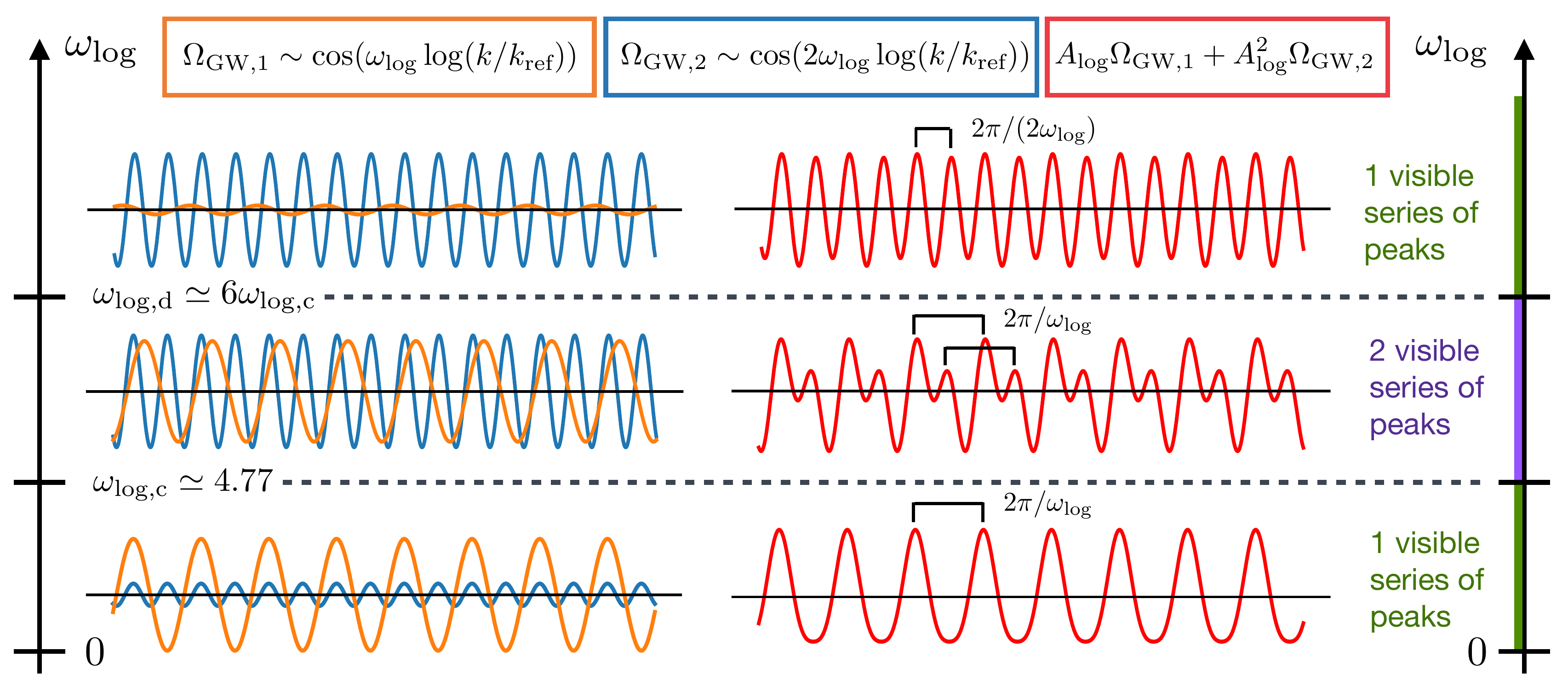}
\end{overpic}
\caption{\textit{Summary of the different regimes of peak structures visible in $\OGW$ for a resonant feature. Here $\Omega_{\textrm{GW},1}$ and $\Omega_{\textrm{GW},2}$ are the various oscillatory contributions to $\OGW$ as defined via \eqref{eq:OGW-analyt-expansion} and $A_\textrm{log} \Omega_{\textrm{GW},1}+ A_\textrm{log}^2 \Omega_{\textrm{GW},2}$ is the complete oscillatory part. For a power spectrum with frequency $\omegalog < \omegalogc$ the oscillatory part of $\OGW$ exhibits a single series of peaks, which becomes two series of peaks for $\omegalogc<\omegalog \lesssim \omegalogd$. In both cases, the maxima are displaced from one another in a manner consistent with a frequency $\omegalog$. For $\omegalog \gtrsim \omegalogd$ the oscillatory part of $\OGW$ again exhibits a single series of peaks, but now the peaks are distributed with frequency $2 \omegalog$. This also leads to a `frequency gap' as there cannot be a series of distinguishable peaks in $\Omega_{\mathrm{GW}}$ consistent with a frequency $\omegalogd \lesssim \omegalog \lesssim 2 \omegalogd$.}}
\label{fig:summary}
\end{figure}

Our main results regarding the oscillatory part of $\OGW$ can be summarised as follows, see also the figure \ref{fig:summary}, and where we use  fig.~\ref{fig:P_OGW_ratios_res_LN1p0_2p5wc_7p5wc_a0p5} from the introduction for illustration:
\begin{itemize}
    \item For $\omegalog < \omegalogc$, with $\omegalogc$ defined in \eqref{eq:omegalogc-def}, the oscillatory contribution to $\OGW$ with frequency $\omegalog$ dominates over the one with $2 \omegalog$, and $\OGW$ effectively exhibits a periodic modulation with frequency $\omegalog$. See sec.~\ref{sec:resonant-peak-structure} and \ref{sec:resonant-templates}.
    \item For $\omegalog \gtrsim \omegalogc$ the two sinusoidal pieces in $\OGW$ have comparable amplitude, resulting in a complicated oscillatory pattern in the GW spectrum. This is e.g.~the case for the two examples with $\omegalog=2.5 \omegalogc$ in fig.~\ref{fig:P_OGW_ratios_res_LN1p0_2p5wc_7p5wc_a0p5} (red and green curves) which may not be easily identifiable by eye as a superposition of two sinusoidal terms.
    \item Increasing $\omegalog$ further, the oscillatory piece with frequency $\omegalog$ becomes more and more suppressed compared to the one with frequency $2\omegalog$. Thus for $\omegalog > \omegalogd$ the modulation in $\OGW$ can generically be well-approximated by a single oscillatory piece with frequency $2\omegalog$. This can be e.g.~seen in fig.~\ref{fig:P_OGW_ratios_res_LN1p0_2p5wc_7p5wc_a0p5} for the oscillatory part of the example with $\omegalog=7.5 \omegalogc$. 
    The exact value of $\omegalogd$ is to some extent a matter of choice as the transition between the various regimes is continuous, but a reasonable value is $\omegalogd \simeq 6 \omegalogc$. One consequence of this is a gap in the observable frequency of oscillations in $\OGW$. Oscillations with $\omegalog > \omegalogd$ in $\Pzeta$ only lead to modulations in $\OGW$ with $\omegalog > 2 \omegalogd$ and hence a single sinusoidal modulation with $\omegalogd < \omegalog < 2 \omegalogd$ cannot occur in $\OGW$.\footnote{In \cite{Braglia:2020taf}, numerical computations in an inflationary model with a resonant feature power spectrum result in a $\log(k)$-periodic modulation of the GW spectrum with frequency $\omegalog \simeq 10 \omegalogc$, which appears to lie in the frequency gap so-defined. In fact, for narrow-peaked power spectra the frequency gap slightly shifts and the model can be understood as already falling into the ``doubled-frequency'' regime.}
    \item At the same time, a larger value of $\omegalog$ generically leads to a smaller amplitude of oscillations in $\OGW$. This can be seen e.g~by comparing the two examples in fig.~\ref{fig:P_OGW_ratios_res_LN1p0_2p5wc_7p5wc_a0p5} with $A_\textrm{log}=1$ and $\omegalog = 2.5 \omegalogc$ or $\omegalog = 7.5 \omegalogc$ (red and blue curves), where the amplitude of modulations is visibly reduced for the case with $\omegalog = 7.5 \omegalogc$. A fit of the template \eqref{eq:resonant_template} to these two examples gives $\mathcal{A}_{\textrm{log},1}=14 \%$, $\mathcal{A}_{\textrm{log},2}=22 \%$ for $\omegalog = 2.5 \omegalogc$ and $\mathcal{A}_{\textrm{log},1}= 0.3 \%$, $\mathcal{A}_{\textrm{log},2}= 7 \%$ for $\omegalog = 7.5 \omegalogc$, illustrating the overall reduction of the amplitude for larger $\omegalog$ as well as the relative suppression of the piece with frequency $\omegalog$. 
    \item The values of $\omegalog$ where the transitions between the three regimes described above occur also depend on the amplitude of oscillations $A_\textrm{log}$ in $\Pzeta$. A reduction in $A_\textrm{log}$ suppresses the contribution with frequency $2 \omegalog$ more than that with $\omegalog$, pushing the transitions between these three regimes to larger values of $\omegalog$. However, a reduction in $A_\textrm{log}$ also reduces the overall amplitude of oscillations in $\OGW$ (e.g.~compare the red and green curves in fig.~\ref{fig:P_OGW_ratios_res_LN1p0_2p5wc_7p5wc_a0p5}). If $A_\textrm{log}$ is too small the amplitude of modulations in $\OGW$ will eventually become too small to be experimentally detectable. This is described in appendix \ref{app:reduce-A-log}.
\end{itemize}

Important open questions concern the detectability of this contribution to the SGWB background due to a resonant feature. Firstly, the scalar fluctuations need to be sufficiently enhanced for the GW energy fraction to be above the detection thresholds of the various GW observatories, typically expressed for $h^2 \OGW$ where $h$ denotes the Hubble parameter in units of $100 \textrm{s}^{-1} \textrm{Mpc}^{-1}$. To compare with these detection thresholds, the GW fraction in \eqref{eq:OmegaGW-i} thus needs to be multiplied by a factor $c_g \Omega_{r,0} h^2$ which also ensures that this corresponds to the energy fraction in GWs today, see the discussion at the beginning of sec.~\ref{sec:features}. As in \cite{Fumagalli:2020nvq}, here we take $c_g \Omega_{r,0} h^2=1.6 \cdot 10^{-5}$. The largest value for $\OGW$ due to a peak in the power spectrum is observed to occur for broad peaks (see e.g.~fig.~\ref{fig:Broad_Narrow_top_LN} in this work) giving an upper bound $h^2 \OGW^{\textrm{max}} \lesssim 1.3 \cdot 10^{-5} \cdot \mathcal{P}_\textrm{max}^2$ where we used \eqref{eq:OmegaGW-i-broad}. To give an example, for $h^2 \OGW \gtrsim 10^{-13}$, as is approximately required for detection with LISA (see e.g.~\cite{Caprini:2019pxz,Flauger:2020qyi}), one requires $\mathcal{P}_\textrm{max} \gtrsim 10^{-4}$. An important task on the theory side is then to determine to what extent models with such a large amplitude of primordial fluctuations are still under theoretical control as far as their backreaction and their perturbative treatment during inflation are concerned. These issues can be most directly addressed in explicit models of features (see a first investigation in our work \cite{Fumagalli:2020nvq}), thus motivating the search for new explicit realisations of features.

Apart from backreaction and perturbativity issues, feature models are also constrained through their associated spectrum of primordial black holes, which can be produced with high abundance if scalar fluctuations are sufficiently enhanced. The mass spectrum of these however depends on the full probability density function (PDF) of fluctuations and not just the power spectrum, and it is hence sensitive to primordial non-Gaussianities. The primordial black hole spectrum thus also tests different aspects of a model than the SGWB contribution, which can be used to further constrain the space of feature models. 

Experimental identification of a GW signal due to resonant (and also sharp) features however also requires a successful reconstruction of the characteristic oscillations. A resolution of the oscillations in frequency space further raises the detection threshold value of $h^2 \OGW$ compared to a signal without modulations \cite{Caprini:2019pxz}. What is not known so far is how large the amplitude of modulations has to be for them to still be visible experimentally. This is particularly important as the GW spectrum due to features frequently exhibits modulations that are at most $\mathcal{O}(10 \%)$ as in the examples in fig.~\ref{fig:P_OGW_ratios_res_LN1p0_2p5wc_7p5wc_a0p5}. This motivates a dedicated investigation regarding the possibility of reconstruction of such signals. The templates in \eqref{eq:resonant_template} and \eqref{eq:OGW_sharp_template} can be used for generating mock data for such an analysis and, moreover, as templates for extracting such a signal from real data.

Another direction for future work concerns the contributions to the SGWB induced \emph{during} inflation, while here we so far only considered the part that is produced in the post-inflationary era. Quite generally, the former are $\epsilon^2$-suppressed compared to the latter, owing to the source being directly related to the scalar fields fluctuations, and not to the curvature perturbation itself. However, in models like sharp features that excite sub-Hubble modes, this signal is also enhanced, leading to a contribution that is potentially comparable to the post-inflationary one. Moreover, a sharp feature also leads to oscillatory modulations of the inflationary-era contribution itself \cite{GWinfSharp} (see also \cite{An:2020fff} for a study of oscillations generated via phase transitions during inflation).

In this work on resonant features and in the publication \cite{Fumagalli:2020nvq} on sharp features we presented a detailed description of how features in the scalar power spectrum are translated into corresponding properties of the GW spectrum. Vice versa, given a primordial contribution to the SGWB, our findings allow for extracting information about the scalar power spectrum. In particular, if the SGWB contains an oscillatory piece, our quantitative results linking the amplitude and frequency of the oscillation in $\Pzeta$ to corresponding parameters in $\OGW$, and in particular the analytical expressions for the oscillatory parts due to a resonant feature (as described in sec.~\ref{sec:resonant-templates}) enable one in principle to identify the corresponding oscillatory feature in $\Pzeta$. A dedicated analysis concerning to what extent the shape of the power spectrum of primordial fluctuations can be reconstructed in full generality from SGWB data is thus an important target for future work. Once accomplished, this would further strengthen the SGWB as an important observable for learning about the very early universe.

\acknowledgments
We are grateful to Sebastian Garcia-Saenz, Sadra Jazayeri, Mauro Pieroni, Lucas Pinol and Caner {\"U}nal for interesting discussions. 
J.F, S.RP, and L.T.W are supported by the European Research Council under the European Union's Horizon 2020 research and innovation programme (grant agreement No 758792, project GEODESI).

\appendix
\section{Gravitational wave spectrum due to peaks in $\Pzeta$}
\label{app:peaks}
In this appendix we record details on how to derive the approximate expressions for $\OGW$ due to a broad or narrow peak in $\Pzeta$ shown in sec~\ref{sec:broad-narrow}. We further explain that $\mathcal{O}(1)$ oscillations in $\Pzeta$ can be modelled as a series of narrow peaks providing support for the treatment of oscillatory power spectra described in sec.~\ref{sec:multiple}. 

\subsection{Broad-peak case}
\label{app:broad}
The definition of a broad peak was given in \eqref{eq:k-broad} and we focus on values of $k$ in the central region of the peak, i.e~$k_\textrm{min} < k < k_\textrm{max}$. Inserting \eqref{eq:k-broad} into \eqref{eq:s-d-constraints} the main observation is that for a broad peak the constraints due to the finite extension of the peak are `weak' in the following sense: the restrictions \eqref{eq:s-d-constraints} correspond to a `large' square, see the left panel of fig.~\ref{fig:int-ranges}, that overlaps with the red integration range over a significant area. 

Most importantly, the blue region contains the part of integration domain over $d$ and $s$ that gives the dominant contribution to $\Omega_\textrm{GW}$. To see this we use that the integration kernel $\TRD(d,s)$ in \eqref{eq:OmegaGW-i} is singular along the line $s=1$ and smooth otherwise, and further drops off as $s^{-4} \ln^2 s$ for $s \gg 1$. As a result, the integral in \eqref{eq:OmegaGW-i} receives its dominant contribution from the integration over values in the vicinity of $s=1$. For a broad peak this part of the integration domain is in essence contained entirely by the blue region. Consider for example the part of the red region that does not overlap with the blue square on the top right in the left panel of fig.~\ref{fig:int-ranges}. For a broad peak ($k_\textrm{max}/ k \gg 1$) this is at $s \gg 1$ and hence would only give a subdominant contribution to the integral. There is also the small corner on the left of the red strip that is not contained in the blue region, see again the LHS plot in fig.~\ref{fig:int-ranges}. However, for $k_\textrm{min}/ k \ll 1$ this area is very small (and it does not contain the line $s=1$). Again, this corner would only give a small correction to the full result if it was included. In contrast, the line $s=1$ and its immediate vicinity, where it overlaps with the red region, is entirely contained in the blue region. 

In practice, this implies that the constraints \eqref{eq:s-d-constraints} due to the finite size of the peak are effectively irrelevant for the computation of $\Omega_\textrm{GW}(k)$ for $k_\textrm{min} < k < k_\textrm{max}$ and we could have equally integrated over the whole range given in \eqref{eq:d-s-int-range} without changing the result by much. Applied to \eqref{eq:OmegaGW-i-for-peak} this implies
\begin{align}
\label{eq:OmegaGW-i-broad-app}
    \Omega_{\textrm{GW}, \textrm{broad}} \approx \int_0^{\frac{1}{\sqrt{3}}} \textrm{d} d \int_{\frac{1}{\sqrt{3}}}^\infty \textrm{d} s \, \TRD(d,s) \, \mathcal{P}_\textrm{peak} \bigg(\frac{\sqrt{3}k}{2}(s+d)\bigg) \mathcal{P}_\textrm{peak} \bigg(\frac{\sqrt{3}k}{2}(s-d)\bigg) \, .
\end{align}
Further assuming that the peak is sufficiently broad to be taken as constant over the central region of the peak, i.e.~$\mathcal{P}_\textrm{peak}(k) \sim \mathcal{P}_\textrm{max} = \textrm{const}$, we arrive at the result \eqref{eq:OmegaGW-i-broad} in the main text.\footnote{For $k < k_\textrm{min}$ or $k > k_\textrm{max}$ one can further confirm that the corresponding value of $\Omega_\textrm{GW}(k)$ will be generically suppressed compared to \eqref{eq:OmegaGW-i-broad}, as in both cases the line $s=1$ will not lie within the blue region.}

\subsection{Narrow-peak case}
\label{app:narrow}
The definition of a narrow peak was given in \eqref{eq:narrowdef} in the main text.
To be specific, consider values of $k \sim k_\star$ as this will be the range of $k$ where $\Omega_\textrm{GW}$ will also peak. For these values of $k$ the side length of the square in $(s-d)$-$(s+d)$-space defined by the constraints \eqref{eq:s-d-constraints} is $\sim \Delta k / k_\star \ll 1$. As a result, in the narrow-peak case the region demarcated by \eqref{eq:s-d-constraints} (the blue square in the right panel of fig.~\ref{fig:int-ranges}) only overlaps with the domain of integration (red strip in fig.~\ref{fig:int-ranges}) over a small area.\footnote{This also holds for $k > k_\star$ and the findings described here can also be applied for $k > k_\star$. For $k \ll k_\star$ the overlap between the square defined by \eqref{eq:s-d-constraints} and the red strip in fig.~\ref{fig:int-ranges} is not necessarily small anymore, even for a narrow peak. However, for $k \ll k_\star$ the constraints \eqref{eq:s-d-constraints} also imply $s \gg 1$, in which case we only get subdominant contributions to $\OGW(k)$ due to the suppressed value of $\TRD(d,s)$ for $s \gg 1$.}

In particular, in this case the conditions \eqref{eq:s-d-constraints} are so restrictive that the variable $d$ is now constrained as
\begin{align}
\label{eq:depsilondef-app}
    d < d_\epsilon = \frac{\Delta k}{\sqrt{3} k} \, ,
\end{align}
which is the maximal value of $d$ at the top left corner of the blue square. For a narrow peak and for $k \sim k_\star$ it then follows that $d_\epsilon \ll 1/ \sqrt{3}$.

We can then use $d_\epsilon$ as a small parameter and compute $\Omega_\textrm{GW}$ in powers of $d_\epsilon$. This is done by expanding the integrand in \eqref{eq:OmegaGW-i} for small $d$ and integrating each term from $0$ to $d_\epsilon$. This gives 
\begin{align}
\label{eq:OmegaGW-i-d0-app}
    \Omega_{\textrm{GW}}(k) = \frac{1}{2}  \int_0^{d_\epsilon} \textrm{d} d \int_{\frac{1}{\sqrt{3}}}^\infty \textrm{d} s \, \bigg[ \TRD(0,s) \, \Pzeta^2 \bigg(\frac{\sqrt{3}k}{2}s\bigg) + \mathcal{O}(d^2) \bigg] \,.
\end{align}
The factor $1/2$ accounts for the fact that the integration over $d$ is over a triangular region (see fig.~\ref{fig:int-ranges}) of height $d_\epsilon$, while here we have integrated over a box of height $d_\epsilon$. After performing the $d$-integration, this becomes
\begin{align}
\label{eq:OmegaGW-i-s-only-app}
    \Omega_{\textrm{GW}}(k) = \frac{1}{2} \, d_\epsilon \int_{\frac{1}{\sqrt{3}}}^\infty \textrm{d} s \, \TRD(0,s) \, \Pzeta^2 \bigg(\frac{\sqrt{3}k}{2}s\bigg) + \mathcal{O}(d_\epsilon^3) \, ,
\end{align}
and we are left with a single integral over $s$. In the following, we will focus on the leading term and suppress the $\mathcal{O}(d_\epsilon^3)$-corrections in what follows.

\begin{figure}[t]
\centering
\begin{overpic}[width=0.7\textwidth]{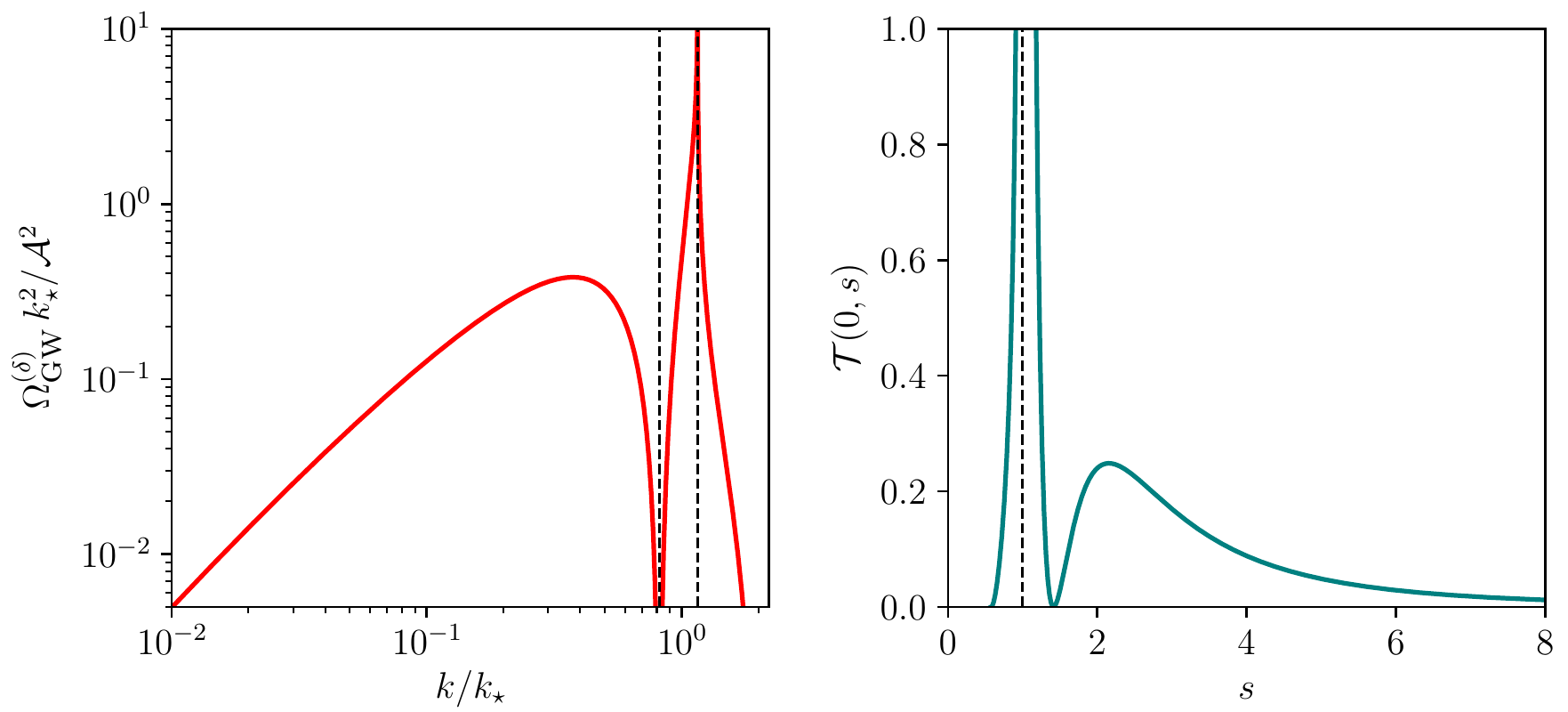}
\end{overpic}
\caption{\textit{\textbf{Left panel:} GW spectrum $\Omega_{\textrm{GW}}^{(\delta)}$ for a $\delta$-peak scalar power spectrum $\Pzeta(k)= \mathcal{A} \delta (k-k_\star)$ in \eqref{eq:OmegaGW-delta} plotted vs.~$k/k_\star$. The dashed vertical lines denote the dip in $\Omega_{\textrm{GW}}^{(\delta)}$ at $k=\sqrt{2/3} k_\star$ and the divergence at $k=2/\sqrt{3} k_\star$. \textbf{Right panel:} $\TRD(0,s)$ vs.~$s$ where $\TRD(d,s)$ is defined in \eqref{eq:TRDds}. The dashed vertical line denotes the singular locus $s=1$ where $\TRD(0,s)$ diverges.}}
\label{fig:TRDs}
\end{figure}

From the picture in the right panel of fig.~\ref{fig:int-ranges} it is apparent that the integration range in $s$ is also limited by the finite size of the peak. We can now proceed differently, depending on the precise width of the peak in $\Pzeta$. In the limit of an infinitely narrow peak the scalar power spectrum can be modelled by a $\delta$-distribution and the corresponding GW spectrum has been computed in \cite{Ananda:2006af, Kohri:2018awv}. In this case we should have replaced $\Pzeta$ by a $\delta$-distribution before performing the $d$-integral. Still, we can reproduce the correct result from \cite{Ananda:2006af,Kohri:2018awv} by replacing the factor $\Pzeta^2$ in \eqref{eq:OmegaGW-i-s-only-app} by 
\begin{align}
\Pzeta^2 \bigg(\frac{\sqrt{3}k}{2}s\bigg) \longrightarrow \frac{\mathcal{A}^2}{\Delta k} \delta \bigg(\frac{\sqrt{3}k}{2}s - k_\star \bigg) \, , \quad \textrm{with} \quad \mathcal{A} \equiv \int_0^\infty \textrm{d} k \, \Pzeta(k) \, ,
\end{align}
which gives
\begin{align}
\label{eq:OmegaGW-delta-app}
    \Omega_{\textrm{GW}}^{(\delta)}(k) = \frac{\mathcal{A}^2}{3 k^2} \TRD \Big(0,\tfrac{2 k_\star}{\sqrt{3} k}\Big) \Theta \Big(2- \tfrac{k}{k_\star} \Big) \, .
\end{align}
This is plotted in the left panel of fig.~\ref{fig:TRDs}. The crucial characteristics of $\Omega_{\textrm{GW}}^{(\delta)}$ are a broad peak at $k \sim 0.5 k_\star$, a dip at $k= \sqrt{2/3} k_\star$ and a principal peak at $k= 2 / \sqrt{3} k_\star$ where $\Omega_{\textrm{GW}}^{(\delta)}$ diverges as a result of resonant amplification. 

The case of a $\delta$-peak is only a toy model, albeit a useful one, as in practice $\Pzeta$ will be of finite width. To account for this finite extension of the peak we return to our expression \eqref{eq:OmegaGW-i-s-only-app}. For a narrow but finite-width peak in $\Pzeta$, the factor $\Pzeta^2$ in \eqref{eq:OmegaGW-i-s-only-app} can be understood to effect a form of running average of the expression $\TRD(0,s)$ with the averaging scale given by the width of the peak in $\Pzeta$. In particular, approximating $\Pzeta(k) = \mathcal{A} / \Delta k \, \Theta (k- k_\textrm{min}) \Theta ( k_\textrm{max} -k)$ and inserting into \eqref{eq:OmegaGW-i-s-only-app} this can be rewritten as
\begin{align}
    \label{eq:OmegaGW-narrow-average-app}
    \Omega_{\textrm{GW}, \textrm{narrow}}(k) = \frac{\mathcal{A}^2}{3 k^2} \frac{1}{\Delta k} \int_{k_\textrm{min}}^{k_\textrm{max}} \textrm{d}q \, \TRD \Big(0,\tfrac{2 q}{\sqrt{3} k} \Big) \, \Theta \Big(2- \tfrac{k}{q} \Big) \, ,
\end{align}
which can be identified as an average over an interval $\Delta k$ of the $\delta$-peak result in \eqref{eq:OmegaGW-delta-app}, but with $k_\star$ the variable to be averaged over. The observation that $\OGW$ for a narrow peak of finite width in $\Pzeta$ can be obtained by averaging over the $\delta$-peak expression has been already made in \cite{Fumagalli:2020nvq,Byrnes:2018txb,Pi:2020otn}.\footnote{However, in these works the averaging is done over the variable $k$ and not $k_\star$. However, for $k \sim k_\star$ the two procedures can be shown to be equivalent up to correction suppressed in $\Delta k / k_\star$.}
Here we have provided further justification for this procedure. The effect of the averaging is that the resonance peak at $k=2 k_\star / \sqrt{3}$ has finite amplitude. Overall, this leads to the well-known result that for a narrow peak at $k_\star$ in $\Pzeta$ the GW spectrum $\OGW$ exhibits a broad peak at $k \sim k_\star/2$ and a narrower principal peak at $k \simeq 2 k_\star / \sqrt{3}$, separated by a dip. 

In the derivation of \eqref{eq:OmegaGW-narrow-average-app} it was assumed that $\Pzeta$ was sufficiently narrow to be well-approximated by a peak of constant amplitude over an interval $\Delta k$. However, we can also envisage situations where the $k$-dependence of the peak in $\Pzeta$ is sufficiently important that it cannot be ignored. In this case we find that there is a different way of approximating $\OGW(k)$ that will apply as long as the peak in $\Pzeta$ is not too narrow. The observation here is that the integration kernel $\TRD(0,s)$ itself is narrowly-peaked between $s=0.6$ and $1.4$ with a divergence at $s=1$ (see the right panel in fig.~\ref{fig:TRDs}). In addition, there is a lower broad peak at larger values of $s$ with maximum near $s\simeq 2.2$. Most importantly, the integral over $s \in [1/\sqrt{3},\infty]$ is finite, with the bulk of the contribution coming from the peak encompassing the singular locus $s=1$:
\begin{align}
  \nonumber & \int_{1/\sqrt{3}}^\infty \textrm{d}s \, \TRD(0,s) = 2.29956 = I_\textrm{tot} \, , \qquad
  && \int_{0.6}^{1.4} \textrm{d}s \, \TRD(0,s) = 1.68998 = 0.73 \, I_\textrm{tot} \, , \\
  \nonumber & \int_{0.8}^{1.2} \textrm{d}s \, \TRD(0,s) = 1.6275 = 0.71 \, I_\textrm{tot} \, , 
  \qquad
  &&\int_{0.9}^{1.1} \textrm{d}s \, \TRD(0,s) = 1.40725 = 0.61 \, I_\textrm{tot} \, .
\end{align}
The peak between $s=0.6$ and $1.4$ is thus responsible for 73\% of the total area under $\TRD(0,s)$ and the region $s \in [0.8,1.2]$ still for 71\%. This is reminiscent of the properties of a resolution of a $\delta$-function. Of course, here there is no parametric limit in which $\TRD(0,s)$ approaches a $\delta$-distribution exactly. Still, given its properties, when included as an integration kernel with an integrand that varies sufficiently slowly in $s$, the effect of $\TRD(0,s)$ is to approximately pin the integrand to $s=1$. In particular, the integrand should not vary too fast over the interval $s \in [0.8,1.2]$ responsible for the bulk of the contribution from $\TRD(0,s)$. In \eqref{eq:OmegaGW-i-s-only-app} this implies that the peak in $\Pzeta$ should not be too narrow. For values of $k \sim k_\star$ we find that this requirement is satisfied for $\Delta k / k_\star \gtrsim 0.3$, corresponding to moderately narrow peaks, but which are formally still consistent with \eqref{eq:narrowdef}. If this is the case, we expect the GW spectrum over the principal peak to be well-approximated by
\begin{align}
\label{eq:OmegaGW-as-Psquared-app}
    \Omega_{\textrm{GW}, \textrm{narrow}}(k) \approx \frac{1}{2} \, d_\epsilon \, \gamma \, \Pzeta^2 \bigg(\frac{\sqrt{3}k}{2}\bigg) \, , \quad \textrm{for} \quad k \sim k_\star \, ,
\end{align}
where $\gamma$ a numerical factor arising from the integration over $s$. This should not be regarded as the result of a formal approximation scheme, but rather a heuristic technique for obtaining an approximate expression for the spectral shape of $\OGW$ that applies to certain types of peaks in $\Pzeta$. If it applies, the spectral shape of $\Omega_{\textrm{GW}}(k)$ is given through the relation $\Omega_{\textrm{GW}}(k) \propto \Pzeta^2(\sqrt{3}k/2)$. Note that there is a further dependence of $\Omega_{\textrm{GW}, \textrm{narrow}}(k)$ on $k$ through the factor $d_\epsilon$. However, this only depends weakly on $k$ and for a narrow peak will not affect the result much. Thus, when employing this result in practice we will not use \eqref{eq:OmegaGW-as-Psquared-app}, but the simpler form recorded in \eqref{eq:OmegaGW-as-Psquared-fit} in the main text, where this additional $k$-dependence is ignored.

\subsection{Oscillatory features as multiple peaks -- integration domains}
\label{app:multiple}
In this appendix we will argue for the applicability of the formula \eqref{eq:kmaxij-def} to predict the location of peaks in the GW spectrum due to a sharp or resonant feature. Here we assume that $A_\textrm{lin}, A_\textrm{log}$ in \eqref{eq:P_of_k_sharp}, \eqref{eq:P_of_k_resonant} are of order $\mathcal{O}(1)$ so that the scalar power spectrum can be modelled as a series of individual peaks. We define a peak as the interval between two intersections between $\Pzeta(k)$ with $\Pbar(k)$ containing a maximum, i.e.~the width of a single peak and the separation between two peaks are each given by half a period of oscillation.

We will use the insight from sec.~\ref{sec:general-observations} that the effect of a peak in $\Pzeta(k)$ is to restrict the range of integration over $(d,s)$ in the expression \eqref{eq:OmegaGW-i} for $\OGW(k)$. That is, a peak in $\Pzeta(k)$ defines a finite region in $(d,s)$-space where the integration in \eqref{eq:OmegaGW-i} receives its dominant contribution. This was shown in blue in fig.~\ref{fig:int-ranges}. The extent of this region in $(d,s)$ varies with $k$, but for every value of $k$ such a region can be defined. 

\begin{figure}[t]
		\centering
		\begin{tikzpicture}[scale=0.65]
		\fill[blue!30!white] (1.0,1.0) rectangle (1.25,1.25);
		\fill[blue!30!white] (1.0,1.5) rectangle (1.25,1.75);
		\fill[blue!30!white] (1.0,2.0) rectangle (1.25,2.25);
		\fill[blue!30!white] (1.0,2.5) rectangle (1.25,2.75);
		\fill[blue!30!white] (1.0,3.0) rectangle (1.25,3.25);
		\fill[blue!30!white] (1.0,3.5) rectangle (1.25,3.75);
		\fill[blue!30!white] (1.0,4.0) rectangle (1.25,4.25);
		\fill[blue!30!white] (1.0,4.5) rectangle (1.25,4.75);
		\fill[blue!30!white] (1.5,1.0) rectangle (1.75,1.25);
		\fill[blue!30!white] (1.5,1.5) rectangle (1.75,1.75);
		\fill[blue!30!white] (1.5,2.0) rectangle (1.75,2.25);
		\fill[blue!30!white] (1.5,2.5) rectangle (1.75,2.75);
		\fill[blue!30!white] (1.5,3.0) rectangle (1.75,3.25);
		\fill[blue!30!white] (1.5,3.5) rectangle (1.75,3.75);
		\fill[blue!30!white] (1.5,4.0) rectangle (1.75,4.25);
		\fill[blue!30!white] (1.5,4.5) rectangle (1.75,4.75);
		\fill[blue!30!white] (2.0,1.0) rectangle (2.25,1.25);
		\fill[blue!30!white] (2.0,1.5) rectangle (2.25,1.75);
		\fill[blue!30!white] (2.0,2.0) rectangle (2.25,2.25);
		\fill[blue!30!white] (2.0,2.5) rectangle (2.25,2.75);
		\fill[blue!30!white] (2.0,3.0) rectangle (2.25,3.25);
		\fill[blue!30!white] (2.0,3.5) rectangle (2.25,3.75);
		\fill[blue!30!white] (2.0,4.0) rectangle (2.25,4.25);
		\fill[blue!30!white] (2.0,4.5) rectangle (2.25,4.75);
		\fill[blue!30!white] (2.5,1.0) rectangle (2.75,1.25);
		\fill[blue!30!white] (2.5,1.5) rectangle (2.75,1.75);
		\fill[blue!30!white] (2.5,2.0) rectangle (2.75,2.25);
		\fill[blue!30!white] (2.5,2.5) rectangle (2.75,2.75);
		\fill[blue!30!white] (2.5,3.0) rectangle (2.75,3.25);
		\fill[blue!30!white] (2.5,3.5) rectangle (2.75,3.75);
		\fill[blue!30!white] (2.5,4.0) rectangle (2.75,4.25);
		\fill[blue!30!white] (2.5,4.5) rectangle (2.75,4.75);
		\fill[blue!30!white] (3.0,1.0) rectangle (3.25,1.25);
		\fill[blue!30!white] (3.0,1.5) rectangle (3.25,1.75);
		\fill[blue!30!white] (3.0,2.0) rectangle (3.25,2.25);
		\fill[blue!30!white] (3.0,2.5) rectangle (3.25,2.75);
		\fill[blue!30!white] (3.0,3.0) rectangle (3.25,3.25);
		\fill[blue!30!white] (3.0,3.5) rectangle (3.25,3.75);
		\fill[blue!30!white] (3.0,4.0) rectangle (3.25,4.25);
		\fill[blue!30!white] (3.0,4.5) rectangle (3.25,4.75);
		\fill[blue!30!white] (3.5,1.0) rectangle (3.75,1.25);
		\fill[blue!30!white] (3.5,1.5) rectangle (3.75,1.75);
		\fill[blue!30!white] (3.5,2.0) rectangle (3.75,2.25);
		\fill[blue!30!white] (3.5,2.5) rectangle (3.75,2.75);
		\fill[blue!30!white] (3.5,3.0) rectangle (3.75,3.25);
		\fill[blue!30!white] (3.5,3.5) rectangle (3.75,3.75);
		\fill[blue!30!white] (3.5,4.0) rectangle (3.75,4.25);
		\fill[blue!30!white] (3.5,4.5) rectangle (3.75,4.75);
		\fill[blue!30!white] (4.0,1.0) rectangle (4.25,1.25);
		\fill[blue!30!white] (4.0,1.5) rectangle (4.25,1.75);
		\fill[blue!30!white] (4.0,2.0) rectangle (4.25,2.25);
		\fill[blue!30!white] (4.0,2.5) rectangle (4.25,2.75);
		\fill[blue!30!white] (4.0,3.0) rectangle (4.25,3.25);
		\fill[blue!30!white] (4.0,3.5) rectangle (4.25,3.75);
		\fill[blue!30!white] (4.0,4.0) rectangle (4.25,4.25);
		\fill[blue!30!white] (4.0,4.5) rectangle (4.25,4.75);
		\fill[blue!30!white] (4.5,1.0) rectangle (4.75,1.25);
		\fill[blue!30!white] (4.5,1.5) rectangle (4.75,1.75);
		\fill[blue!30!white] (4.5,2.0) rectangle (4.75,2.25);
		\fill[blue!30!white] (4.5,2.5) rectangle (4.75,2.75);
		\fill[blue!30!white] (4.5,3.0) rectangle (4.75,3.25);
		\fill[blue!30!white] (4.5,3.5) rectangle (4.75,3.75);
		\fill[blue!30!white] (4.5,4.0) rectangle (4.75,4.25);
		\fill[blue!30!white] (4.5,4.5) rectangle (4.75,4.75);
		\fill[black] (1.125,1.125) circle (1.5pt);
		\fill[black] (1.625,1.125) circle (1.5pt);
		\fill[black] (2.125,1.125) circle (1.5pt);
		\fill[black] (2.625,1.125) circle (1.5pt);
		\fill[black] (3.125,1.125) circle (1.5pt);
		\fill[black] (3.625,1.125) circle (1.5pt);
		\fill[black] (4.125,1.125) circle (1.5pt);
		\fill[black] (4.625,1.125) circle (1.5pt);
		
		\fill[black] (1.125,1.625) circle (1.5pt);
		\fill[black] (1.625,1.625) circle (1.5pt);
		\fill[black] (2.125,1.625) circle (1.5pt);
		\fill[black] (2.625,1.625) circle (1.5pt);
		\fill[black] (3.125,1.625) circle (1.5pt);
		\fill[black] (3.625,1.625) circle (1.5pt);
		\fill[black] (4.125,1.625) circle (1.5pt);
		\fill[black] (4.625,1.625) circle (1.5pt);
		
		\fill[black] (1.125,1.625) circle (1.5pt);
		\fill[black] (1.625,1.625) circle (1.5pt);
		\fill[black] (2.125,1.625) circle (1.5pt);
		\fill[black] (2.625,1.625) circle (1.5pt);
		\fill[black] (3.125,1.625) circle (1.5pt);
		\fill[black] (3.625,1.625) circle (1.5pt);
		\fill[black] (4.125,1.625) circle (1.5pt);
		\fill[black] (4.625,1.625) circle (1.5pt);
		
		\fill[black] (1.125,2.125) circle (1.5pt);
		\fill[black] (1.625,2.125) circle (1.5pt);
		\fill[black] (2.125,2.125) circle (1.5pt);
		\fill[black] (2.625,2.125) circle (1.5pt);
		\fill[black] (3.125,2.125) circle (1.5pt);
		\fill[black] (3.625,2.125) circle (1.5pt);
		\fill[black] (4.125,2.125) circle (1.5pt);
		\fill[black] (4.625,2.125) circle (1.5pt);
		
		\fill[black] (1.125,2.625) circle (1.5pt);
		\fill[black] (1.625,2.625) circle (1.5pt);
		\fill[black] (2.125,2.625) circle (1.5pt);
		\fill[black] (2.625,2.625) circle (1.5pt);
		\fill[black] (3.125,2.625) circle (1.5pt);
		\fill[black] (3.625,2.625) circle (1.5pt);
		\fill[black] (4.125,2.625) circle (1.5pt);
		\fill[black] (4.625,2.625) circle (1.5pt);
		
		\fill[black] (1.125,3.125) circle (1.5pt);
		\fill[black] (1.625,3.125) circle (1.5pt);
		\fill[black] (2.125,3.125) circle (1.5pt);
		\fill[black] (2.625,3.125) circle (1.5pt);
		\fill[black] (3.125,3.125) circle (1.5pt);
		\fill[black] (3.625,3.125) circle (1.5pt);
		\fill[black] (4.125,3.125) circle (1.5pt);
		\fill[black] (4.625,3.125) circle (1.5pt);
		
		\fill[black] (1.125,3.625) circle (1.5pt);
		\fill[black] (1.625,3.625) circle (1.5pt);
		\fill[black] (2.125,3.625) circle (1.5pt);
		\fill[black] (2.625,3.625) circle (1.5pt);
		\fill[black] (3.125,3.625) circle (1.5pt);
		\fill[black] (3.625,3.625) circle (1.5pt);
		\fill[black] (4.125,3.625) circle (1.5pt);
		\fill[black] (4.625,3.625) circle (1.5pt);
		
		\fill[black] (1.125,4.125) circle (1.5pt);
		\fill[black] (1.625,4.125) circle (1.5pt);
		\fill[black] (2.125,4.125) circle (1.5pt);
		\fill[black] (2.625,4.125) circle (1.5pt);
		\fill[black] (3.125,4.125) circle (1.5pt);
		\fill[black] (3.625,4.125) circle (1.5pt);
		\fill[black] (4.125,4.125) circle (1.5pt);
		\fill[black] (4.625,4.125) circle (1.5pt);
		
		\fill[black] (1.125,4.625) circle (1.5pt);
		\fill[black] (1.625,4.625) circle (1.5pt);
		\fill[black] (2.125,4.625) circle (1.5pt);
		\fill[black] (2.625,4.625) circle (1.5pt);
		\fill[black] (3.125,4.625) circle (1.5pt);
		\fill[black] (3.625,4.625) circle (1.5pt);
		\fill[black] (4.125,4.625) circle (1.5pt);
		\fill[black] (4.625,4.625) circle (1.5pt);
		
		\fill[red!100!white, opacity=0.2] (1, 1) -- (0, 2.0) -- (6.5-2, 6.5) -- (6.5, 6.5) -- cycle;
		\draw[thick,->] (0,0) -- (6.5,0) node[anchor=north west] {$s-d$};
        \draw[thick,->] (0,0) -- (0,6.5) node[anchor=south east] {$s+d$};
        \draw[thick,->] (0,0) -- (-2.5,2.5) node[anchor=south west] {$d$};
        \draw[thick,->] (0,0) -- (6.5,6.5) node[anchor=south east] {$s$};
        \draw [line width=0.6pt,dash pattern=on 5pt off 5pt] (6.5-2, 6.5) -- (-1.0,1.0) node[anchor=east, rotate=45] {$\frac{1}{\sqrt{3}}$};
        \draw [line width=0.6pt,dash pattern=on 5pt off 5pt] (6.5-2, 6.5) -- (-1.0,1.0) node[anchor=east, rotate=45] {$\frac{1}{\sqrt{3}}$};
        \draw [line width=1.0pt,dash pattern=on 5pt off 5pt] (1,1)-- (2,0) node[anchor=north, rotate=45] {$\frac{1}{\sqrt{3}}$};
        \draw [line width=1.0pt,dash pattern=on 5pt off 5pt] (1.0,1.0)-- (-2.5,4.5);
        \draw [line width=1.0pt,dash pattern=on 5pt off 5pt] (1/0.57735,1/0.57735)-- (2/0.57735,0) node[anchor=north, rotate=45] {$1$};
        \draw [line width=1.0pt,dash pattern=on 5pt off 5pt] (-2.5,2/0.57735+2.5)-- (1/0.57735,1/0.57735);
		\end{tikzpicture}
		\quad
		\begin{tikzpicture}[scale=0.65]
		%\fill[yellow!20!white] (2, 0) -- (-2.5, 2+2.5) -- (-2.5, 6.5) -- (6.5, 6.5) -- (6.5,0) -- cycle;
		\fill[blue!30!white] (0.877,0.877) rectangle (1.14,1.14);
		\fill[blue!30!white] (0.877,1.48) rectangle (1.14,1.93);
		\fill[blue!30!white] (0.877,2.51) rectangle (1.14,3.27);
		\fill[blue!30!white] (0.877,4.26) rectangle (1.14,5.54);
		\fill[blue!30!white] (1.48,0.877) rectangle (1.93,1.14);
		\fill[blue!30!white] (1.48,1.48) rectangle (1.93,1.93);
		\fill[blue!30!white] (1.48,2.51) rectangle (1.93,3.27);
		\fill[blue!30!white] (1.48,4.26) rectangle (1.93,5.54);
		\fill[blue!30!white] (2.51,0.877) rectangle (3.27,1.14);
		\fill[blue!30!white] (2.51,1.48) rectangle (3.27,1.93);
		\fill[blue!30!white] (2.51,2.51) rectangle (3.27,3.27);
		\fill[blue!30!white] (2.51,4.26) rectangle (3.27,5.54);
		\fill[blue!30!white] (4.26,0.877) rectangle (5.54,1.14);
		\fill[blue!30!white] (4.26,1.48) rectangle (5.54,1.93);
		\fill[blue!30!white] (4.26,2.51) rectangle (5.54,3.27);
		\fill[blue!30!white] (4.26,4.26) rectangle (5.54,5.54);
		
		\fill[black] (1.0085,1.0085) circle (2pt);
		\fill[black] (1.0085,1.705) circle (2pt);
		\fill[black] (1.0085,2.89) circle (2pt);
		\fill[black] (1.0085,4.9) circle (2pt);
		\fill[black] (1.705,1.0085) circle (2pt);
		\fill[black] (1.705,1.705) circle (2pt);
		\fill[black] (1.705,2.89) circle (2pt);
		\fill[black] (1.705,4.9) circle (2pt);
		\fill[black] (2.89,1.0085) circle (2pt);
		\fill[black] (2.89,1.705) circle (2pt);
		\fill[black] (2.89,2.89) circle (2pt);
		\fill[black] (2.89,4.9) circle (2pt);
		\fill[black] (4.9,1.0085) circle (2pt);
		\fill[black] (4.9,1.705) circle (2pt);
		\fill[black] (4.9,2.89) circle (2pt);
		\fill[black] (4.9,4.9) circle (2pt);
		
		\fill[red!100!white, opacity=0.2] (1, 1) -- (0, 2.0) -- (6.5-2, 6.5) -- (6.5, 6.5) -- cycle;
		\draw[thick,->] (0,0) -- (6.5,0) node[anchor=north west] {$s-d$};
        \draw[thick,->] (0,0) -- (0,6.5) node[anchor=south east] {$s+d$};
        \draw[thick,->] (0,0) -- (-2.5,2.5) node[anchor=south west] {$d$};
        \draw[thick,->] (0,0) -- (6.5,6.5) node[anchor=south east] {$s$};
        \draw [line width=0.6pt,dash pattern=on 5pt off 5pt] (6.5-2, 6.5) -- (-1.0,1.0) node[anchor=east, rotate=45] {$\frac{1}{\sqrt{3}}$};
        \draw [line width=0.6pt,dash pattern=on 5pt off 5pt] (6.5-2, 6.5) -- (-1.0,1.0) node[anchor=east, rotate=45] {$\frac{1}{\sqrt{3}}$};
        \draw [line width=1.0pt,dash pattern=on 5pt off 5pt] (1,1)-- (2,0) node[anchor=north, rotate=45] {$\frac{1}{\sqrt{3}}$};
        \draw [line width=1.0pt,dash pattern=on 5pt off 5pt] (1.0,1.0)-- (-2.5,4.5);
        \draw [line width=1.0pt,dash pattern=on 5pt off 5pt] (1/0.57735,1/0.57735)-- (2/0.57735,0) node[anchor=north, rotate=45] {$1$};
        \draw [line width=1.0pt,dash pattern=on 5pt off 5pt] (-2.5,2/0.57735+2.5)-- (1/0.57735,1/0.57735);
		\end{tikzpicture} 
		\caption{\textit{Constraints (blue boxes) on the integration domain in expression \eqref{eq:OmegaGW-i} (red band) for a scalar power spectrum with a sharp feature with eight peaks (left panel) and a resonant feature over four peaks with $\omegalog=2.5 \omegalogc$ (right panel).}}
		\label{fig:int-ranges-resonant-sharp}
	\end{figure}
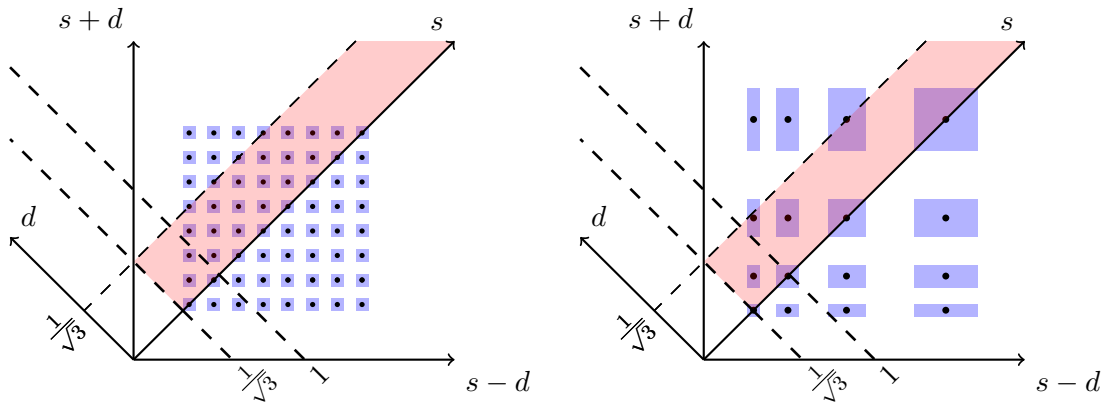
	
This can be generalised to a series of peaks, as we consider here, in which case one will find multiple such regions in $(d,s)$-space. Repeating the analysis in sec.~\ref{sec:general-observations} for multiple peaks, for a scalar power spectrum with $N$ independent peaks one finds $N^2$ such (disconnected) regions in $(d,s)$-space, which arises from the fact that \eqref{eq:OmegaGW-i} contains two factors of $\Pzeta$. This is again best illustrated graphically: For a fixed value of $k$, the constraints on the domain of integration due to a sharp or resonant feature are shown in fig.~\ref{fig:int-ranges-resonant-sharp} as the blue boxes on the $(s-d)$-$(s+d)$-plane, with the left panel displaying the result for a sharp feature (with 8 peaks in $\Pzeta$) and the right panel is for the resonant feature case (with 4 peaks in $\Pzeta$). For a sharp feature, the blue boxes all have the same width, height and separation, which is due to the periodicity in $k$ of $\Pzeta$. The changing size and separation in the resonant feature case is consistent with the $\log(k)$-periodicity of $\Pzeta$. The red region once more denotes the domain of integration over $(d,s)$ in expression \eqref{eq:OmegaGW-i}, and hence only the blue boxes that overlap with the red band will contribute to $\OGW$. The black circles in the centre of the blue regions are the images of the maxima of the peaks on the $(s+d)$-$(s-d)$-plane. Denoting the loci of the maxima by $k_{\star i}$, these are given by
\begin{align}
\label{eq:max-image}
    s-d = \frac{2 k_{\star i}}{\sqrt{3} k} \, , \quad s+d = \frac{2 k_{\star j}}{\sqrt{3} k} \, , \quad  \textrm{with} \quad i,j=1, \ldots, N \, .
\end{align}

The boxes on the diagonal (i.e.~on the $s$-axis) correspond to the constraints due to the $N$ individual peaks in $\Pzeta$, while off-diagonal boxes are regions where the constraints on $(d,s)$ from different peaks overlap. This is what was referred to by `interactions' of peaks in the main text. Note that only half the area of the boxes on the diagonal contributes to the integral, while off-diagonal boxes that lie within the red band contribute with their full area. This factor-2-mismatch correctly reproduces the combinatorial factors that arise when substituting for the two factors of $\Pzeta$ in \eqref{eq:OmegaGW-i} with a sum over individual peaks and then expanding. 

As explained in sec.~\ref{sec:multiple}, the individual peaks in $\Pzeta$ due to a feature will in general be narrow in the sense \eqref{eq:narrowdef}, implying that the blue regions will in general have a small area. For a sharp feature (left panel in fig.~\ref{fig:int-ranges-resonant-sharp}) this applies to all of the blue regions. For a resonant feature (right panel in fig.~\ref{fig:int-ranges-resonant-sharp}) this is only the case for the blue regions at smaller values of $s$, as with increasing $s$ the area of the boxes is growing by default. However, as the dominant contribution to the GW spectrum arises from the region near $s=1$, it is only the size of the blue boxes in this region that is of relevance. At the same time, for both sharp and resonant features one typically has many of the blue boxes overlapping with the red region. 

The fact that each blue region (of relevance) is small allows us to proceed as in sec.~\ref{app:narrow} to estimate the contribution to $\OGW$ from each blue region. That is, we can treat any such blue box in the same way as we have done with a single narrow peak. Here we sketch the main steps, which are the same as in sec.~\ref{app:narrow}. Firstly, note that such a blue region only spans a small interval in $d$, which we can use to trivialise the $d$-integration, by taking the integrand as effectively constant in $d$ in the blue box. One is then left with the integral over $s$ with an integration kernel $\mathcal{T}_\textrm{RD}(d_{ij}, s)$, where $d_{ij}$ is the value of $d$ at the centre of the $ij$-th box. This integration kernel diverges at $s=1$, but the integral over it is finite. Performing the $s$-integral, the effect of this divergence is to produce a tall narrow resonance peak in $\OGW$. This occurs for values of $k$ when the blue region and in particular the images of the maxima overlap with the line $s=1$.

Given \eqref{eq:max-image}, we can predict the loci of resonance peaks in $\OGW(k)$. In particular, the points \eqref{eq:max-image} intersect the line $s=1$ when
\begin{align}
    k= \frac{1}{\sqrt{3}}(k_{\star i}+ k_{\star j}) \, ,
\end{align}
which is what we have recorded in \eqref{eq:kmaxij-def} as the loci of resonance peaks. However, the images of the maxima have to intersect $s=1$ within the domain of integration (the red band), which is bounded in $d$ as $0< d< 1/\sqrt{3}$. Applied to \eqref{eq:max-image} this gives
\begin{align}
k > |k_{\star i}- k_{\star j}| \, ,
\end{align}
which we identify as the constraint in \eqref{eq:kmaxij-def}. 

To summarise, we can predict the peak-structure of $\OGW(k)$ due to a sharp or resonant feature by modelling the scalar power spectrum as a series of individual peaks and computing their corresponding resonance peaks in the GW spectrum via \eqref{eq:kmaxij-def}. This has been used successfully to analyse the peak-structure of $\OGW(k)$ due to a sharp feature in \cite{Fumagalli:2020nvq}. In sec.~\ref{sec:resonant-peak-structure} of this work we apply this to the resonant feature case.

\section{A resonant feature with a Gaussian-like peak}
\label{app:Gausslike}
In this appendix we present a further test of our analytical expressions \eqref{analyticalO1}, \eqref{analyticalO2b} for the oscillatory part of $\OGW$, or rather the templates \eqref{eq:OmegaGW1-fit}, \eqref{eq:OmegaGW2-fit} that generalise the analytical results to power spectra with nontrivial envelope $\Pbar$. In the main text we provided checks by comparing with numerical results for scalar power spectra with envelopes with top-hat or lognormal peak profiles. What is common to these examples is that these peaks in the envelope are mirror-symmetric in $\log(k)$ about their maximum at $k=\kref$. To dispel any doubt about whether the good agreement between the numerical results and our analytical templates in sec.~\ref{sec:resonant-templates} has anything to do with this symmetry of the peaks, we will here consider the case of a scalar power spectrum that does not exhibit a mirror-symmetry in $\log(k)$. 

\begin{figure}[t]
\centering
\begin{overpic}[width=0.8\textwidth]{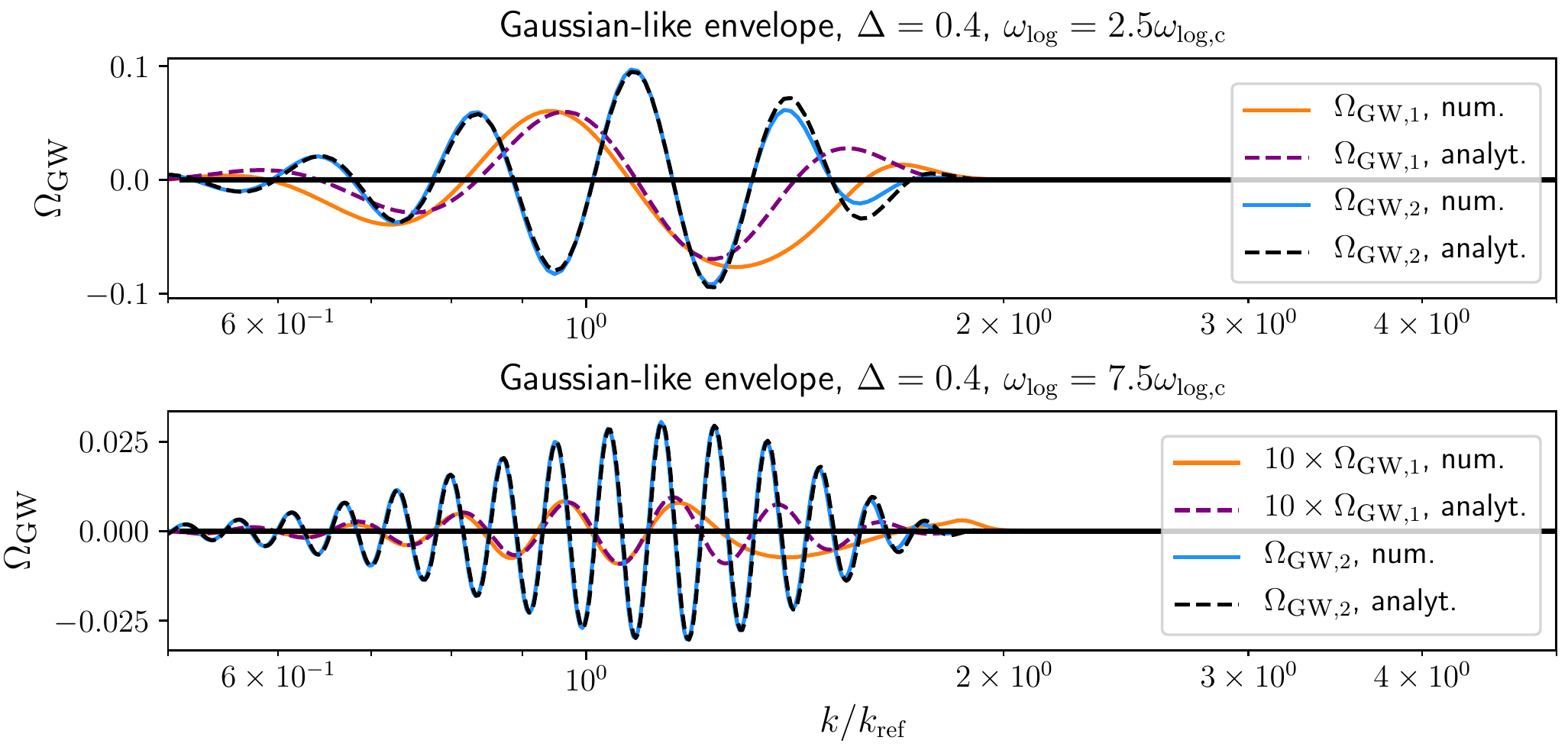}
\end{overpic}
\caption{\textit{Numerical result and analytical templates \eqref{eq:OmegaGW1-fit}, \eqref{eq:OmegaGW2-fit} for $\Omega_{\textrm{GW},1}$ and $\Omega_{\textrm{GW},2}$ for a scalar power spectrum \eqref{eq:P_of_k_resonant} with a lognormal envelope given in \eqref{eq:Pbar-Gausslike} with $\Delta=0.4$ and $\omegalog =2.5 \omegalogc$ (upper panel) and $\omegalog =7.5 \omegalogc$ (lower panel). The parameters $\tilde{\gamma}_{1,2}$ and $\phi_{1,2}$ in the templates \eqref{eq:OmegaGW1-fit} and \eqref{eq:OmegaGW2-fit} are chosen to best match the numerical curves.}}
\label{fig:OGW_1_2_GL0p4_2p5_7p5_vs_analyt}
\end{figure}

\begin{figure}[t]
\centering
\begin{overpic}[width=0.8\textwidth]{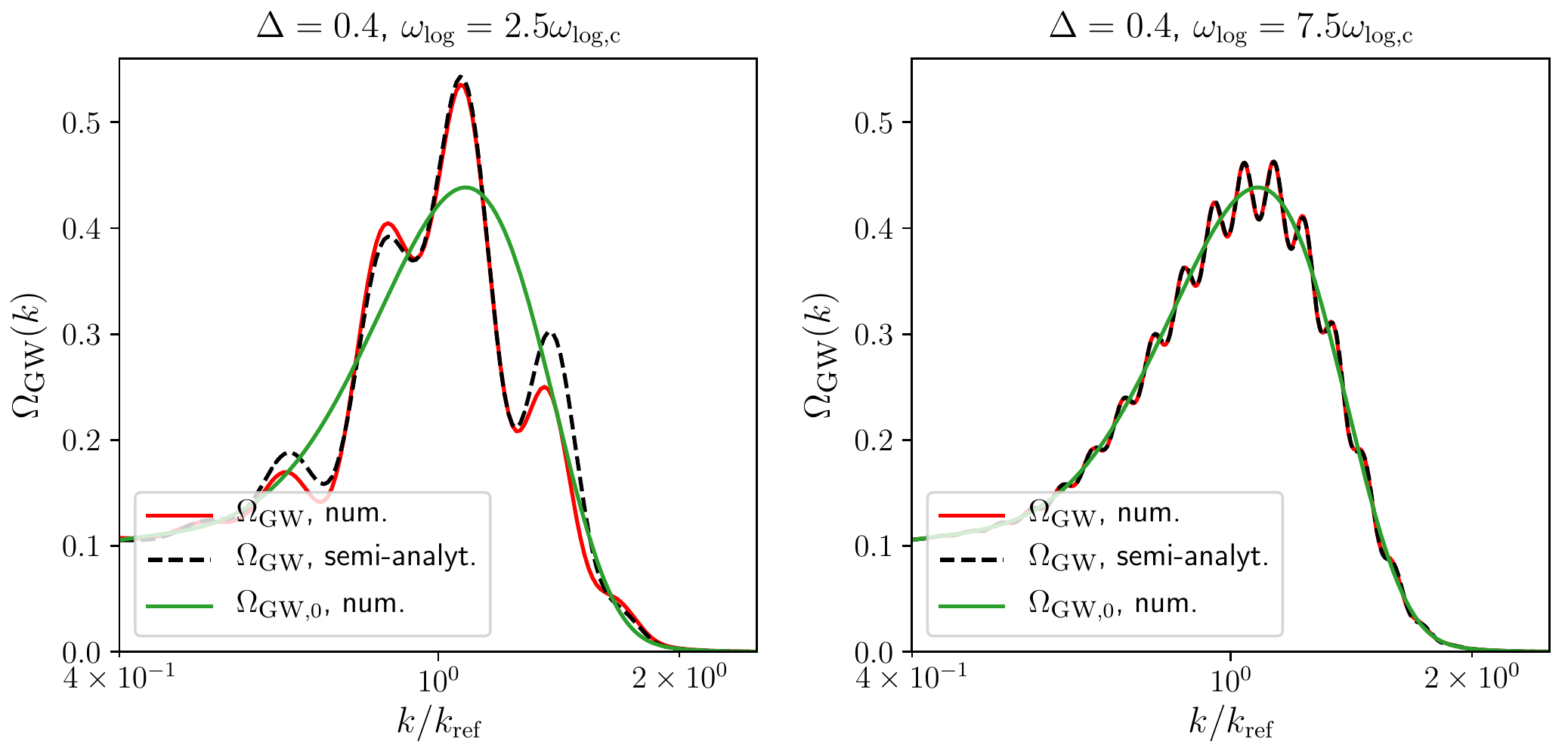}
\end{overpic}
\caption{\textit{Numerical result for $\OGW$ (red) and $\Omega_{\textrm{GW},0}$ (green) vs.~$k/\kref$ for a scalar power spectrum \eqref{eq:P_of_k_resonant} with $A_\textrm{log}=1$ and envelope $\Pbar$ given by a broad Gaussian-like peak \eqref{eq:Pbar-Gausslike} with $\Delta=0.4$. The left panel is for $\omegalog =2.5 \omegalogc$ and the right panel for $\omegalog =7.5 \omegalogc$. The black dashed curve corresponds to a semi-analytical approximation as described in appendix \ref{app:Gausslike}. We find a very good agreement between the numerical and semi-analytical result over the central part of the peak of $\OGW$.}}
\label{fig:OGW_GL0p4_2p5_7p5_vs_analyt}
\end{figure}

To be explicit, here we will consider scalar power spectra of the resonant type \eqref{eq:P_of_k_resonant} with an envelope $\Pbar$ given by
\begin{align}
\label{eq:Pbar-Gausslike}
    \Pbar(k) = \mathcal{P}_\textrm{max} \, \exp \bigg( -\frac{1}{2 \Delta^2} \frac{(k-\kref)^2}{(\kref^2-(k-\kref)^2)} \bigg) \, , \quad \textrm{with} \quad \mathcal{P}_\textrm{max}=1 \, ,
\end{align}
where the range of validity of this expression is $0 < k < 2 \kref$. For $k > 2 \kref$ we set $\Pbar(k)=0$.
Near its maximum at $k=\kref$ this resembles a Gaussian peak, but away from the maximum it has suppressed tails compared to a Gaussian. Still, for brevity we refer to this as a Gaussian-like peak. In particular, the denominator in the exponential was included to ensure that the power spectrum drops sufficiently fast for $k \rightarrow 0$. Note that \eqref{eq:Pbar-Gausslike} is still mirror-symmetric about $k=\kref$, but now this is for reflections in the variable $k$ and not $\log(k)$. Compared to a lognormal peak \eqref{eq:Pbar-LN} with the same value of $\Delta$, the Gaussian-like peak has more power at small values $k < \kref$, but then drops off faster for $k > \kref$.

Here we will show explicit results for an example with a Gaussian-like envelope with $\Delta=0.4$. The reason for this choice is that for this value of $\Delta$ the peak is fairly narrow and thus represents a non-trivial test for the templates \eqref{eq:OmegaGW1-fit}, \eqref{eq:OmegaGW2-fit}. In addition, this will allow for direct comparisons with the examples with a lognormal envlope and the same choice $\Delta=0.4$ (see figs.~\ref{fig:OGW_1_2_LN0p4_2p5_7p5_vs_analyt} and \ref{fig:OGW_LN0p4_2p5_7p5_vs_analyt}). In fig.~\ref{fig:OGW_1_2_GL0p4_2p5_7p5_vs_analyt} we display the numerical results for $\Omega_{\textrm{GW},1}$, $\Omega_{\textrm{GW},2}$ and fits of the analytical templates \eqref{eq:OmegaGW1-fit}, \eqref{eq:OmegaGW2-fit} for two examples with a Gaussian-like envelope with $\Delta=0.4$ and $\omegalog =2.5 \omegalogc$ (upper panel) and $\omegalog =7.5 \omegalogc$ (lower panel). The parameters $\tilde{\gamma}_{1,2}$ and $\phi_{1,2}$ in the templates \eqref{eq:OmegaGW1-fit} and \eqref{eq:OmegaGW2-fit} are chosen to best match the numerical curves.\footnote{For $\omegalog=2.5 \omegalogc$ we choose $\tilde{\gamma}_1=0.072$, $\tilde{\gamma}_2=0.098$, $\phi_1=2.14$, $\phi_2=1.42$ and for $\omegalog=7.5 \omegalogc$ we use $\tilde{\gamma}_1=0.0010$, $\tilde{\gamma}_2=0.0305$, $\phi_1=-0.18$, $\phi_2=1.12$.} We observe that the result for $\Omega_{\textrm{GW},2}$ is very-well matched by the template for both values of $\omegalog$, capturing both the oscillation as well as the envelope. For $\Omega_{\textrm{GW},1}$ the template provides a reasonably good approximation for $k \lesssim \kref$, but departs from the numerical result for $k > \kref$, where the signal cannot be modelled anymore by a sinusoidal oscillation with frequency $\omegalog$. The upshot is that the templates can still be used to model the oscillatory part of $\OGW$, albeit with some restrictions. For smaller values of $\omegalog$ the analytical fit works best for $k \lesssim \kref$ due to the difficulty of matching the result for $\Omega_{\textrm{GW},1}$. However, for larger values of $\omegalog$ the contribution $\Omega_{\textrm{GW},1}$ is suppressed compared to $\Omega_{\textrm{GW},2}$ and hence the oscillatory part is again reproduced to a high accuracy by the superposition of the two templates. 

This can be seen in fig.~\ref{fig:OGW_GL0p4_2p5_7p5_vs_analyt} where we show the full numerical result for $\OGW$ (red curve) and compare to our fits based on the analytical templates \eqref{eq:OmegaGW1-fit} and \eqref{eq:OmegaGW2-fit}. Here we however proceed differently than in the main text. There we used analytical expressions for the smooth background spectrum $\Omega_{\textrm{GW},0}$ based on the broad-peak or narrow-peak approximations collected in section \ref{sec:broad-narrow}. Here we will follow a different strategy to which one can revert when no analytical result for $\Omega_{\textrm{GW},0}$ is available. That is, here we will use the numerical result for the smooth background $\Omega_{\textrm{GW},0}$, shown as the green curve, and only model the oscillatory part with the analytical templates \eqref{eq:OmegaGW1-fit} and \eqref{eq:OmegaGW2-fit}. This semi-analytical fit is shown as the black dashed line in fig.~\ref{fig:OGW_GL0p4_2p5_7p5_vs_analyt}. For $\omegalog=2.5 \omegalogc$ the semi-analytical fit provides a good match to the full signal for $k \lesssim \kref$, as we expect from the discussion above. For $\omegalog=7.5 \omegalogc$ the semi-analytical fit is near-indistinguishable from the full result over the whole range of the peak. To summarise, the description of the oscillatory part of $\OGW$ as a superposition of two sinusoidal oscillations modulating a common envelope as given by the templates \eqref{eq:OmegaGW1-fit}, \eqref{eq:OmegaGW2-fit} does not appear to hinge on a particular type of shape for the envelope $\Pbar$. The templates \eqref{eq:OmegaGW1-fit}, \eqref{eq:OmegaGW2-fit} provide good fits for the top-hat and lognormal envelopes discussed in the main text and the Gaussian-like envelope presented here (despite some inaccuracies observed for $k >\kref$ and smaller values of $\omegalog$), suggesting that they are universally applicable. 

\section{Oscillatory part of $\OGW$ for $A_\textrm{log}<1$}
\label{app:reduce-A-log}
One key result of this paper is that for a resonant feature in the scalar power spectrum of type \eqref{eq:P_of_k_resonant} with frequency of oscillation $\omegalog$, the corresponding GW energy density spectrum exhibits a superposition of modulations with frequencies $\omegalog$ and $2\omegalog$. The relative importance of these two contributions depends on $\omegalog$ itself, but also on the amplitude $A_\textrm{log}$ of the oscillations in $\Pzeta$. In the main text of this work we mainly focussed on examples with $A_\textrm{log} \simeq 1$, as required for the resonance analysis in sec.~\ref{sec:resonant-peak-structure}, and by choice in the description of our analytic results in sec.~\ref{sec:resonant-templates-analytical} and in the numerical examples in sec.~\ref{sec:resonant-templates-numerical}, motivated by the observability of the signal. In this appendix we consider the effect of reducing $A_\textrm{log}$ on the pattern of oscillations in $\OGW$ and argue that the case with $A_\textrm{log} \simeq 1$ correctly captures the main effects in all situations of relevance.

\begin{figure}[t]
\centering
\begin{overpic}[width=0.8\textwidth]{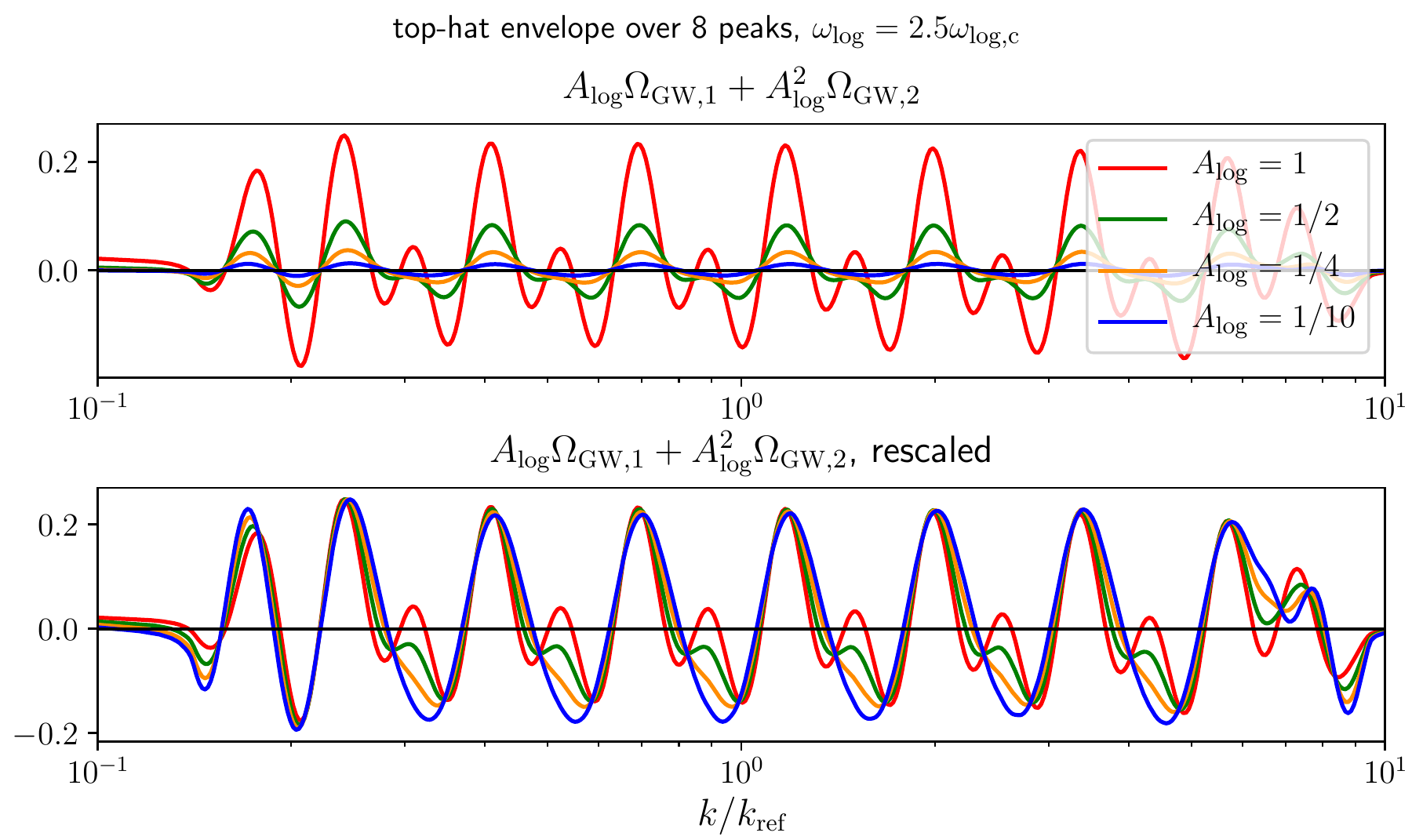}
\end{overpic}
\caption{\textit{Numerically obtained plots of the oscillatory parts $A_\textrm{log} \Omega_{\textrm{GW},1}+A_\textrm{log}^2 \Omega_{\textrm{GW},2}$ of the GW spectrum as defined via \eqref{eq:OGW-analyt-expansion} vs.~$k/\kref$ for a scalar power spectrum \eqref{eq:P_of_k_resonant} with frequency $\omegalog = 2.5 \omegalogc$ and $A_\textrm{log}= 1, \, 1/2, \, 1/4, \, 1/10$. The envelope $\Pbar$ is given by a top-hat with $\mathcal{P}_\textrm{max}=1$ and geometric mean at $k=\kref$ stretching over 8 periods of oscillation. The complete result for $\OGW$ for $A_\textrm{log}= 1, \, 1/2$ is given in fig.~\ref{fig:P_OGW_2p5wc_8}. The upper panel shows $A_\textrm{log} \Omega_{\textrm{GW},1}+A_\textrm{log}^2 \Omega_{\textrm{GW},2}$ for the four examples. In the lower panel we rescaled the plots for $A_\textrm{log}<1$ from the upper panel for better visibility by requiring that the global maxima in the plotted range coincide with that of the example with $A_\textrm{log}=1$. When $A_\textrm{log}$ is reduced, the contribution $\Omega_{\textrm{GW},2}$ with frequency $2 \omegalog$ is suppressed relative to the term $\Omega_{\textrm{GW},1}$ with frequency $\omegalog$, resulting in changes to the periodicity pattern of the oscillations as described in the main text.}}
\label{OGW_1_2_differentAs_2p5wc}
\end{figure}

\begin{figure}[t]
\centering
\begin{overpic}[width=0.8\textwidth]{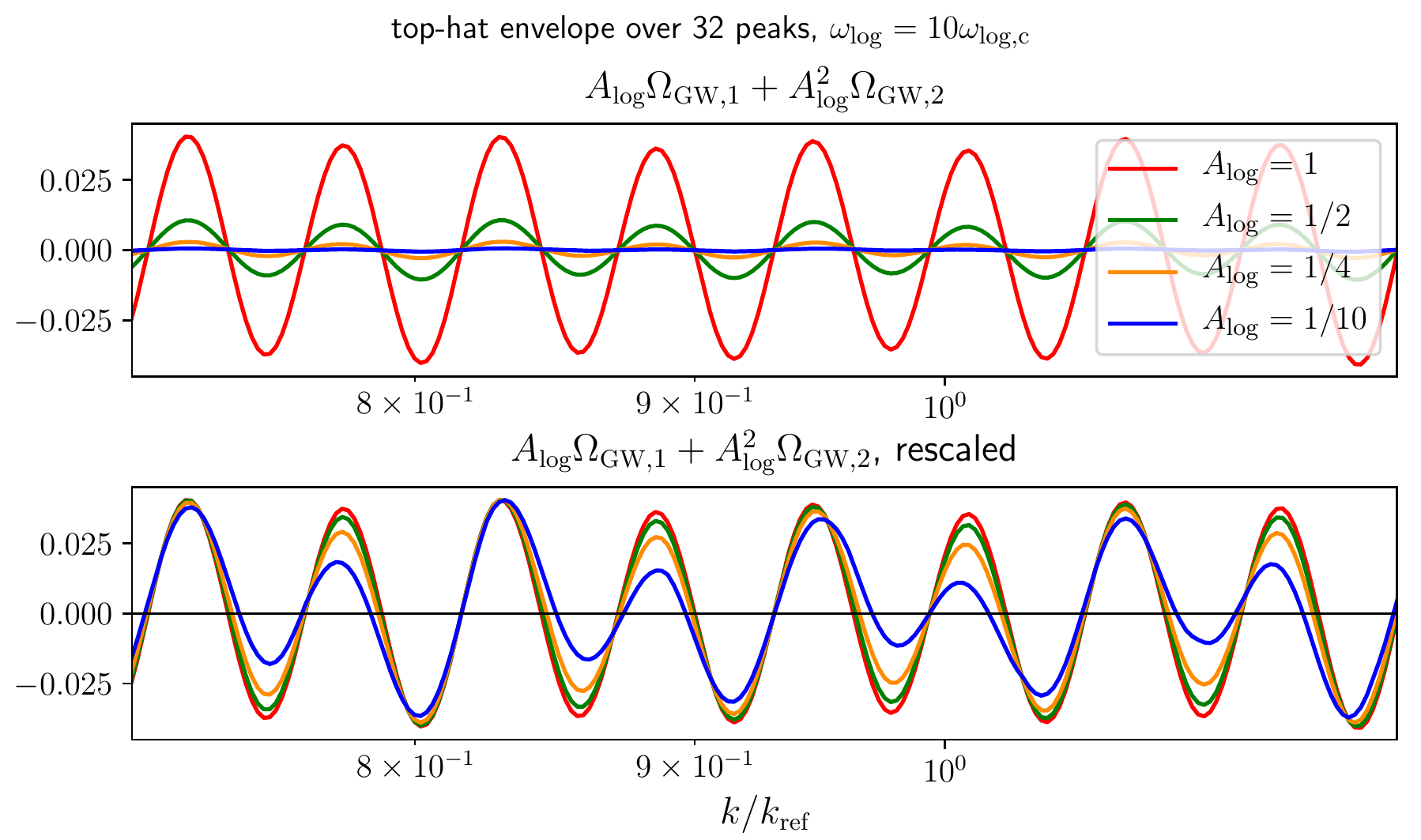}
\end{overpic}
\caption{\textit{This figure has the same content as fig.~\ref{OGW_1_2_differentAs_2p5wc} above, but for an example with $\omegalog = 10 \omegalogc$. The envelope $\Pbar$ is also unchanged compared to fig.~\ref{OGW_1_2_differentAs_2p5wc} and now stretches over 32 periods of oscillation. For better visibility we restrict the plotted range in $k$ to the central region $k \in [0.7 \kref, 1.2 \kref]$. The change in the peak pattern with $A_\textrm{log}$ is described in the main text.}}
\label{OGW_1_2_differentAs_10wc}
\end{figure}

In this work we established that the oscillatory part of $\OGW$ can exhibit one of three different qualitative peak structures depending on the value of $\omegalog$. For $A_\textrm{log} \simeq 1$ this is graphically summarised in fig.~\ref{fig:summary} and described at the end of sec.~\ref{sec:resonant-templates-analytical} and in the conclusions. In this case, for $\omegalog \lesssim \omegalogc$ the oscillations in $\OGW$ are dominated by the oscillatory part with frequency $\omegalog$ while for $\omegalog \gtrsim \omegalogd \simeq 6 \omegalogc$ it is the piece with frequency $2 \omegalog$ that dominates. In the intermediate regime, $\omegalogc \lesssim \omegalog \lesssim \omegalogd$ the two oscillatory pieces are of comparable amplitudes resulting in a complicated periodicity structure consisting of a superposition of two series of peaks.     

To investigate how this is modified by changing the value of $A_\textrm{log}$ we will consider the oscillatory part of $\OGW$ given by $A_\textrm{log} \Omega_{\textrm{GW},1}+ A_\textrm{log}^2 \Omega_{\textrm{GW},2}$ as in \eqref{eq:OGW-analyt-expansion}. Here we use the notation from sec.~\ref{sec:resonant-templates-analytical} and denote by $\Omega_{\textrm{GW},1}$ the oscillatory piece with frequency $\omegalog$ and by $\Omega_{\textrm{GW},2}$ the part with frequency $2\omegalog$. We then compute $A_\textrm{log} \Omega_{\textrm{GW},1}+ A_\textrm{log}^2 \Omega_{\textrm{GW},2}$ numerically for two example power spectra of type \eqref{eq:P_of_k_resonant} with $\omegalog = 2.5 \omegalogc$ (fig.~\ref{OGW_1_2_differentAs_2p5wc}) and $\omegalog = 10 \omegalogc$ (fig.~\ref{OGW_1_2_differentAs_10wc}). As we are mainly interested in the oscillatory piece we choose an envelope $\Pbar$ that is constant (with amplitude $\mathcal{P}_\textrm{max}=1$) over a broad range of scales in $k$.\footnote{This is again the top-hat envelope used throughout this work, e.g.~in figs.~\ref{fig:Broad_Narrow_top_LN}, \ref{fig:P_OGW_2p5wc_8}, \ref{fig:P_OGW_7p5wc_24}, \ref{fig:OGW_2p5_7p5_vs_analyt} and \ref{fig:OGW_1_2_2p5_vs_analyt}.} In both figs.~\ref{OGW_1_2_differentAs_2p5wc} and \ref{OGW_1_2_differentAs_10wc} we then plot numerical results for $A_\textrm{log} \Omega_{\textrm{GW},1}+ A_\textrm{log}^2 \Omega_{\textrm{GW},2}$ vs.~$k/\kref$ for the four choices $A_\textrm{log}=1$, $1/2$, $1/4$ and $1/10$. The upper panels display $A_\textrm{log} \Omega_{\textrm{GW},1}+ A_\textrm{log}^2 \Omega_{\textrm{GW},2}$ as computed numerically. In the lower panels we rescaled the plots for the examples with $A_\textrm{log} < 1$ for better visibility. 

We make the following observations. According to our analysis, for $\omegalog = 2.5 \omegalogc$ and $A_\textrm{log}=1$ the oscillatory part of $\OGW$ should exhibit two series of peaks, resulting from $\Omega_{\textrm{GW},1}$ and $\Omega_{\textrm{GW},2}$ contributing with a comparable amplitude, and this is what we see numerically in fig.~\ref{OGW_1_2_differentAs_2p5wc} (red curve). Reducing the value $A_\textrm{log}$, the relative contribution of $\Omega_{\textrm{GW},2}$ is lowered resulting in one series of peaks becoming attenuated until for $A_\textrm{log}=1/10$ (blue curve) the oscillation is well-described by a single sinusoidal modulation with frequency $\omegalog$ (see the lower panel of fig.~\ref{OGW_1_2_differentAs_2p5wc}). For $A_\textrm{log}=1$ this oscillatory pattern could only have been observed for $\omegalog \lesssim \omegalogc$, but here we have made it appear for $\omegalog = 2.5 \omegalogc$ by sufficiently reducing $A_\textrm{log}$. However, this comes at a price. Lowering $A_\textrm{log}$ also leads to a decrease in the overall amplitude of the oscillatory part (see the upper panel of fig.~\ref{OGW_1_2_differentAs_2p5wc}). For example, the factor used for rescaling the result for $A_\textrm{log}=1/10$ from the upper to the lower panel is $\sim 19$. As a result, by lowering $A_\textrm{log}$ the oscillations in $\OGW$ quickly become so small that it remains questionable whether they could ever be detected experimentally. For the example used here the oscillatory part modulates a smooth GW spectrum $\Omega_{\textrm{GW},0} \sim 0.823$ for values $k \sim \kref$ (see e.g.~fig.~\ref{fig:OGW_2p5_7p5_vs_analyt} and the corresponding discussion). For $A_\textrm{log}=1/10$ the amplitude of oscillation is thus only of $\%$-level. 

An even more severe situation arises for the example with $\omegalog = 10 \omegalogc$. For $A_\textrm{log}=1$ our analysis predicts that the oscillation in $\OGW$ should be well-described by a single sinusoidal piece with frequency $2 \omegalog$ and this what is observed in fig.~\ref{OGW_1_2_differentAs_10wc} (red curve). Again, we can change the qualitative behaviour of the periodicity structure by sufficiently lowering $A_\textrm{log}$, and for $A_\textrm{log}=1/10$ (blue curve) we instead observe two distinguishable series of peaks. However, for a larger value of $\omegalog$ the maximal amplitude of the oscillatory piece is already reduced compared to examples with smaller values of $\omegalog$, independently of the value of $A_\textrm{log}$ (e.g.~compare the absolute amplitudes in the upper panels of figs.~\ref{OGW_1_2_differentAs_2p5wc} and \ref{OGW_1_2_differentAs_10wc}). Thus, for this larger value of $\omegalog$ the oscillations very quickly become unobservably small unless $A_\textrm{log} \simeq 1$.

The upshot is that a change in the value of $A_\textrm{log}$ affects the oscillatory piece of $\OGW$ in two ways. Firstly, a lower value of $A_\textrm{log}$ leads to a decrease in the amplitude of oscillations in $\OGW$. Secondly, it modifies the peak pattern by shifting the relative importance of the pieces with frequencies $\omegalog$ and $2 \omegalog$. However, to lead to large qualitative effects, $A_\textrm{log}$ typically needs to be lowered so much as to render the oscillations unobservably small. Thus, in situations of principal interest, i.e.~for visible oscillations in $\OGW$, the qualitative behaviour of the periodicity pattern of $\OGW$ is by and large captured by our analysis for $A_\textrm{log}=1$. 

\bibliographystyle{apsrev4-1}
\bibliography{Biblio-2020}
\end{document}